\documentclass[fleqn,usenatbib]{mnras}

\usepackage{newtxtext,newtxmath}
\usepackage[T1]{fontenc}

\DeclareRobustCommand{\VAN}[3]{#2}
\let\VANthebibliography\thebibliography
\def\thebibliography{\DeclareRobustCommand{\VAN}[3]{##3}\VANthebibliography}

\usepackage{graphicx}
\usepackage{grffile}	% Including figure files
\usepackage{amsmath}	% Advanced maths commands
\usepackage{float}
\usepackage{tablefootnote}
\title[Confusion and identification in CTA data]{Testing source confusion and identification capability in Cherenkov Telescope Array data}

\author[Mestre, Torres, \& de O{\~n}a Wilhelmi]{
Enrique Mestre$^{1,2}$\thanks{E-mail: emestre@ujaen.es},
Diego F. Torres$^{1,3,4}$\thanks{E-mail: dtorres@ice.csic.es},
Emma de O{\~n}a Wilhelmi$^{5}$\thanks{E-mail: emma.de.ona.wilhelmi@desy.de},
Josep Mart{\'i}$^{2}$
\\
$^{1}$Institute of Space Sciences (ICE, CSIC), Campus UAB, Carrer de Can Magrans s/n, 08193 Barcelona, Spain\\
$^{2}$ Departamento de F{\'i}sica (EPSJ), Universidad de Ja{\'e}n, Campus Las Lagunillas s/n, A3, E-23071 Ja{\'e}n, Spain. \\
$^{3}$Institució Catalana de Recerca i Estudis Avançats (ICREA), E-08010 Barcelona, Spain \\
$^{4}$Institut d’Estudis Espacials de Catalunya (IEEC), 08034 Barcelona, Spain\\
$^{5}$Deutsches Elektronen-Synchrotron DESY, Platanenallee 6, 15738 Zeuthen, Germany
}

\date{Accepted 2022 October 6. Received 2022 October 6; in original form 2022 July 27}

\pubyear{2022}

\begin{document}
\label{firstpage}
\pagerange{\pageref{firstpage}--\pageref{lastpage}}
\maketitle

\begin{abstract}
The Cherenkov Telescope Array will provide the deepest survey of the Galactic Plane performed at very-high-energy gamma-rays. Consequently, this survey will unavoidably face the challenge 
of source confusion, i.e., the non-unique attribution of signal to a source due to multiple overlapping sources.
Among the known populations of Galactic gamma-ray sources and given their extension and number, pulsar wind nebulae (PWNe, and PWN TeV halos) will be the most affected. 
We aim to probe source confusion of TeV PWNe in forthcoming CTA data. For this purpose, we performed and analyzed simulations of artificially confused PWNe with CTA. As a basis for our simulations, we applied our study to TeV data collected from the H.E.S.S. Galactic Plane Survey for ten extended and two point-like firmly identified PWNe, probing various configurations of source confusion involving different projected separations, relative orientations, flux levels, and extensions among sources. Source confusion, defined here to appear when the sum of the Gaussian width of two sources is larger than the separation between their centroids, occurred in $\sim$30\% of the simulations. For this sample and $0.5\degr{}$ of average separation between sources, we found that CTA can likely resolve up to 60\% of those confused sources above 500 GeV.
% With an upper limit of 23\% at 95\% CL
% We discuss how this could worsen due to the detection of dimmer sources by CTA and review the limitations of the analysis performed. 
%
Finally, we also considered simulations of isolated extended sources to see how well they could be matched to a library of morphological templates.
The outcome of the simulations indicates a remarkable capability (more than 95\% of the cases studied) to match a simulation with the correct input template in its proper orientation. 
\end{abstract}

\begin{keywords}
instrumentation: detectors -- ISM: supernova remnants
\end{keywords}

%%%%%%%%%%%%%%%%%%%%%%%%%%%%%%%%%%%%%%%%%%%%%%%%%%

%%%%%%%%%%%%%%%%% BODY OF PAPER %%%%%%%%%%%%%%%%%%

\section{Introduction}

The Cherenkov Telescope Array\protect\footnote{\url{https://www.cta-observatory.org}} (CTA, \citealt{2019scta.book.....C}) is the next major ground-based observatory for very-high-energy (VHE, E > 100 GeV) gamma-ray astronomy. CTA is to be located on both hemispheres, a northern location in La Palma (Spain) and a southern one in Paranal (Chile). Each site will host an array of Small-, Medium-, and Large-Size Telescopes (SSTs, MSTs, and LSTs) sensitive to different energy ranges from 20 GeV to more than 300 TeV. SSTs are currently planned to be installed only at the southern site. CTA will improve the sensitivity of current Imaging Atmospheric Cherenkov Telescopes (IACTs) by a factor from five to ten depending on the energy range \citep{2019APh...111...35A}. The first Large-Sized Telescope (LST-1, \citealt{2021arXiv210806005M}), built on-site (currently finalizing the commissioning phase) and located at the Observatorio del Roque de los Muchachos (La Palma), has been operating since October 2018. 

The CTA main goals are categorized in various Key Science Projects (KSPs), including surveys (e.g., the Galactic Plane Survey, GPS) targeting various types of sources.
The CTA GPS \citep{2013APh....43..317D,2019scta.book.....C} will cover the full Galactic plane, especially the inner region (i.e., $|l| < 60\degr{}$), using both the northern and southern arrays at point-source sensitivities of a few mCrab\protect\footnote{The Crab unit is the flux of the source having a spectrum similar to that measured for the Crab nebula and pulsar between 500 GeV and 80 TeV with HEGRA ($d\rm{N}/d\rm{E} = 2.83 \times 10^{-11} (\rm{E}/1 {\rm TeV})^{-2.62}$ cm$^{-2}$ s$^{-1}$ TeV$^{-1}$, \citealt{2004ApJ...614..897A}) above a certain energy. For example, 1 mCrab $= 5.07 \times 10^{-13}$ ph cm$^{-2}$ s$^{-1}$ for an energy threshold of 125 GeV.}. 
Its main goal is to provide a census of Galactic VHE gamma-ray sources, identify promising targets for follow-up observations, and characterize the Galactic plane diffuse emission properties.

The High Energy Spectroscopic System (H.E.S.S., \citealt{2003APh....20..111B}) experiment released the most comprehensive survey of the Galactic plane at VHE gamma-rays up to date (HGPS, \citealt{HESSGPS_paper}). The HGPS comprises about 2700 hours of data (after quality selection) for longitudes from $l = 250\degr{}$ to $65\degr{}$ and latitudes $|b| \leq 3\degr{}$. The resulting HGPS catalog contains 78 VHE sources with 31 firmly identified pulsar wind nebulae (PWNe), supernova remnants (SNRs), composite SNRs, or gamma-ray binaries. The future CTA GPS catalog is expected to achieve the detection at 5$\sigma$ significance of $\sim 500$ VHE sources at energies from 70 GeV to 200 TeV, including more than 200 PWNe \citep{2021arXiv210903729R}. It is more than six times the objects in the HGPS or the third High Altitude Water Cherenkov catalog (3HAWC, with 65 TeV sources detected, \citealt{2020ApJ...905...76A}).

The source confusion problem, i.e., the difficulty in discriminating sources that overlap in crowded regions, is a problem that the CTA GPS will need to address. Particularly given the large extension and number of some gamma-ray source classes and the relatively low angular resolution of IACTs at tens of GeV to TeV energies (i.e., $\gtrsim 0.05\degr{}$).
Initial studies led to an approximate lower limit to the amount of source confusion of $13-24$\% at 100 GeV and $9-18$\% at 1 TeV in the region defined by $|l| < 30 \degr{}$ and $|b| < 2\degr{}$ \citep{2019scta.book.....C}. 
% In the cited study, a position in the sky is defined as confused if there is more than one simulated source within a radius of 1.3 times the CTA's angular resolution.

Our main goal is to probe, through simulations, the source confusion problem in forthcoming CTA data and the instrument's identification capabilities. We are particularly interested in constraining the source confusion regarding the Galactic population of TeV PWNe, which is expected to be the most numerous population of sources. For this purpose, we tested if future CTA data can be directly compared to a library of empirical (e.g., here based on HGPS data) morphological source templates. Furthermore, we aim to probe whether the cross-match with such a library could provide hints to unravel source confusion and, if so, to what extent. In the future, this work  will pave the way to explore the possibility of employing simulated morphological templates instead (from magneto-hydrodynamical; MHD, hydrodynamical; HD, or HD+B results, see e.g.,  \cite{Volpi2008,Kolb2017,Olmi2020}).

%%%%%%%%%%%%%%%%%%%%%%%%%%%%%%%%%%%%%%%%%%%%%%%%%%%
\section{Simulations and methods}
\label{sec:simmeths}
%%%%%%%%%%%%%%%%%%%%%%%%%%%%%%%%%%%%%%%%%%%%%%%%%%%
\subsection{The library of source templates}
\label{sec:srclibrary} 
%%%%%%%%%%%%%%%%%%%%%%%%%%%%%%%%%%%%%%%%%%%%%%%%%%%%%%%%%%%%%%%%%%%%%%%

We first built a library of spatial templates in the form of sky maps, which describe the morphology of different PWNe by depicting the intensity distribution in a region containing each source of interest. We obtained the above mentioned templates from the H.E.S.S. Galactic plane survey \citep{HESSGPS_paper}. 
% Which provides a catalog of VHE sources and Galactic sky maps
The HGPS events maps (see Section 3.1 of the HGPS paper) are built from the reconstructed positions of the primary gamma-ray photons from all events observed by the H.E.S.S. survey covering the region defined by $l = 70\degr{}$ to $250\degr{}$ and $b = \pm5\degr{}$ (with spatial bin size of $0.02\degr{}$). We used slices of these events maps, centered on the different sources of interest, as templates for our simulations. 

Our sources of interest are 12 firmly identified PWNe among the HGPS sources, consisting of ten extended and two compatible with being point-like (HESS J1747-281 and HESS J1818-154) PWNe. 
These sources and their properties are listed in Table \ref{tab:HESS_sources} (see also Tables 3, 10, and 11 in \citealt{HESSGPS_paper}). 
We have restricted the source library to these firmly identified PWNe for two practical reasons.
Firstly, the sample is sufficient in size to simulate a significant amount of different PWNe combined in pairs (i.e., artificially confused) in a limited computational time. 
% i.e., in timescales of a few weeks. 
Secondly, to facilitate the analysis of the performance of the simulations and the interpretation of the results, we prefer to have the input sources as best characterized as possible from the outset. 

\begin{table*}
    \caption{
    Summary of firmly identified extended PWNe among the HGPS sources. The columns respectively list the following characteristics as reported in HGPS: name, related object, galactic longitude ($l$) and latitude ($b$), source extension ($\sigma$) in Gaussian sigmas, square root of the Test Statistic (TS), the parameters of the spectral model, as well as the integral flux over 1 TeV$^{\rm a}$.
    }
    \centering
    \begin{tabular}{p{0.11\linewidth}p{0.09\linewidth}p{0.03\linewidth}p{0.04\linewidth}p{0.08\linewidth}p{0.03\linewidth}p{0.12\linewidth}p{0.03\linewidth}p{0.07\linewidth}p{0.08\linewidth}p{0.1\linewidth}}
    \hline
    Name & Object & $l$ & $b$ & Size ($\sigma$) & $\sqrt{\rm TS}$ & N$_{0}$ $\times 10^{-12}$ & E$_{0}$ & $\Gamma$ & E$_{\rm cutoff}$ & F$_{ > \rm 1 TeV}$ $\times 10^{-12}$ \\
    \hline \\
    & & [deg] & [deg] & [deg] &  & [cm$^{-2}$ s$^{-1}$ TeV$^{-1}$] & [TeV] & & [TeV] & [cm$^{-2}$ s$^{-1}$] \\
    \hline
    \hline
    HESS J0835-455 & Vela X & 263.96 & -3.05 & $0.58 \pm 0.052$ & 39.4 & $6.41 \pm 0.33$ & 1.70 & $1.35 \pm 0.08$ & $12.3\pm 1.7$ & $17.43 \pm 1.40$\\
    HESS J1303-631 & G304.10-0.24 & 304.24 & -0.35 & $0.18 \pm 0.015$ & 54.5 & $7.45 \pm 0.24$ & 0.95 & $2.04\pm 0.06$ & $15.12 \pm 3.2$& $5.21 \pm 0.35$\\
    HESS J1356-645 & G309.92-2.51 & 309.79 & -2.50 & $0.23 \pm 0.020 $ & 17.3 & $0.57 \pm 0.05$ & 2.74 & $2.20\pm 0.08$ & & $4.39 \pm 0.39$\\
    HESS J1418-609 & G313.32+0.13 & 313.24 & 0.14 & $0.11 \pm 0.011 $ & 21.9 & $0.83 \pm 0.05$ & 1.87 & $2.26\pm 0.05$ & & $2.69 \pm 0.15$\\
    HESS J1420-607 & G313.54+0.23 & 313.58 & 0.27 & $0.08 \pm 0.006$ & 27.6 & $0.84 \pm 0.04$ & 1.87 & $2.20\pm 0.05$ & & $2.77 \pm 0.15$\\
    HESS J1514-591 & MSH 15-52 & 320.32 & -1.19 & $0.14 \pm 0.026$ &  42.0 & $7.95\pm 0.31$ & 0.95 & $2.05\pm 0.06$ & $19.20\pm5.0$ & $5.72 \pm 0.42$\\
    HESS J1554-550 & G327.15-1.04 & 327.16 & -1.08 & $0.02 \pm 0.009$ & 9.1 & $0.058 \pm 0.011$ & 2.26 & $2.19\pm 0.17$ &  &  $0.29 \pm 0.06$\\
    HESS J1825-137 & G18.00-0.69 & 17.53 & -0.62 & $0.46 \pm 0.032$ & 76.5 & $69.5 \pm 2.9$ & 0.65 & $2.15\pm 0.06$ & $13.57\pm 3.9$ & $19.15 \pm 1.85$\\
    HESS J1837-069 & G25.24-0.19 & 25.15 & -0.09 & $0.36 \pm 0.031 $ & 41.5 & $20.0 \pm 0.7$ & 0.95 & $2.54\pm 0.04$ & & $11.55 \pm 0.49$\\ 
    HESS J1849-000 & G32.64+0.53 & 32.61 & 0.53 & $0.09 \pm 0.015$ & 9.1 & $0.077 \pm 0.010$ & 2.74 & $1.97\pm0.09$ & & $0.58 \pm 0.07$\\ 
    \hline
    HESS J1747-281 & G0.87+0.08 & $0.87$ & $0.08$ & Point-like & - & $0.84 \pm 0.13$ & $1.0$ & $2.4 \pm 0.11$ &  &  $0.60 \pm 0.13$\\
    HESS J1818-154 & G15.4+0.1 & $15.41$ & $0.16$ & Point-like & 5.6 & $0.11 \pm 0.02$ & $1.54$ & $2.21 \pm 0.15$ &  & $0.7 \pm 0.2$\\
    \hline
    \hline
    \end{tabular}
    \begin{flushleft}
    \footnotesize{ $^{\rm a}$ See Tables 3, 10, and 11 in \citealt{HESSGPS_paper} for further detail.}
    \end{flushleft}
    \label{tab:HESS_sources}
\end{table*}

All source templates depict a square region of $3\degr{}$ side in $151\times 151$ spatial bins. 
% The angular resolution (68\% PSF containment radius) of H.E.S.S. is $\sim 0.08\degr{}$ on average, with approximately 10\% variations for different survey regions \citep{HESSGPS_paper}. 
The bin size of the templates, i.e., a box of size $0.02\degr{}$, is comparable to the best angular resolution predicted for the CTA performance\textsuperscript{\ref{footnote1}}. Other H.E.S.S. sources, not of our interest but lying in the field of view of the templates, were masked and excluded from the simulations.
Finally, to account for various orientations of the sources, we appended ten different rotations (in steps of 36\degr{}) of each template to the library of source templates. 
% The physical distance to a source can also be modified by simply rescaling its template's flux and spatial dimensions. We fixed, however, the distances to the sources for the simulations we present in this paper.
\begin{table}
    \caption{Pearson's correlation coefficient ($C_{i,j}$) computed for each template compared to a rotation of itself.}
    \centering
    \begin{tabular}{clcl}
    \hline
    Name & $\langle C_{i,j}\rangle$ \\
    \hline
    \hline
    HESS J0835-455 & $0.72 \pm 0.03$\\
    HESS J1303-631 & $0.88 \pm 0.01$\\
    HESS J1356-645 & $0.44 \pm 0.03$\\
    HESS J1418-609 & $0.58 \pm 0.02$\\
    HESS J1420-607 & $0.71 \pm 0.01$\\
    HESS J1514-591 & $0.89 \pm 0.02$\\
    HESS J1554-550 & $0.87 \pm 0.02$\\
    HESS J1825-137 & $0.75 \pm 0.06$\\
    HESS J1837-069 & $0.67 \pm 0.03$\\ 
    HESS J1849-000 & $0.77 \pm 0.03$\\ 
    \hline
    \hline
    \end{tabular}
    \label{tab:pears_coeff}
\end{table}

It is interesting to quantify the degree of similarity between the templates and rotations of themselves. The latter gives us an estimation of how well we can describe the source morphology.
For instance, matching particular morphological features of the templates with data must be ruled out if we cannot determine their orientation compared to data.
We chose Pearson's correlation coefficient ($C_{i,j}$), defined by Equation \ref{eq:pearson_coeff}, as a measurement of the morphological auto-correlation degree of the templates. We calculated the mean and deviation of the statistic from nine different rotations of each template in steps of 36\degr{}. Table \ref{tab:pears_coeff} summarizes the results.
The coefficient is computed as:

\begin{equation}
C_{i,j} = \dfrac{\sum\limits_{m}\sum\limits_{n}(Z_{i;\ m,n}-\Bar{Z_{i}}) \times (Z_{j;\ m,n}-\Bar{Z_{j}})}{\sqrt{\left( \sum\limits_{m}\sum\limits_{n}Z_{i;\ m,n}-\Bar{Z_{i}}\right)^{2} \left( \sum\limits_{m}\sum\limits_{n}Z_{j;\ m,n}-\Bar{Z_{j}}\right)^{2}}} 
\label{eq:pearson_coeff}
\end{equation}

Where $i$ and $j$ stand for the i$^{\rm{th}}$ and j$^{\rm{th}}$ rotation of one template, $Z_{m,n}$ indicates spatial bins in image coordinates, for templates with dimensions $M\times N$, and $\Bar{Z_{i}}$ corresponds to the i$^{\rm{th}}$ template's average. Pearson's correlation coefficient of two nearly identical templates approaches $C_{i,j} = 1$. 
%%%%%%%%%%%%%%%%%%%%%%%%%%%%%%%%%%%%%%%%%%%%%%%%%%%%%%

%%%%%%%%%%%%%%%%%%%%%%%%%%%%%%%%%%%%%%%%%%%%%%%%%%%
\subsection{Simulation tools}
\label{sec:simtools}
%%%%%%%%%%%%%%%%%%%%%%%%%%%%%%%%%%%%%%%%%%%%%%%%%%%

The simulations we performed were implemented with {\sc ctools} (version 1.7.4) software package. {\sc ctools}\protect\footnote{http://cta.irap.omp.eu/ctools/about.html} \citep{2016A&A...593A...1K} has been developed for the scientific analysis of data obtained with the existing and future Cherenkov telescopes, such as H.E.S.S., MAGIC, VERITAS, and CTA. It is based on {\sc GammaLib} (consisting of a C++ library and a {\sc python} module, \citealt{2016A&A...593A...1K}), which provides a framework for the analysis of astronomical gamma-ray data.
% Not attached to any specific gamma-ray telescope 

\begin{figure*}
    \centering
    \includegraphics[width=0.498\linewidth]{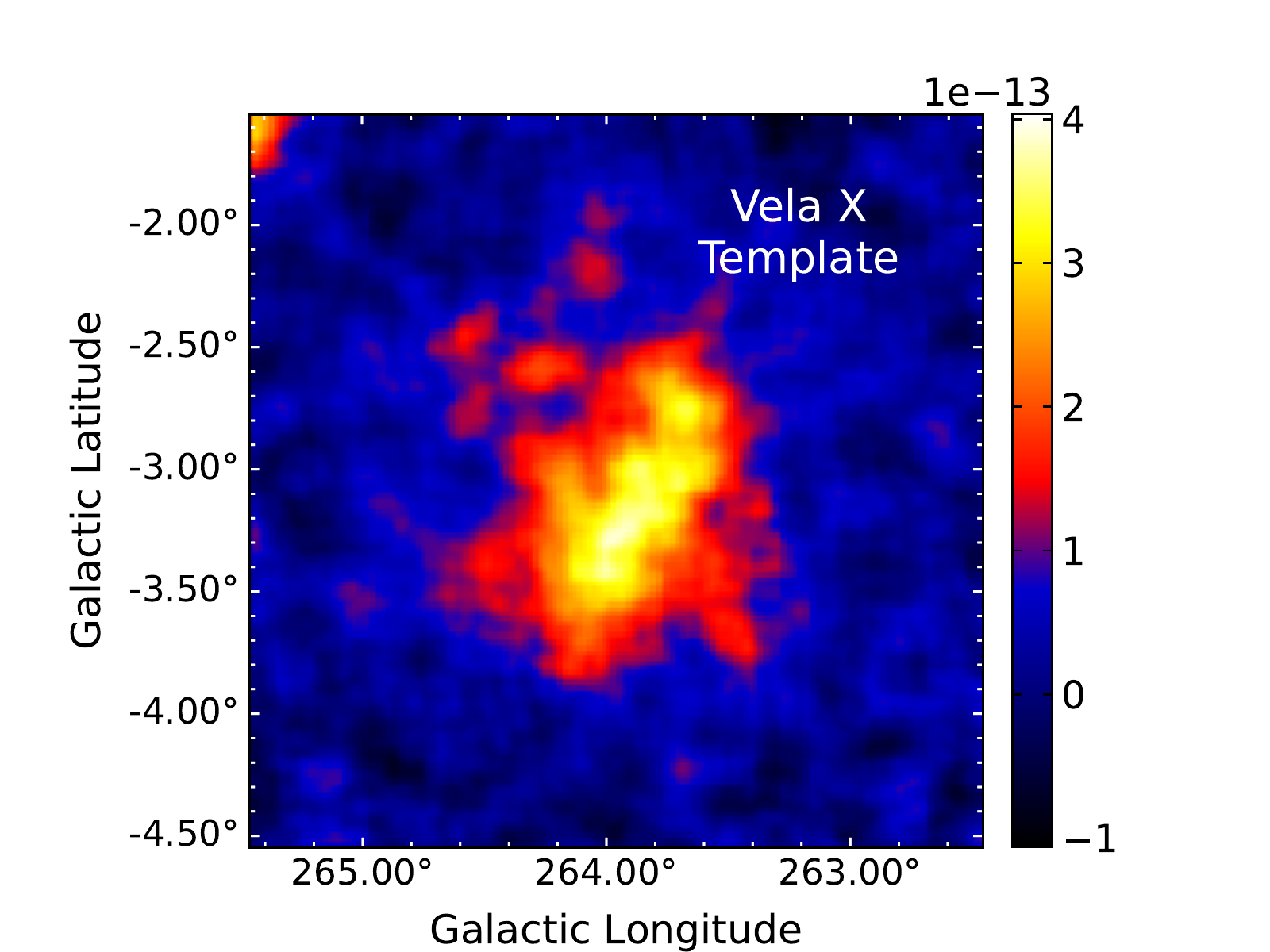}
    \includegraphics[width=0.498\linewidth]{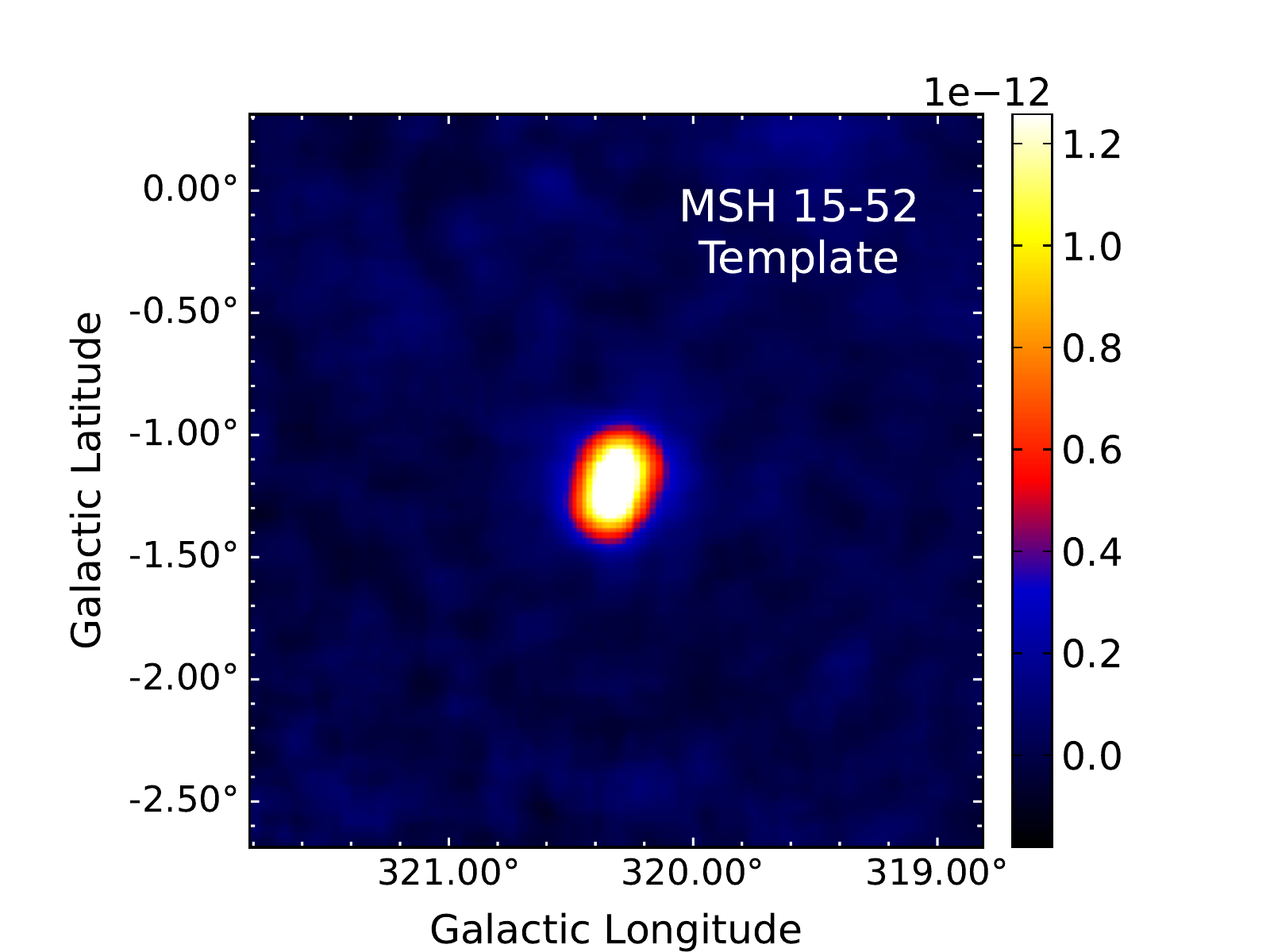} 
    \includegraphics[width=0.498\linewidth]{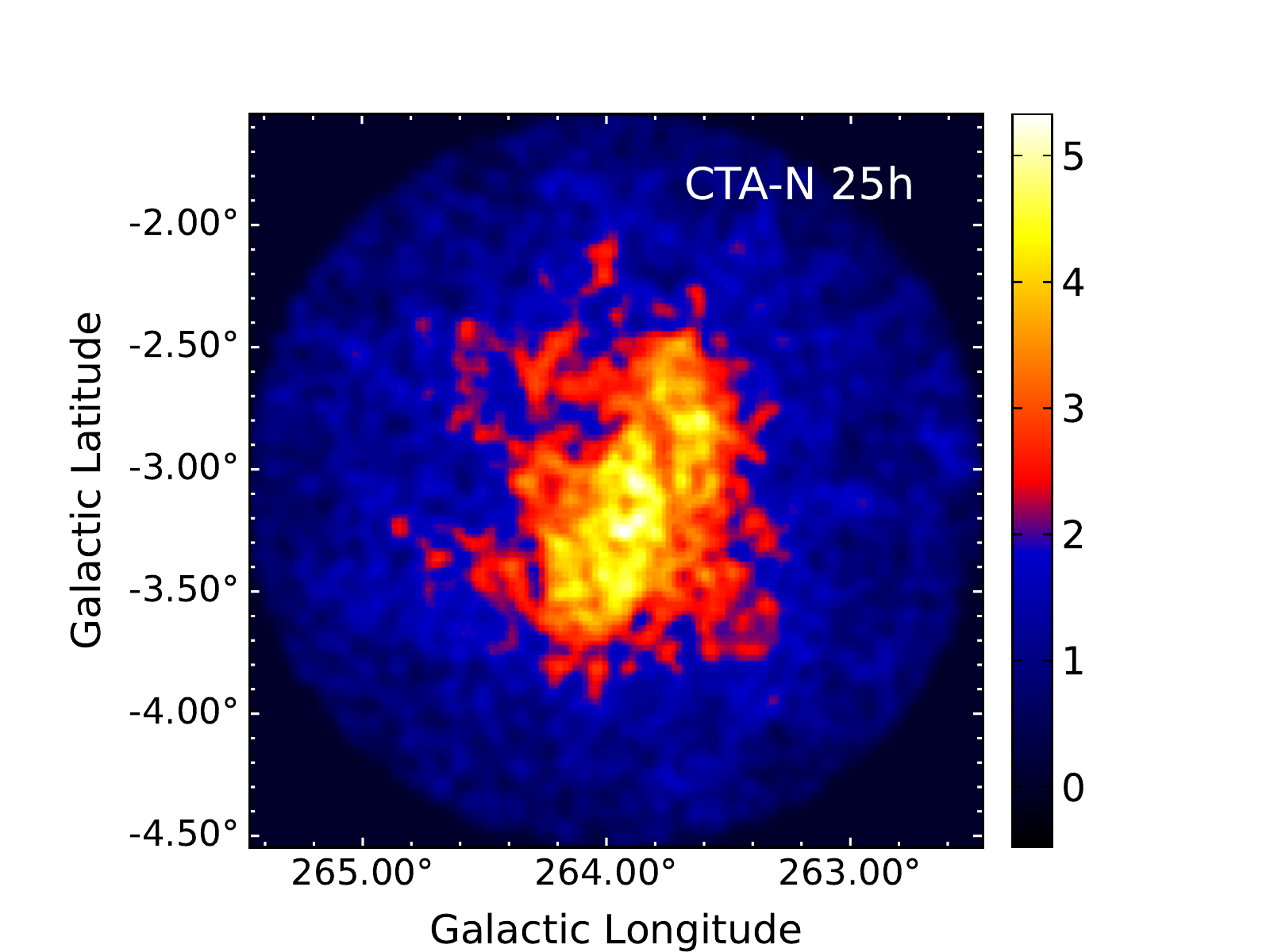}
    \includegraphics[width=0.498\linewidth]{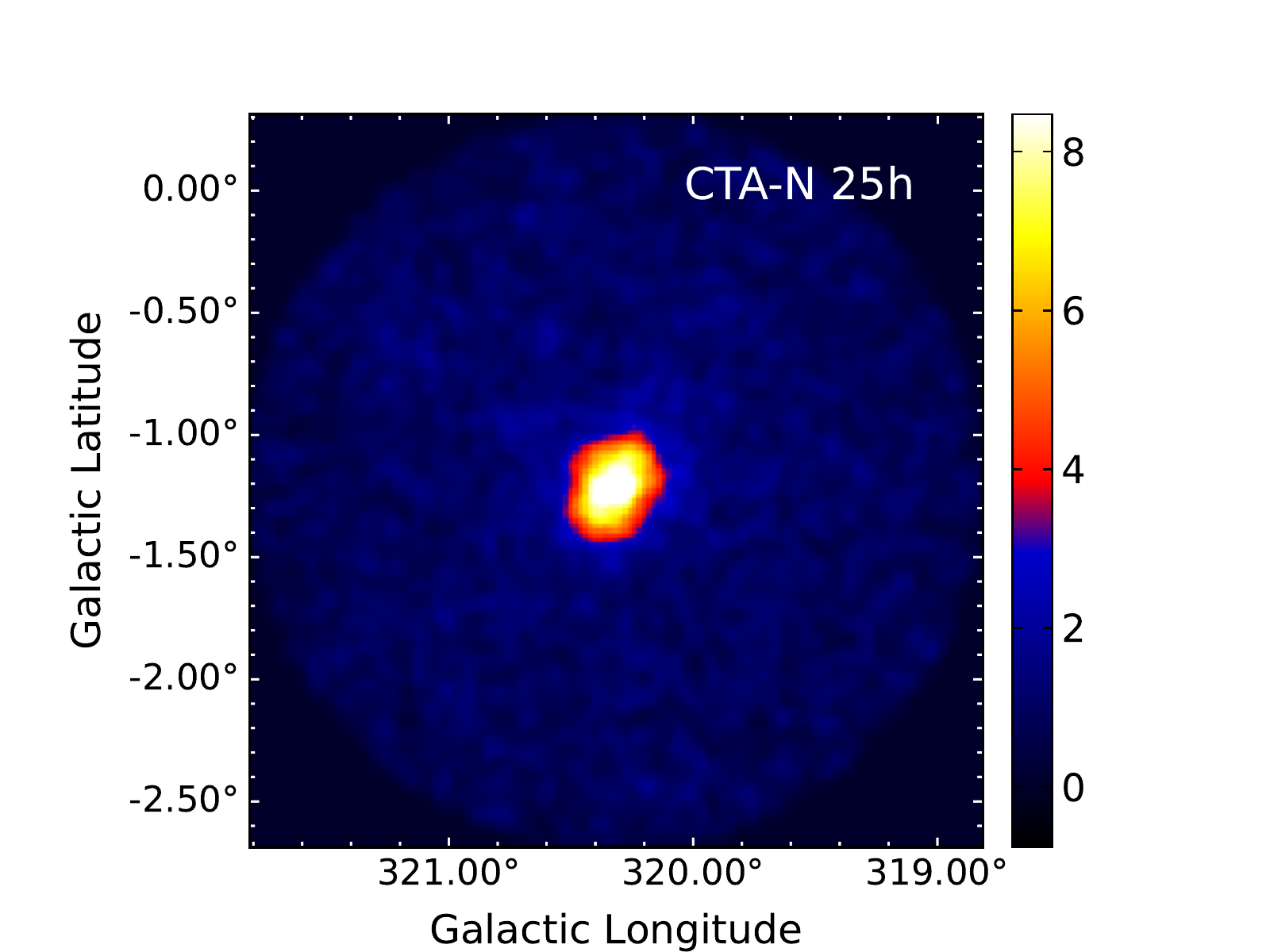}  
    \includegraphics[width=0.498\linewidth]{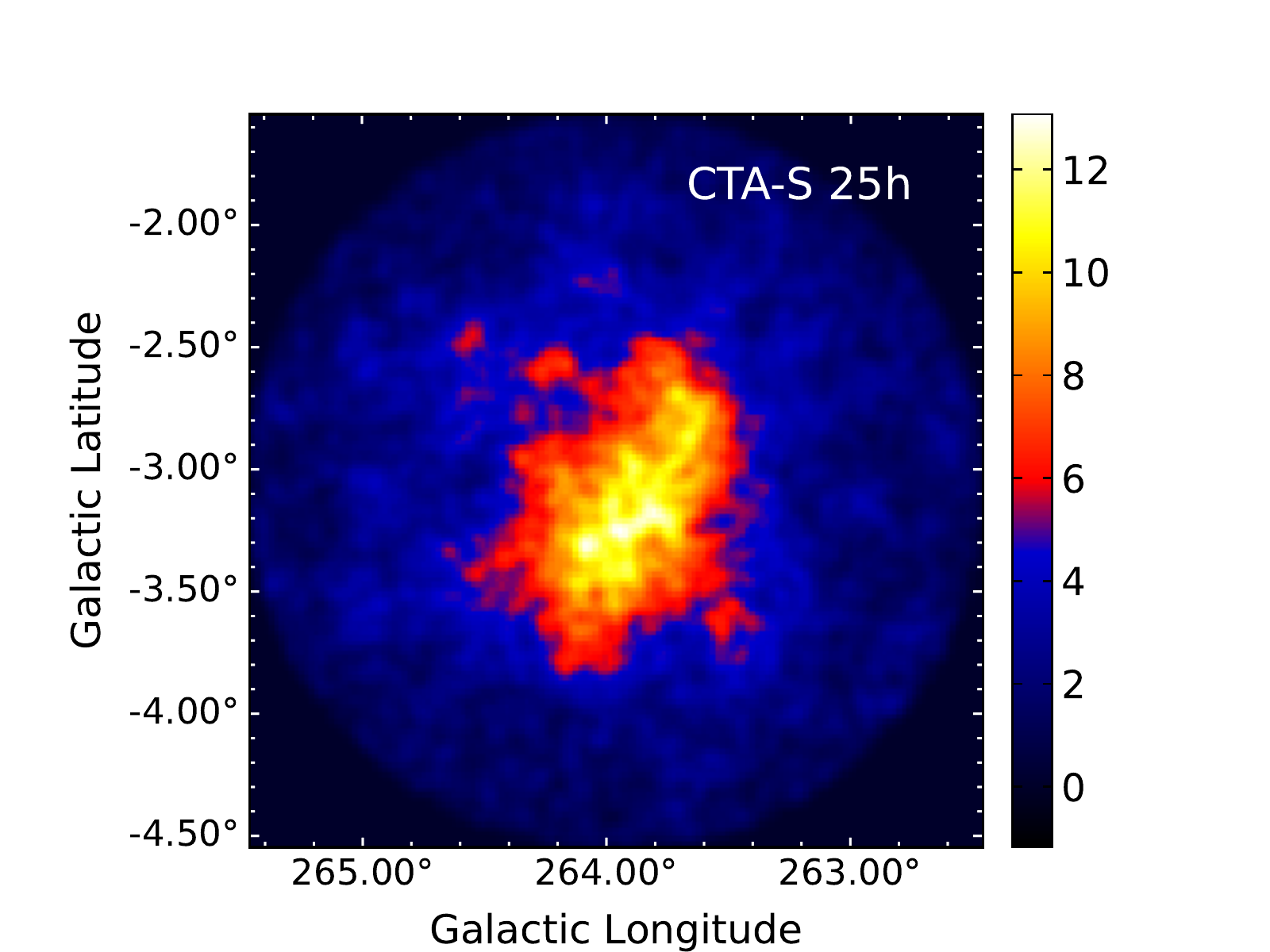}
    \includegraphics[width=0.498\linewidth]{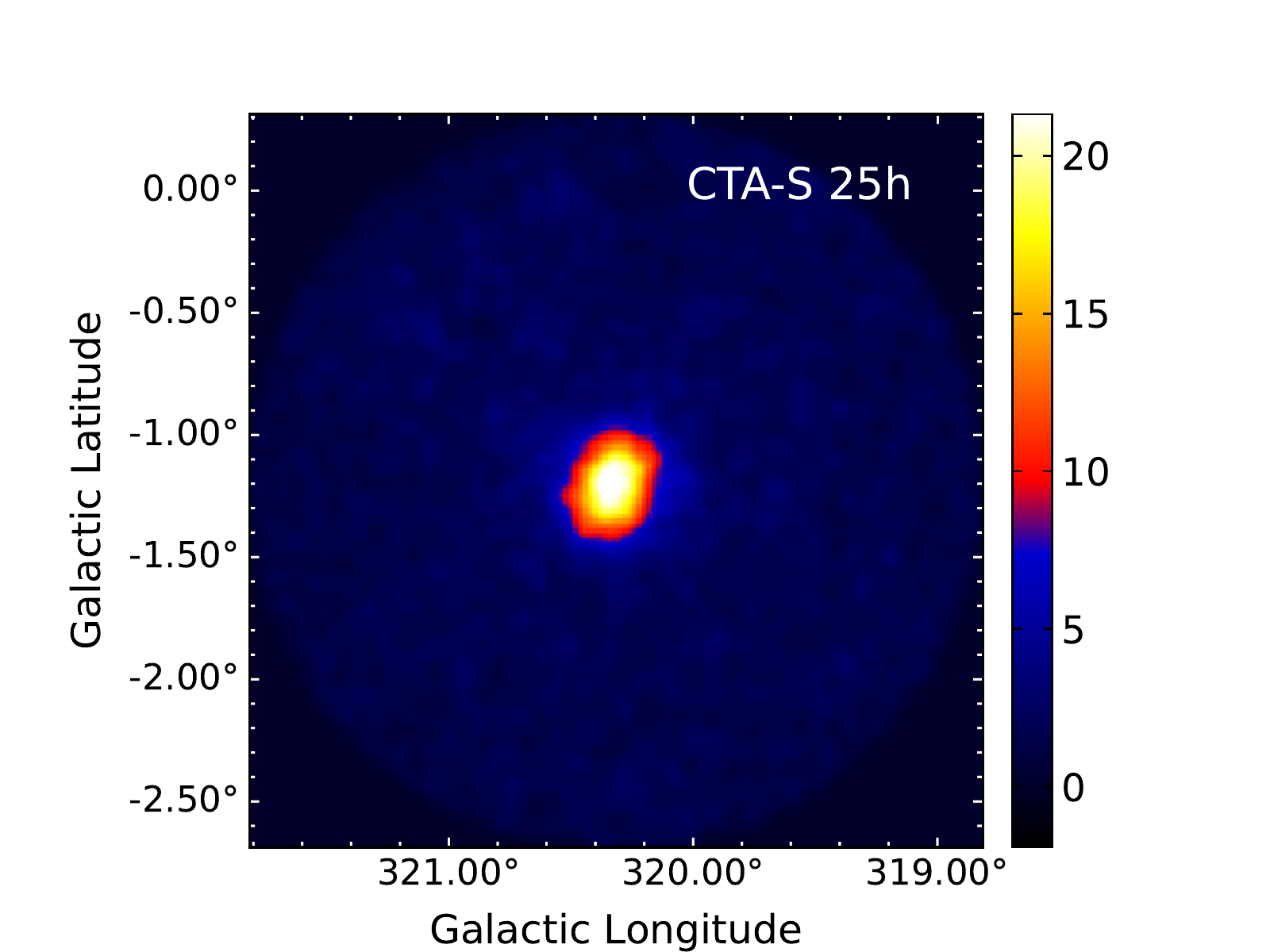}
    \caption{The top panels depict the morphological templates of Vela X (left) and MSH 15-52 (right), obtained from the H.E.S.S. Galactic plane Survey. 
    In the central and lower panels: the sky maps of Vela X and MSH 15-52 for the 25-hour simulated observations with the northern (CTA-N) and southern (CTA-S) arrays of CTA above 0.5 TeV.
    The plots have been smoothed with a Gaussian kernel of $\sigma \approx 0.04\degr{}$. 
    The color bars located at the side of the templates are in units of integral flux (cm$^{-2}$ s$^{-1}$) over 1 TeV (normalized by the factor shown on top of the bar), while those at the side of the simulated CTA sky maps show the number of events (counts) per pixel of $0.02\degr{}$ without background subtraction.} 
    \label{fig:VelaXmaps}
\end{figure*}

The simulations were carried out with the {\sc ctobssim} tool, which creates simulated events lists using the instrument characteristics (specified by the Instrument Response Functions, IRFs) and an input model comprising a list of sources with specific spectral and spatial models. The simulated events lists can be represented in a sky map with the tool {\sc ctskymap} (see, e.g., Figure \ref{fig:VelaXmaps}). 
By default, the sky maps of the simulated observations are shown in celestial coordinates with Cartesian projection. The exposure time of each observation is 25 hours. The simulations were analyzed with the tool {\sc ctlike}, performing an unbinned maximum likelihood fit to the data. 
% The model to fit is also specified with a list of sources described with a given spectral and spatial model.
The following parameters of the fitting model were left free to vary: (1) all parameters of the source power-law ($d{\rm N}/d{\rm E} = {\rm N}_{0} \times ({\rm E}/{\rm E}_{0})^{-\Gamma}$) or exponentially cutoff power-law ($d{\rm N}/d{\rm E} = {\rm N}_{0} \times ({\rm E}/{\rm E}_{0})^{-\Gamma} \times \exp{(-{\rm E}/{\rm E}_{\rm cutoff})}$) spectra, except the reference energy (${\rm E}_{0}$), (2) the norm and tilt of the background spectral model, and (3) the position and width ($\sigma$) of the sources with the spatial model corresponding to a radial Gaussian. The background component, provided by the CTA IRFs, is modeled by a template predicting the background rates as a function of position in the field of view and energy. The model is multiplied by a spectral power-law component, such that the energy distribution of the background is determined by fitting its amplitude and slope (tilt). The spatial model can be also specified using a template sky map containing the source of interest and describing an arbitrary intensity distribution. In the latter case, the spatial model does not include free parameters.

The characteristics of the CTA performance\protect\footnote{\label{footnote1} https://www.cta-observatory.org/science/ctao-performance/}, extensively studied with Monte Carlo (MC) simulations of the CTA instrument \citep{2019APh...111...35A} based on the CORSIKA air shower code \citep{1998cmcc.book.....H} and telescope simulation tool {\sc sim\_telarray} \citep{2008APh....30..149B}, are provided by the Instrument Response Functions (version {\sc prod5} v0.1, \citealt{cherenkov_telescope_array_observatory_2021_5499840}). 
The IRFs were calculated for the planned southern and northern arrays and for different sub-arrays of telescopes that observe an object at three zenith angles (i.e., $20\degr{}$, $40\degr{}$, and $60\degr{}$). Different analysis cuts are applied for each IRF, considering observation times of 0.5, 5, and 50 hours.
The {\sc prod5} v0.1 version of the IRFs we employed assumes the CTA arrays in the dubbed Alpha configuration, i.e., accounting for 4 LSTs and 9 MSTs in the northern array and 14 MSTs and 37 SSTs in the southern one (spread over an area approximately of $0.25$ and $3$ km$^{2}$, respectively). We referred the results to the northern and southern arrays at $20\degr{}$ zenith angle with the IRFs referenced to 50\,h.

%%%%%%%%%%%%%%%%%%%%%%%%%%%%%%%%%%%%%%%%%%%%%%%%%%%%%%%%%%%%%%%%%%%%%%%
\subsection{Simulating CTA observations}
\label{sec:obssimuls}
%%%%%%%%%%%%%%%%%%%%%%%%%%%%%%%%%%%%%%%%%%%%%%%%%%%%%%

We simulated a 25-hour observation of each source, isolated and in different orientations, with the northern and southern arrays of CTA.
We consider 25 h to be a reasonable estimate of the average observation time for the selected targets during the CTA GPS since 1620 h of total observation time (1020 h with the southern array and 600 h with the northern one) are requested for the GPS during a ten-year programme \citep[see][]{2019scta.book.....C}. In comparison, the average livetime for the H.E.S.S. observations of the PWNe listed in Table \ref{tab:HESS_sources} is $\sim 50$ h, comprising the HGPS 2864 h of total observation time \citep{HESSGPS_paper}.
We performed all simulations with an energy threshold of 500 GeV (corresponding approximately to those of the templates, see Table 11 in \citealt{HESSGPS_paper}), taking the input positions and spectral models of the sources from the HGPS source catalog. 
% The source confusion problem will be prominent at lower energies due to the poorer angular resolution of IACTs at tens of GeV. The angular resolution predicted for CTA at $100$ GeV is $\sim 0.15\degr{}$, approximately a factor two larger than that predicted at $500$ GeV.
Figure \ref{fig:VelaXmaps}, e.g., shows some simulations corresponding to Vela X and MSH 15-52 in their unaltered orientation compared to the templates of the sources (in the top panels). 

\begin{table}
\caption{The square root of the TS resulting from fitting to the CTA simulated observations of MSH 15-52 and Vela X (artificially confused): (1) The Vela X template morphology (top left panel in Fig. \ref{fig:VelaXmaps}). (2) A Gaussian source model. (3) A source with the morphology of the MSH 15-52 and Vela X confused templates (see Fig. \ref{fig:confused_templates}). The hypotheses listed correspond to: ${\rm H}_{\rm Temp. 1,\alpha\ =\ 0\degr{}}$, ${\rm H}_{\rm Gauss}$, and ${\rm H}_{\rm Conf.}$ in Eq. \ref{eq:deltaTS}, respectively.}
\centering
%\scriptsize
\footnotesize
\begin{tabular}{llllll}
\hline
Parameter & CTA Array & Vela-X & Gaussian & Two sources\\
\hline
\hline
\\
{\it $0.52$\degr{} separation}\\
{\it $36$\degr{} orientation}\\
\\
$\sqrt{\rm TS}$ & North &  $75.5$ & $81.0$  & $90.9$\\
$\sqrt{\rm TS}$ & South &  $121.0$ & $127.3$ & $139.7$\\
\hline
\\
{\it $0.54$\degr{} separation} \\
{\it $180$\degr{} orientation}\\
\\
$\sqrt{\rm TS}$ & North &  $72.5$ & $76.8$ & $83.7$\\
$\sqrt{\rm TS}$ & South &  $116.2$ & $122.7$ & $131.4$ \\
\hline
\hline
\end{tabular}
\label{tab:velaXmsh1552}
\end{table}

\begin{table*}
\caption{The best-fitting parameters to the CTA 25-hour simulations of Vela X (central and bottom left panels of Fig. \ref{fig:VelaXmaps}) compared to the source's input model.}
\centering
%\scriptsize
\footnotesize
\begin{tabular}{lllllll}
\hline
Parameter & Fixed?  &  Model & CTA-N result & CTA-S result\\
\hline
\hline
\\
{\it Template fit:}\\
\\
$\sqrt{\rm TS}$ (Vela X template) &  &  & $102.3$ & $167.1$ \\
\\
{\it Spectrum results}\\
\\
$\rm{N}_{0} [10^{-12}\rm{cm}^{-2}\rm{s}^{-1}\rm{TeV}^{-1}]$ & No  & $6.41$ & $6.42 \pm 0.12$ & $6.33 \pm 0.08$ \\
$\rm{E}_{0} [\rm{TeV}]$ & Yes & $1.7$ & $1.7$ & $1.7$\\
Index & No & $1.35$ & $1.37 \pm 0.03$ & $1.33 \pm 0.02$ \\
$\rm{E}_{cutoff} [\rm{TeV}]$ & No & $12.3$ & $12.5 \pm 0.6$ & $12.1 \pm 0.3$ \\
\hline
\\
{\it Gaussian fit}\\
\\
Ra & No & $128.887$ & $128.812 \pm 0.011$ & $128.831 \pm 0.006$\\
Dec & No & $-45.659$ & $-45.662 \pm 0.008$ &  $-45.662 \pm 0.004$\\
Size (Gaussian) & No & $0.58$ & $0.542 \pm 0.006$ & $0.541 \pm 0.004$ \\
$\sqrt{\rm TS}$ (Gaussian) &  &  & $95.5$ & $152.9$\\
\hline
\\
%TS (Template 2) &  &  & & \\
%TS (Template 3) &  &  & & \\
% Size and TS taken from the table of HESS galaxy survey paper size = $0.580 \pm 0.052$  TS = $1552.3$.
\hline
\hline
\end{tabular}
\label{tab:velaX}
\end{table*}

The sources were next artificially confused in pairs, resulting in 66 (=12$\times$11/2) possible pairings for the chosen sample, not accounting for different projected separations or rotations. Firstly, we place two templates in the same position, with one of them taken as a reference. We relocate a template by simply modifying its central bin's reference position (preserving the bin size). Next, we introduce between the sources a random separation retrieved from a Gaussian distribution centered in 0.5$\degr{}$ with $0.25\degr{}$ of full width at half maximum (FWHM).

We obtained the mean separation between the sources (i.e., $d_{\rm PWN} \approx 0.5\degr{}$) from an estimation of the projected density of PWNe ($\rho_{\rm PWN}$) in a central region of the Galaxy defined by $|l| < 30\degr{}$ and $|b| < 0.5\degr{}$. We considered; $\rho_{\rm PWN} = \rm{N}_{\rm PWNe}/\rm{A} = 1/(\pi d_{\rm PWN}^{2}/4)$, where A corresponds to the area (i.e., $60$ deg$^{2}$) and N$_{\rm PWNe}$ to the number of sources therein. The latter was computed from a Galactic source distribution model (\citealt{2011MmSAI..82..726R,2022MNRAS.511.1439F}) that has been used in previous works to evaluate the source confusion problem in the CTA Galactic survey. We  obtained from the same $\rm{N}_{\rm PWNe} \sim 188$ (or $\rho_{\rm PWN} = 3.1$ sources per square degree in the cited region) at a sensitivity level of 3 mCrab \citep[i.e., $5.24 \times 10^{-14}$ cm$^{-2}$ s$^{-1}$ of flux above 1 TeV,][]{2013APh....43..317D}. 

In a former, limited study of confusion, see \cite{2019scta.book.....C}, the exercise evaluated whether two sources were positionally coincident within a certain radius defined through the CTA PSF. In particular, the definition adopted posed 
that a position in the sky was confused if there was more than one simulated source within a radius of 1.3 times the CTA's angular resolution. 
In these studies, the sources were taken from different extrapolations of the total source count (as a function of flux, i.e., $\log N - \log S$ diagram), sizes, and spectral indices distributions that were assumed to be consistent with existing data. A specific spatial distribution of sources around the Galactic Centre was assumed, with no diffuse emission except for the Galactic Centre ridge (see Section 6.4.2 of the cited work). The main limitations to consider were the unknown shapes of the sources, the undetermined level of diffuse emission, the high source density in the inner Galaxy, and the dependency of source identification on the analysis methods.

Unlike in \cite{2019scta.book.....C}, we established the following more general criterion: two sources are strictly confused when $\sigma_{1} + \sigma_{2} > d$, where $\sigma_{1}$ and $\sigma_{2}$ are the Gaussian widths of the sources (see Table \ref{tab:HESS_sources}) and $d$ the projected separation between their centroids. We set $\sigma = 0.02\degr{}$ for the point-like sources. CTA's predicted angular resolution (68\% PSF containment radius) at a few TeVs is smaller than $0.05\degr{}$ for both the northern and southern arrays. It is then considerably smaller than the average projected separation of the simulations performed, i.e., $0.5\degr{}$, and small compared to the Gaussian size of most sources in the library. For this reason, we can neglect the CTA PSF in the confusion criterion above. In contrast to the initial studies on source confusion cited in \cite{2019scta.book.....C}, the CTA's identification capabilities in our simulations are then limited by noise rather than by the instrument's PSF. 
% However, we must consider it when performing simulations at smaller projected separations, particularly in the limit of point-like sources. It could be the case, e.g., for the Galactic center's closest regions or in case significantly smaller sensitivities are achieved.

\begin{figure*}
    \centering
     \includegraphics[width=0.498\linewidth]{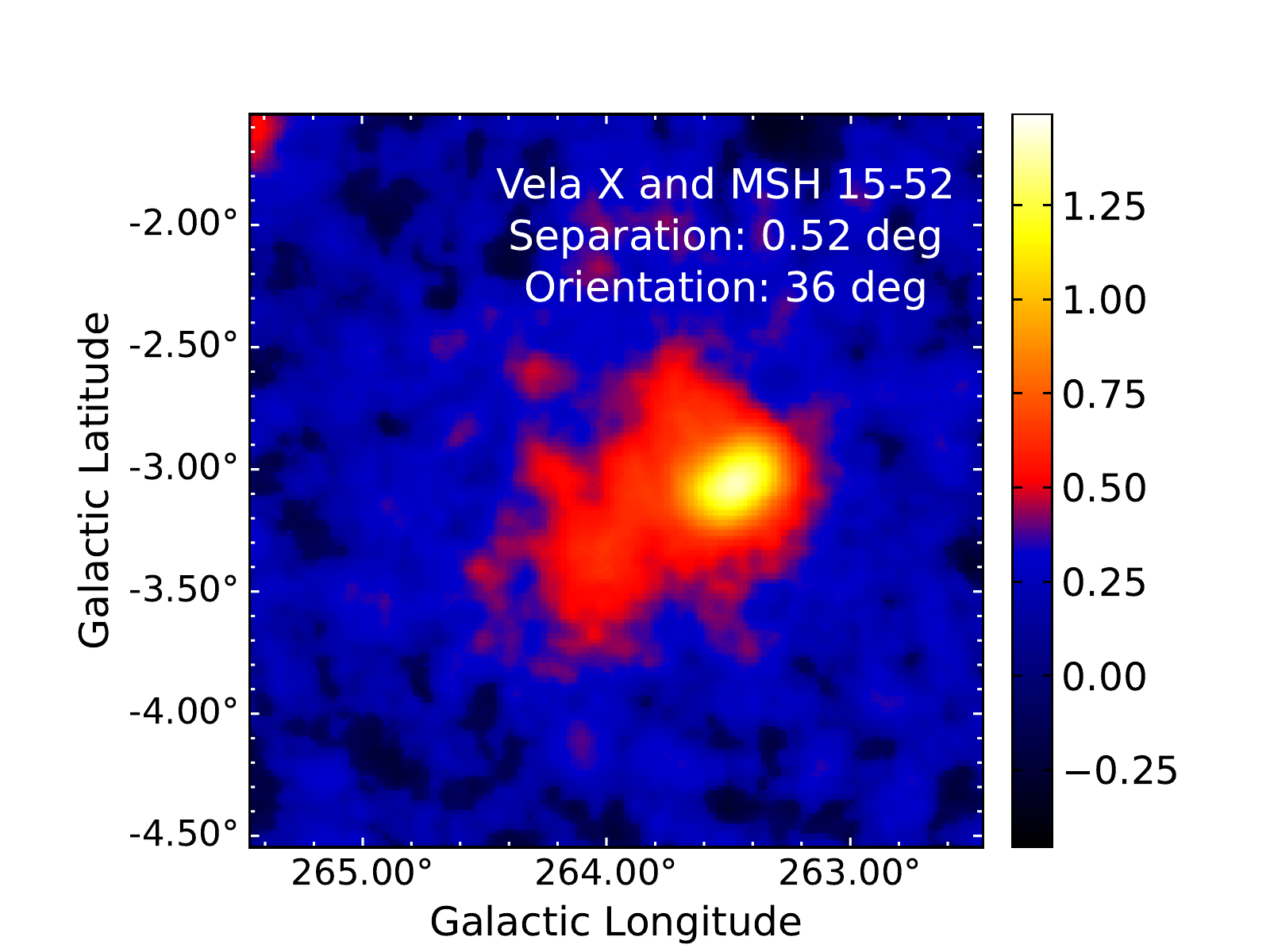} 
     \includegraphics[width=0.498\linewidth]{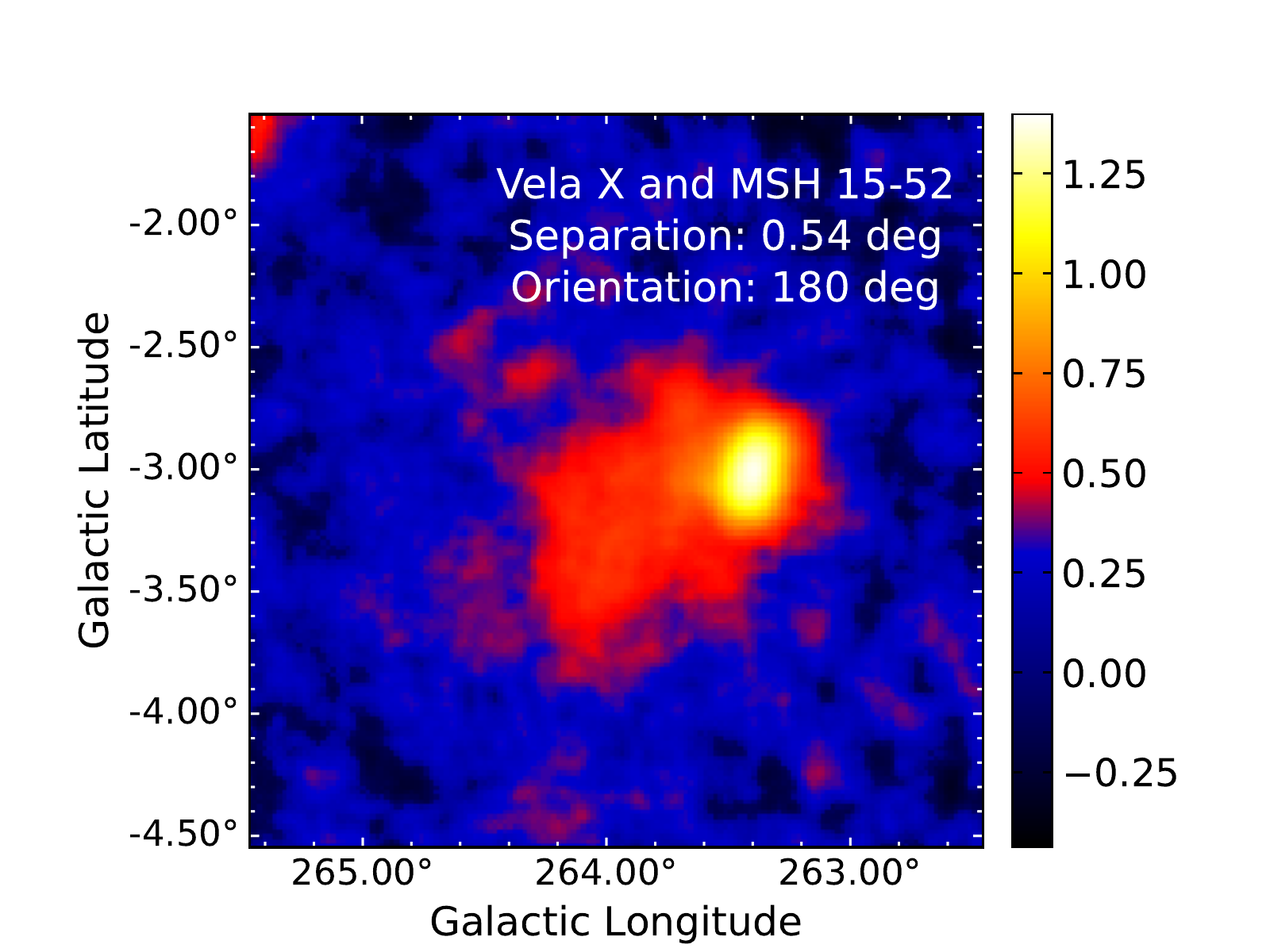} \\
     \includegraphics[width=0.498\linewidth]{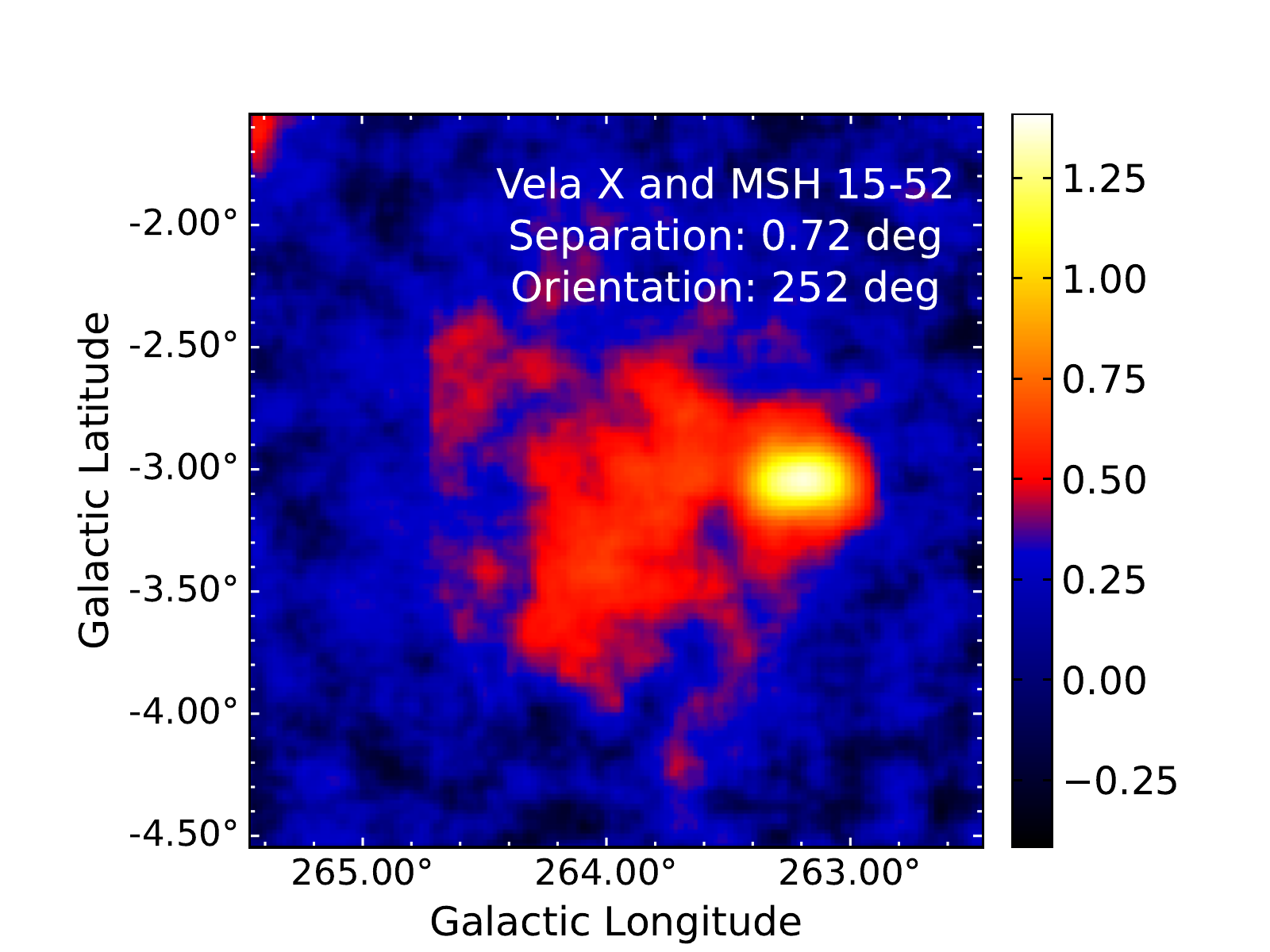}
    \caption{The morphological templates of Vela X and MSH 15-52 artificially confused, i.e., separated by $0.52\degr{}$ (top left), $0.54\degr{}$ (top right), and $0.72\degr{}$ (bottom) with MSH 15-52 rotated clockwise by $36\degr{}$, $180\degr{}$, and $252\degr{}$, respectively, compared to the orientation from H.E.S.S. data. The color bars at the side of the templates are in units of the square root of the integral flux over 1 TeV ($\times 10^{6}$ cm$^{-2}$ s$^{-1}$), preserving the sign in each bin. We applied this transformation to the addition of the templates to better visualize the contribution of both sources. The plots have been smoothed with a small Gaussian kernel ($\sigma \sim 0.04\degr{}$).}
    \label{fig:confused_templates}
\end{figure*}

Finally, we rotate clockwise the template previously shifted by an angle corresponding to a random multiple of $36\degr{}$ (including $360\degr{}$) to explore different relative orientations among the templates. Figure \ref{fig:confused_templates} depicts three simulations of Vela X and MSH 15-52 artificially confused with random separations taken from the probability distribution discussed above and different orientations of MSH 15-52 as an example. In the lower panel, the sources are not strictly confused according to the separation imposed ($0.72\degr{}$). The panels are centered at the position of the Vela X template.

To restrict the computational time, we limited the simulations to two different configurations for each possible pair of confused PWNe out of the source library, i.e., each configuration consisting of an arbitrary projected separation and relative orientation.
The 66 possible pairings we have, with two CTA arrays and using two different configurations for each pair of confused sources, resulted in 264 simulations.
% Of the type shown in Figure \ref{fig:confused_templates}
The total census of templates involved in these simulations amounts to 252 templates, i.e., 120 templates depicting one source (twelve source templates in ten different orientations) and 132 templates of artificially confused sources (66 pairings in two different configurations each).

%%%%%%%%%%%%%%%%%%%%%%%%%%%%%%%%%%%%%%%%%%%%%%%%%%%%%%%%%%%%%%%%%%
\subsection{Analysis of the simulations}
\label{sec:simulanals} 
%%%%%%%%%%%%%%%%%%%%%%%%%%%%%%%%%%%%%%%%%%%%%%%%%%%%%%%%%%%%%%%%%%

We fitted the different source templates (i.e., considering the various orientations) and a Gaussian source to the simulations of the sources as isolated.
We dubbed the cited hypotheses ${\rm H}_{\rm Temp. i, \alpha}$ --where the first subindex identifies the template and $\alpha$ the rotation angle-- and ${\rm H}_{\rm Gauss}$, respectively.
% The parameters regarding the source spectral and background models (the latter taken from the IRFs) are free, together with the source position and Gaussian width in the Gaussian source hypothesis.
The source detection significance (in Gaussian $\sigma$) was approximated as the square root of the Test Statistic ($\sqrt{\rm TS}$). The Test Statistic is defined from the maximum log-likelihood value obtained when fitting the source (together with the background) to the data (${\rm ln}\ L$) and the same if only fitting the background model (${\rm ln}\ L_{0}$), as ${\rm TS} = 2 \times{\rm ln}(L/L_{0})$. The TS was then used to compare the goodness of fit among the different hypotheses employed to model each observation simulation.

To analyze the simulations of confused sources performed, we considered different hypotheses. Firstly, we fitted the following models to each simulation:

\begin{enumerate}
    \item A source described by a template resulting from two confused sources (${\rm H}_{\rm Conf.}$) and exponentially cutoff power-law spectrum. This hypothesis accounts for all pairings of PWNe from the library in the two configurations considered, i.e., ${\rm H}_{\rm Conf.}$ comprises all 132 templates used for simulating artificially confused sources (see Section \ref{sec:obssimuls}). Conceptually, we cannot interpret the latter as a proper 3D fit to the simulated data since the sources have different spectra and each confused template fitted to the simulations has a unique spectrum. The ${\rm H}_{\rm Conf.}$ hypothesis, however, can still represent quantitatively well a given observation simulation.
    \item A source with a radial Gaussian spatial model and exponentially cutoff power-law spectrum (i.e., ${\rm H}_{\rm Gauss}$).
    \item The source templates in the library with all their corresponding rotations (${\rm H}_{\rm Temp. i, \alpha}$). We dubbed the particular cases in which the fitting template is one of those involved in the simulation as ${\rm H}_{\rm Temp. 1, \alpha}$ and ${\rm H}_{\rm Temp. 2, \alpha}$. We fitted the one-source templates considering the spectral shapes listed in Table \ref{tab:HESS_sources}. 
    %\item A source described by each of the individual sources used in the confused template, with their source templates rotated clockwise by an angle $\alpha$, where $\alpha$ ranges from $0\degr{}$ to $360\degr{}$ in steps of $36\degr{}$ (${\rm H}_{\rm Temp. 1, \alpha}$ and ${\rm H}_{\rm Temp. 2, \alpha}$).
    %\item The rest of the source templates in the library with all their corresponding rotations (${\rm H}_{\rm Temp. i, \alpha}$, where the first subindex identifies the template and $\alpha$ the rotation angle).
\end{enumerate}

None of the cited hypotheses, fitted to a given simulation of confused sources, can reproduce the input model. 
However, they can represent (or fit) well the simulated observation. This hence allows us to probe the probability for two confused sources to be well-described by only one source.
% We are only interested in probing how well the morphological models containing only one source template describe the simulations, compared to the actual combination of templates behind the same (see e.g., Table \ref{tab:velaXmsh1552}).

We next compared the goodness-of-fit corresponding to the different hypotheses with the value of the Test Statistic (TS). Note that 
\begin{equation}
{\rm TS}_{\rm{H}_{i}} - {\rm TS}_{\rm{H}_{j}} = 2 \times \ln(L_{i}/L_{j}),
\label{eq:TSdif}
\end{equation}
where $\ln(L_{i})$ is the maximum log-likelihood value corresponding to the hypothesis H$_{i}$. Therefore, if we define:

\begin{equation}
\begin{split}
        \Delta {\rm TS} = \max({\rm TS}_{\rm Conf.}) - \max({\rm TS}_{\rm Gauss}, {\rm TS}_{\rm Temp.\ 1, \alpha}, \\
        {\rm TS}_{\rm Temp.\ 2, \alpha}, {\rm TS}_{\rm Temp. i, \alpha})
        % {\rm TS}_{\rm Temp.\ 1}, {\rm TS}_{\rm Temp.\ 2}
\end{split}
    \label{eq:deltaTS}
\end{equation}
we can conclude that a positive and large $\Delta {\rm TS}$ for a given observational dataset translates into better prospects for resolving the two (confused) sources.
For example, the two cases shown in Table \ref{tab:velaXmsh1552}, satisfy $\Delta {\rm TS} > 0$ for both the CTA northern and southern arrays.
In this case, the MSH 15-52 template (i.e., ${\rm H}_{\rm Temp. 2, \alpha}$ hypothesis in Eq. \ref{eq:deltaTS}), could not explain the simulated data; ${\rm TS}_{\rm Temp. 2, \alpha} \approx 0$.
Alternatively, we can compare the different hypotheses fitted to the simulated data using Akaike's Information Criterion (AIC, \citealt{Akaike1973InformationTA,1100705}) for not nested models. However, we reached the same conclusions after analyzing the simulations as if using Equation \ref{eq:deltaTS} (see Appendix \ref{AppendixA} for a detailed comparison of AIC and TS statistics).

When $\Delta {\rm TS} > 0$ happens, we try to decompose the best-fitting template corresponding to the hypothesis ${\rm H}_{\rm Conf.}$ in the two templates taken from the library, assigning each one to a different source. 
For this purpose, we perform the unbinned maximum-likelihood fit to the given simulation considering the two sources cross-matched with the templates library, each with its corresponding spectrum and source template as the spatial model (plus the background, $\rm{H}_{\rm 2src}$). When fitting the latter hypothesis, we can indeed retrieve the simulated model provided a correct identification of the sources from the cited cross-match. 
As mentioned in Section \ref{sec:simtools}, the free parameters of the fit are the ones of the spectral models (except the reference energies) and those of the background. The TS of the i$^{\rm{th}}$ source (where $\rm{i} \in \{ 1,2 \}$) is designated as TS$_{\rm 2src,i}$. Note that one of the sources has been shifted and rotated. Hence, the position and orientation used to model the latter source are inherited (fixed) from the best-fitting template under the ${\rm H}_{\rm Conf.}$ hypothesis.

%%%%%%%%%%%%%%%%%%%%%%%%%%%%%%%%%%%%%%%%%%%%%%%%%%
\section{Results}
\label{sec:results}
%%%%%%%%%%%%%%%%%%%%%%%%%%%%%%%%%%%%%%%%%%%%%%%%%%

%%%%%%%%%%%%%%%%%%%%%%%%%%%%%%%%%%%%%%%%%%%%%%%%%%
\subsection{Simulations of isolated sources}
\label{subsec:isolated}
%%%%%%%%%%%%%%%%%%%%%%%%%%%%%%%%%%%%%%%%%%%%%%%%%%

We analyzed the simulations of the sources listed in Table \ref{tab:HESS_sources} (see Section \ref{sec:obssimuls}), assuming that the sources are isolated and have different arbitrary orientations.
For this purpose, we fitted all hypotheses (see Section \ref{sec:simulanals}) regarding an isolated source to each simulation. 
As an example, Table \ref{tab:velaX} summarizes the best-fitting model for the simulations of Vela X depicted in Figure \ref{fig:VelaXmaps}. The results of fitting the Vela X template morphology (left panel in the cited figure) and a Gaussian model to the simulated data are summarized in the top and bottom parts of the table, respectively. We can conclude that the fits successfully recover the source's characteristics.

The outcome of the simulations indicates a remarkable capability to match a simulation with both the correct input template and its proper orientation. 
In more than 95\% of the simulations, the best-fitted model (with maximum TS) identified the input template with the correct orientation.
In addition, in more than 90\% of cases the best-fitting template model qualitatively improved a Gaussian source fit: $\rm{TS}_{\rm Template} - \rm{TS}_{\rm Gauss} \gtrsim 25$. 
Furthermore, more than 80\% of the simulations of the faintest and less extended sources in the library (i.e., $\rm{F}_{\rm >1TeV} < 10^{-12}$ cm$^{2}$ s$^{-1}$ and $\sigma < 0.1\degr{}$) are correctly matched with the input template and orientation. These simulations are also seemingly best represented by the template compared to a Gaussian source.
The left panel of Figure \ref{fig:angldeg} depicts some examples of how the detection significance varies with the orientation of the template fitted to a simulation. 
% Provided a correct identification of the isolated source. 
None of the simulations concerning an isolated nebula resulted best represented by a confused template compared to the best-fitting hypothesis regarding one source, i.e., Equation \ref{eq:deltaTS} satisfied $\Delta \rm{TS} < 0$ for all simulations of isolated sources.

\subsection{Simulations of confused sources}
\label{subsec:confused}

\begin{figure*}
\centering
    \includegraphics[width=0.45\linewidth]{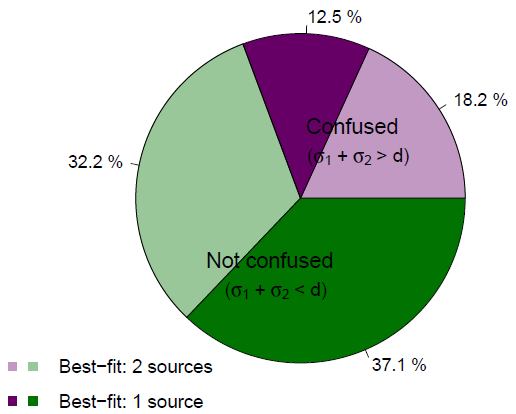}
    \includegraphics[width=0.47\linewidth]{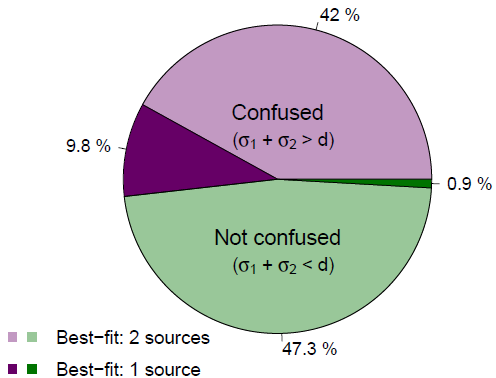}
    \caption{Distribution of the simulations of two sources at a certain distance according to whether the sources simulated are actually confused ($\sigma_{1} + \sigma_{2} > d$, in purple) or not (in green) and to whether the best-fitting model assumes one source (in dark tonalities) or two (in light tonalities). The right panel excludes the simulations involving the two dimmest extended sources (HESS J1554-550 and HESS J1849-000) and those point-like.}
    \label{fig:piecharts}
    %srcdistr.R
\end{figure*}

Figure \ref{fig:piecharts} summarizes the results. The ratios are computed from the 264 simulations with the 12 PWNe listed in Table \ref{tab:HESS_sources}, with an uncertainty estimated of $\sim$ [1 - 2]\%. We calculated the latter error through bootstrapping with the method further explained in Section \ref{subsec:dimsources} and applied in Figure \ref{fig:fracresfracdim}. Note that Table \ref{tab:velaXmsh1552} and Figure \ref{fig:confused_templates} already exemplified a particular result among those accounted for in Figure \ref{fig:piecharts}.
Since the separations among each simulated pair of sources are obtained randomly from a probability distribution, it is not guaranteed that the confusion criterion holds for each simulation. 
Secondly, independently of whether the sources are strictly confused or not, we can classify them according to the result of Equation \ref{eq:deltaTS}, i.e., in simulations best fitted to the hypothesis ${\rm H}_{\rm Conf.}$ (i.e., $\Delta {\rm TS} > 0$) or any other hypothesis considered ($\Delta {\rm TS} < 0$). 
Note that, in most simulations, the sources are not strictly confused (see the dark and light green sectors in the left chart of Figure \ref{fig:piecharts}), although this does not exclude one of the sources lying on top of the extended emission of the other. The latter is possible given the angular resolution of VHE gamma-ray telescopes, of few arcminutes, and the various extensions of the sources considered, from point-like to very extended ones, as e.g., HESS J0835-455. 

All sources in the library would be detected above 5$\sigma$ after 25\,h, both with the CTA northern and southern arrays when isolated. The latter was expected, since H.E.S.S. detected all of them in observation times of $\sim [20 -  50]$ hours. In this case, the input model from the simulation can be retrieved with small statistical errors, see, e.g., Table \ref{tab:velaX}. However, detecting all sources in the library when artificially placed on top of (or very close to) each other is no longer guaranteed. Note in Table \ref{tab:HESS_sources} that the integrated TeV flux of any two sources taken from the library may differ by a factor larger than fifty. 

From the simulations performed, approximately 30 per cent were strictly confused. Only about 18 per cent of the simulations corresponded to sources strictly confused with the data best described by the $\rm{H}_{\rm Conf.}$ hypothesis. It is important to note that the population of point-like and/or dim unresolved nebulae in the source library crucially affects the latter proportions. 
For example, compare the right chart of Figure \ref{fig:piecharts}, in which the point-like sources together with the two dimmest extended nebulae (i.e., HESS J1554-550 and HESS J1849-000) were excluded of the source library, with the complete results of the simulations (at the left). Hence, the results obtained for our simulations are only valid if the source library is representative of the TeV population of Galactic PWNe.

Since the population of point-like and/or dim nebulae that CTA will detect is surely underestimated by our library --which is based on H.E.S.S. data--, we must regard the cases in which the source confusion problem would be likely resolved by CTA (i.e., 18\%) as an optimistic prospect. We will discuss the issues brought by dim sources more in-depth below.

Once we probe the presence of two sources ($\Delta {\rm TS} > 0$) in a simulation, we can perform a joint fit of both sources (each with its corresponding template from the library), retrieving the input model of the simulation with high precision (see Figures \ref{fig:two_src_TS}, \ref{fig:amplitudes}, \ref{fig:indices}, \ref{fig:Ecuts}, and \ref{fig:TS_vs_flux}). 
The sources not detected at $5\sigma$ significance in all simulations performed are only those in Table \ref{tab:HESS_sources} with integral flux at energies above 1 TeV smaller than $\sim 10^{-12}$ cm$^{-2}$ s$^{-1}$ (see Figure \ref{fig:TS_vs_flux}). However, in some cases, these dim and small sources could be detected in 25 hours, particularly when not artificially confused with a bright (more extended) source.

%\subsection{Impact of the projected separation -- average PWN density}
%\label{subsec:seppwn}
%%%%%%%%%%%%%%%%%%%%%%%%%%%%%%%%%%%%%%%%%%%%%%%%%%%%%%%%%%%%%%%%%%%

The effect of changing the CTA array employed in the simulations onto Equation \ref{eq:deltaTS} is exemplified in the bottom panel of Figure \ref{fig:eq1_stat} (with vertical arrows). 
The simulations indicate that, in general, the detection significance of the confused sources hypothesis compared to the isolated source one, i.e., $\sqrt{|\Delta \rm{TS}|}$ with $\Delta \rm{TS}$ computed according to Equation \ref{eq:deltaTS}, increases approximately linearly with the projected separation between the sources. Figure \ref{fig:sqrtTS_vs_separation} illustrates the general trend mentioned. However, the same is not apparent in the bottom panel of Figure \ref{fig:eq1_stat}, which only portrays two different projected separations.

%%%%%%%%%%%%%%%%%%%%%%%%%%%%%%%%%%%%%%%%%%%%%%%%%%%%%%%
\subsection{Issues brought by dim sources}
\label{subsec:dimsources}
%%%%%%%%%%%%%%%%%%%%%%%%%%%%%%%%%%%%%%%%%%%%%%%%%%%%%%%

\begin{figure*}
    \centering
    \includegraphics[width=0.49\linewidth]{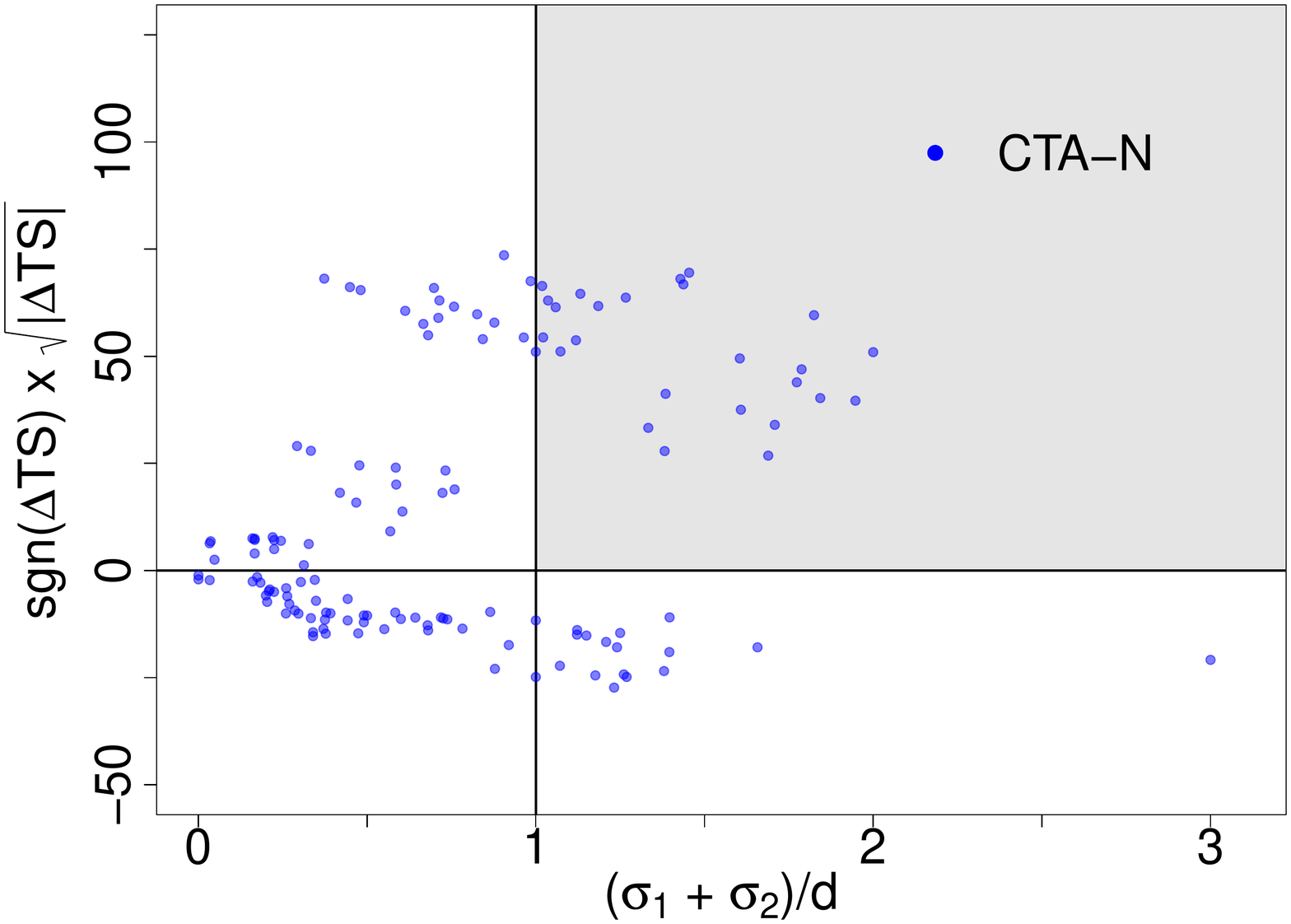}
    \includegraphics[width=0.49\linewidth]{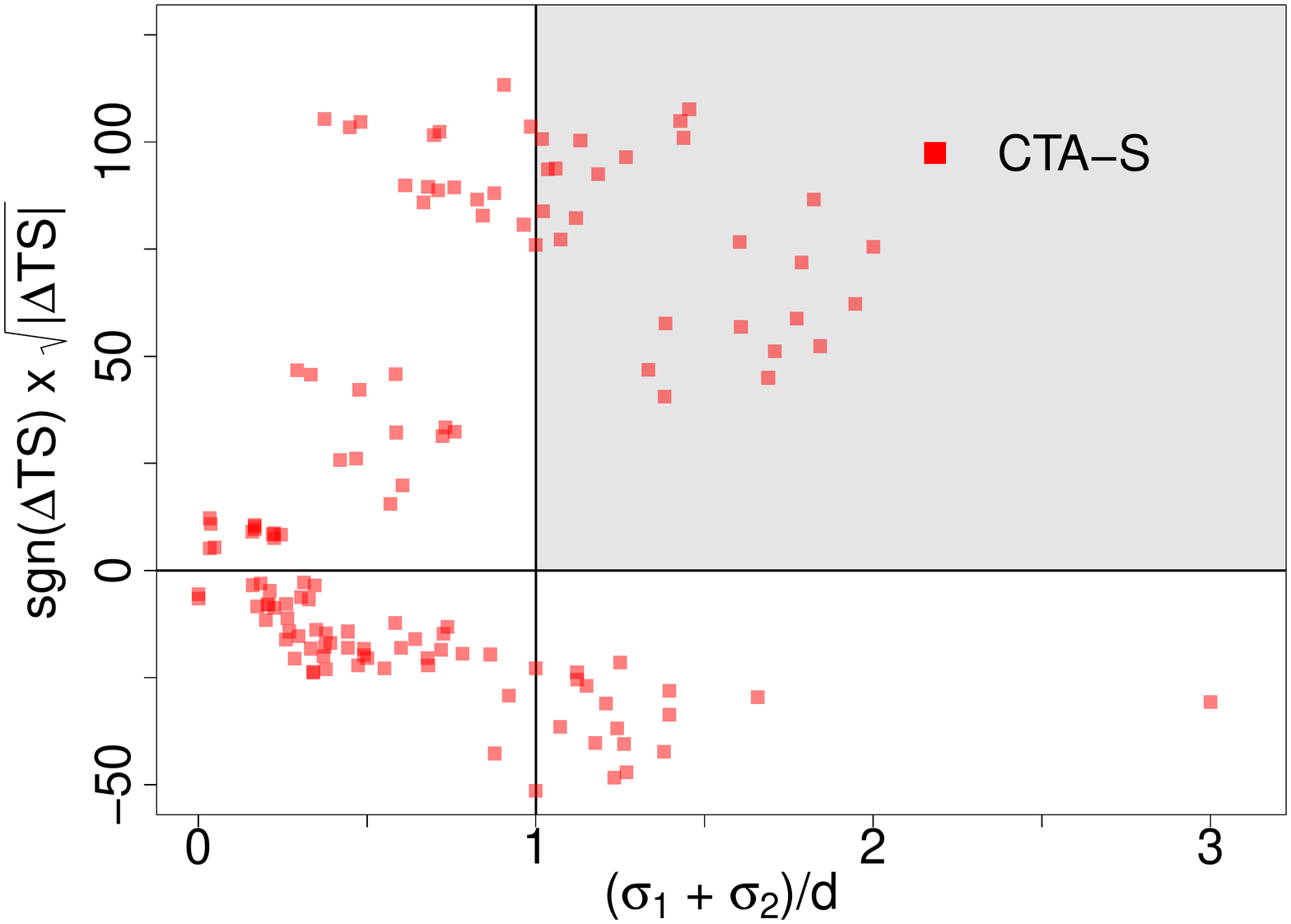}\\
    \includegraphics[width=0.49\linewidth]{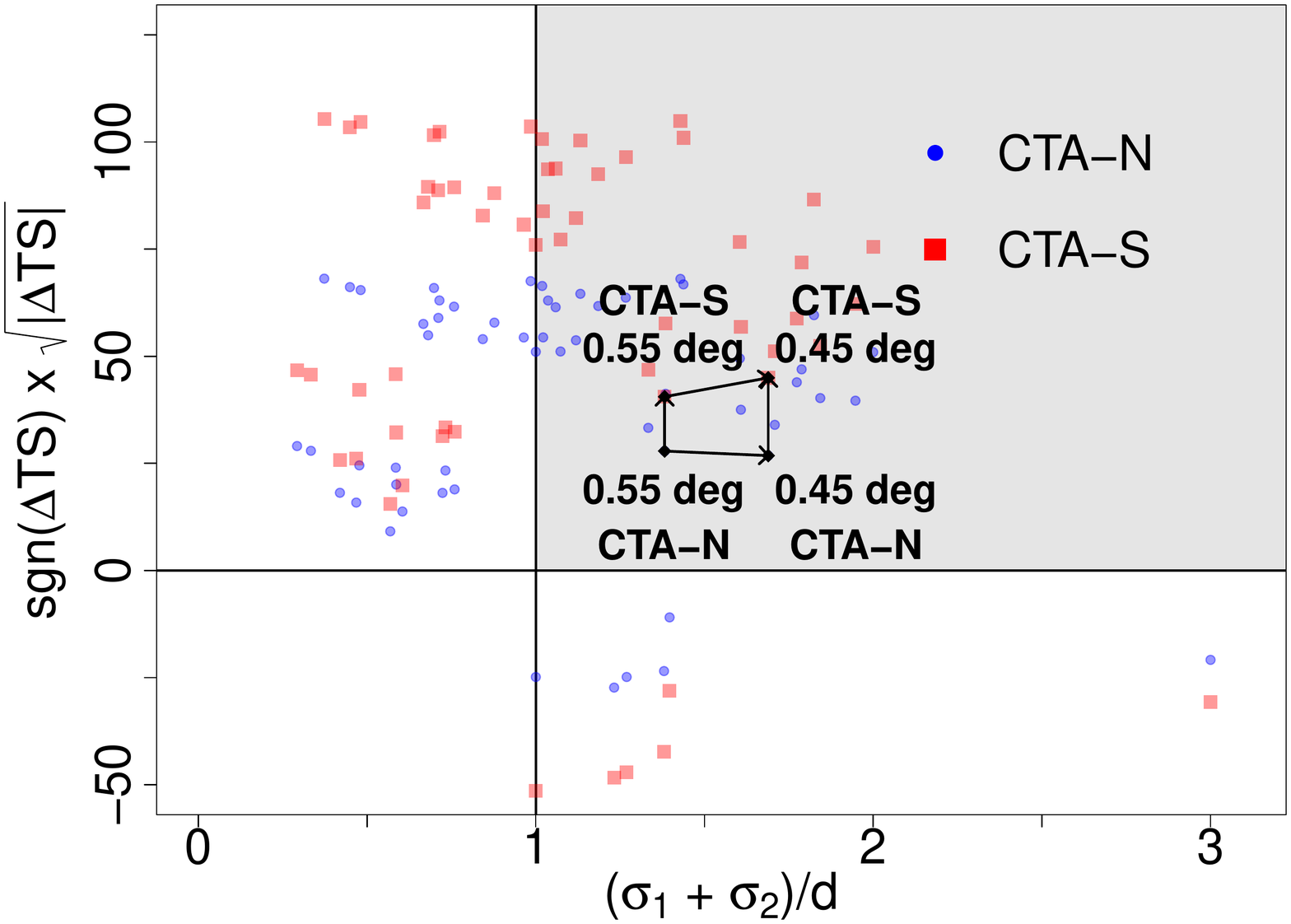}
    %plots_TS_def.R
    \caption{The square root of $\Delta \rm{TS}$ in Equation \ref{eq:deltaTS} (preserving the sign) is plotted for each pair of confused sources against the sum of their Gaussian sizes ($\sigma_{1} + \sigma_{2}$, see the fifth column in Table \ref{tab:HESS_sources}) divided by the separation imposed. 
    % The cases lying at the left of the vertical line, i.e., $\sigma_{1} + \sigma_{2} < d$, are not strictly confused (i.e., green tonalities in Figure \ref{fig:piecharts}). The points below the horizontal line (in the negative part of the Y-axis) correspond to simulations that showed no indication of being best fitted by two confused sources (i.e., dark colors in Figure \ref{fig:piecharts}). 
    In the lower panel we excluded the two dimmest extended sources (HESS J1554-550 and HESS J1849-000), and those point-like. The vertical black arrows exemplify how the results for the simulations (of a given pair of confused sources) move in the diagram when switching from the northern to the southern array. The horizontal arrows illustrate the effect of changing the separation between the sources.}
    \label{fig:eq1_stat}
\end{figure*}

Figure \ref{fig:eq1_stat} shows the results of applying Equation \ref{eq:deltaTS} to the observation simulations. The top left panel corresponds to the northern CTA array and the right to the southern one. The simulations best represented by the ${\rm H}_{\rm Conf.}$ hypothesis have a positive Y-axis (i.e., light colors in Figure \ref{fig:piecharts}), while the sources participating in a simulation are not strictly confused if $\sigma_{1} + \sigma_{2} < d$, i.e., at the left of the vertical line in Figure \ref{fig:eq1_stat} (in green tonalities in Figure \ref{fig:piecharts}). We are thus most interested in the simulations located in the grey shaded area marked. 
As we expected, there are simulations that satisfy the confusion criterion; $\sigma_{1} + \sigma_{2} > d$, but are best fitted to one particular library template (either rotated or not, i.e., $\Delta {\rm TS} < 0$), particularly if the confusion is strong, i.e., $\sigma_{1} + \sigma_{2} >> d$.

It is noticeable from comparing the top left and right panels in Figure \ref{fig:eq1_stat} that considering the southern array instead of the northern one for the same input configuration increases the source's significance obtained in the best-fitting hypothesis (whatever the same is). Typically, $\sqrt{\rm TS}$ multiplies by a factor of $\sqrt{[2-3]}$ due to the improvement of sensitivity. However, the sign of $\Delta$TS (from Equation \ref{eq:deltaTS}) is usually kept compared to that for the northern array. 
% Except if the best-fitting hypothesis switches from any of the one source template hypotheses to $\rm{H}_{\rm Conf.}$ by using the southern array instead of the northern one.

A considerable number of observation simulations, as seen in the bottom left quadrant in the panels at the top of Figure \ref{fig:eq1_stat}, are best described by only one source, i.e., $\Delta {\rm TS} < 0$, despite not being strictly confused ($\sigma_{1} + \sigma_{2} < d$). 
Those cases are related only with four (out of the twelve) sources; HESS J1554-550, HESS J1849-000, HESS J1747-281, and HESS J1818-154, when placed close to a much more extended and/or luminous nebula (see the bottom panel of Figure \ref{fig:eq1_stat}, which does not show the simulations related to the cited sources).

In the cases we referred to, the hypothesis regarding only the most extended and luminous among the sources represents the simulated data better than a combination of both source templates, even when according to the set up of the simulation, the sources are not strictly confused. It is apparent that, in these cases, simply perturbing the brightest source's spectrum may lead systematically to a better representation of data than if also slightly modifying its morphology (spatial template), resulting in $\Delta {\rm TS} < 0$. Figure \ref{fig:piecharts} may also be affected by this spectral effect.
The confusion criterion does not likely hold in these cases due to the extension of one of the sources ($\sigma < 0.1$ or point-like). Consider HESS J1747-281 or HESS J1818-154, e.g., simulated at $0.65\degr{}$ of HESS J0835-455 (more luminous by a factor above 25). 
A joint fit of the two input source models to the observation simulations, in these cases, results in a significance achieved for the dimmest nebula well below $5\sigma$.
These simulations are noticeable in the top panels of Figure \ref{fig:two_src_TS}, below the dashed line ($\rm{TS} < 25$). It is not a surprising result that detecting a given source with CTA (or H.E.S.S.) may depend on the presence or not of bright nearby sources, mainly when the source is dim and/or small (or point-like).

The problem of addressing faint and small gamma-ray nebulae in the vicinity of bright and extended sources can only increase with the advent of CTA. Its improved sensitivity will translate into the detection of numerous dim sources. To account for this, we may, as a first approximation, extrapolate the results to arbitrary populations with different ratios of faint and small nebulae by resampling the simulations through a bootstrapping technique.

We carried out this extrapolation by extracting random samples from the simulations performed. Repeated elements in these samples were allowed, i.e., we subtract one randomly chosen template from the whole sample at a time to generate new samples with varying percentages of dim sources. We used samples of various sizes from 5 to 45 simulations out of the 264 available, extracting $5 \times 10^{5}$ samples of each size. Next, we computed for each sample the ratio of simulations that are best fitted to the $\rm{H}_{\rm Conf}$ hypothesis and the ratio of faint and small sources, treating each of the samples as a different population of sources.

\begin{figure}
    \centering
    \includegraphics[width=0.95\linewidth]{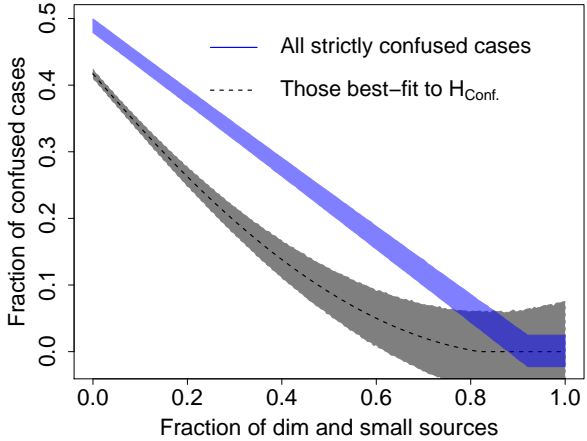}
    \caption{The fraction of simulations strictly confused and best fitted to the $\rm{H}_{Conf.}$ hypothesis (in black) and the total fraction of strictly confused ones (in blue) versus the fraction of faint and not very extended sources in the sample, i.e., with $\rm{S}_{b} < 10^{-10}$ cm$^{-2}$ s$^{-1}$ deg$^{-2}$ and $\sigma < 0.1\degr{}$. The shadowed areas represent the 1-$\sigma$ error regions. 
    % The difference between the grey and the blue shadows in this figure represents sources that are not fitted to a confused case, despite being confused in the input template.
    }
    \label{fig:fracresfracdim}
    %bootstrap_simuls.R
\end{figure}

We present the results in Figure \ref{fig:fracresfracdim}.
The latter extrapolation, for example, resulted in $\sim 9\%$ of all simulations ($\sim 40$\% of those strictly confused) best represented by the $\rm{H}_{\rm Conf.}$ hypothesis (with an upper limit of 17\% at 95\% CL) for a fraction of dim and small sources of $0.5$. The latter means that 50\% of PWNe in the input library would be similar in flux and extension to HESS J1554-550, HESS J1849-000, HESS J1747-281, or HESS J1818-154 ($\rm{S}_{b} < 10^{-10}$ cm$^{-2}$ s$^{-1}$ deg$^{-2}$ and $\sigma < 0.1\degr{}$).
In the cited case, about 22\% of the simulations strictly satisfied the confusion criterion. The simulations best fitted to the $\rm{H}_{\rm Conf.}$ hypothesis (see black dashed line in Figure \ref{fig:fracresfracdim}), however, included wrong associations of templates to the simulated data. Out of the 264 simulations, the cases best represented by two confused sources in which one of the nebulae is incorrectly identified account for 8\%. In only one simulation both of the nebulae were misidentified.

The problem of wrongly associated templates will also increase with the addition of newly discovered faint and small nebulae (and the number of templates in the library). However, the extrapolations explained above to different populations of sources allowed us to predict the relative amount of possible mismatches with the template library in different scenarios. For a source population consisting of 50\% dim sources, the simulations best fitted by the $\rm{H}_{\rm Conf.}$ hypothesis in which one of the nebulae is incorrectly identified were limited to $\sim 26$\% (at a 95\% confidence level). 

The difference between the grey and the blue shadows in Figure \ref{fig:fracresfracdim} represents simulations that are not fitted to two confused sources, despite being confused in the input template. Note also in the figure that, as follows from the average separation between sources (i.e., $0.5\degr{}$), the fraction of confused sources reaches zero when the fraction of dim and small or point-like sources is close to one. Hence, this study cannot characterize the source confusion problem regarding only these small and/or point-like sources.

%%%%%%%%%%%%%%%%%%%%%%%%%%%%%%%%%%%%%%%%%%%%%%%%%%%%%%%%%%%%%%%%
\subsection{Minimal auto-correlation degree to make two templates different }
\label{sec:autocor}
%%%%%%%%%%%%%%%%%%%%%%%%%%%%%%%%%%%%%%%%%%%%%%%%%%%%%%%%%%%%%%%%

An expected degeneracy for $\alpha$ plus $180\degr{}$ in the best-fitting orientation of the templates, derived from the elliptical-like morphology of some of the nebulae templates employed, was observed (see Figure \ref{fig:angldeg}). In some cases, the best-fitting rotation angle results close to the input one (black and red lines in the right panel of Figure \ref{fig:angldeg}). In others, it turns to a rotation angle separated by about $180\degr{}$ of the input value, e.g., see the blue line in the right panel of the cited figure (where the input orientation is closer to the local maximum at $0\degr{}$).
The degeneracy observed in the best-fitting orientation is enhanced when the morphological auto-correlation degree of the simulated template approaches $C_{i,j} = 1$. We are then interested in the smallest variation of morphological auto-correlation degree that leads (for a given source) to a noticeable change in the source detection significance. This would be helpful, for instance, if the template library is built, e.g., from HD, MHD, or HD+B simulations. The smallest variation of $C_{i,j}$ detectable would be related, in this case, to the level of accuracy needed in the template simulations.

We performed simulations of a few nebulae among the population considered but with the source template blurred up to different degrees. To blur the templates, we convolved its central region (a box of 1.5$\sigma$ of side) with a function depending on an index parameter ($\beta$).
This function takes the value $f(\rm{X},\rm{Y},\beta) = 2-\max(|\rm{X}-\rm{X}_{0}|^{\beta},|\rm{Y}-\rm{Y}_{0}|^{\beta})/1.5\sigma$ inside the box, and $f(\rm{X},\rm{Y},\beta) = 1$ outside the box, where X and Y are pixel coordinates in the template and X$_{0}$ and Y$_{0}$ are referred to the template's center. 
The index (ranging from 1 to 3) modulates the harshness of the blurring applied, corresponding the $\beta = 1$ case to the most similar template to the original. 
Figure \ref{fig:templateblurring}, provides an example of different degrees of blurring (quantified by the $\beta$ parameter) applied to the HESS J1825-137 template. The Pearson's correlation coefficient between each blurred template and the unaltered one, where $\Delta C_{i,j} \approx 0$ for $\beta = 1$, is noted in each panel.

\begin{figure}
    \centering
    \includegraphics[width=0.95\linewidth]{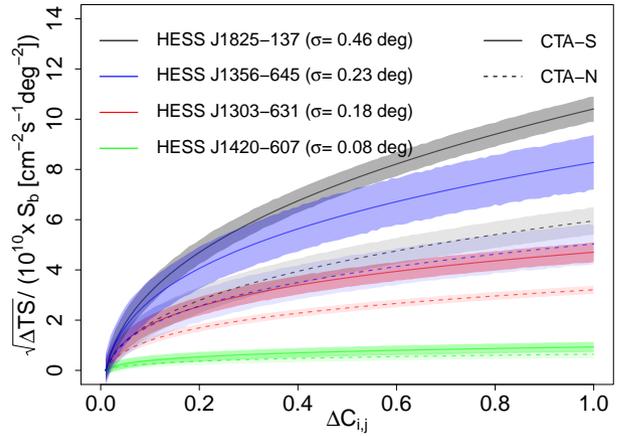}
    \caption{The change of the source detection significance when fitting a template to observation simulations performed with various degrees of blurring and those with the unaltered template, i.e., $\Delta {\rm TS} = {\rm TS}_{\rm Blur.}-{\rm TS}_{\rm Unalt.}$. We normalized $\sqrt{\Delta {\rm TS}}$ by the surface brightness of the source; $\rm{S}_{\rm b} = 10.5$, $6.2$, $11.3$, and $22.6$ in $10^{-10}$ cm$^{-2}$ s$^{-1}$ deg$^{-2}$, respectively, for each of the nebulae (in the same order of the legend). The Pearson’s correlation coefficient ($C_{i,j}$, in Equation \ref{eq:pearson_coeff}) quantifies the similarity between the blurred and unaltered templates.}
    \label{fig:tempblur}
    %fin_templ_degr.R paper_degr_templ.R
\end{figure}

Next, we fitted the original (not altered) source template to the simulated data to probe the sensitivity of the simulations on the morphological auto-correlation degree. Figure \ref{fig:tempblur} shows how the detection significance changes with respect to the variation of the morphological auto-correlation degree for the blurred templates of different nebulae. In the figure, a larger $\Delta C_{i,j}$ indicates a more significant blurring applied.
We normalized each curve by the surface brightness (${\rm S}_{b}$) of the corresponding nebula, derived from the template (in units of cm$^{-2}$ s$^{-1}$ deg$^{-2}$), to compensate for both the differences in flux and morphology of the sources at the same time.
% As expected, the sensitivity of the simulations to a change of $C_{i,j}$ depends on the luminosity of the source. 
Using the surface brightness of particular PWNe, i.e., multiplying the Y-axis of Figure \ref{fig:tempblur} by the surface brightness values specified in the figure's caption, we obtained $\sqrt{\Delta \rm{TS}} \gtrsim 5$ for $C_{i,j} < 0.1$ in all the cases depicted. It means that our simulations are sensitive to such small variations of $C_{i,j}$ for PWNe as HESS J1825-137 or HESS J1303-631, which are among the rather luminous ones of our population.
If the templates come from a theoretical prediction of a model, it would seem that two such templates should certainly be part of the library for $\Delta C_{i,j} > [0.05-0.1]$.
Provided a given surface brightness, the size of the source is the most critical parameter, being the simulations of more extended sources more sensitive to deformations of the template (as highlighted in Figure \ref{fig:tempblur}, see the sources extensions noted in Gaussian $\sigma$).
% Using the surface brightness of particular PWNe (as an example, note the detailed values in the caption of Figure \ref{fig:tempblur}), we obtain that a relatively large $\sqrt{\Delta \rm{TS}}$ already arises for low values of $\Delta C_{i,j}$.
%
% For comparison, note that simply rotating a template can imply a Pearson's coefficient of the same order, i.e., $\Delta C_{i,j} < 0.1$ compared to the unaltered template. The latter is due to the average morphological auto-correlation degree of the nebulae over all the rotations, which is generally high (e.g., $\langle C \rangle_{i,j} \approx 0.9$ for HESS J1303-631, see Table \ref{tab:pears_coeff}).
%
Note that Pearson's coefficient depends on the template's binning. However, the latter compares well with the angular resolution expected for CTA that will ultimately determine the template's bin size.

%%%%%%%%%%%%%%%%%%%%%%%%%%%%%%%%%%%%%%%%%%%%%%%%%%%%%%%%%%%%%%%%%%%

%%%%%%%%%%%%%%%%%%%%%%%%%%%%%%%%%%%%%%%%%%%%%%%%%%%%%%%%%%%%%%%%%%%
\section{Discussion: real data and computational time}
\label{sec:discussion}
%%%%%%%%%%%%%%%%%%%%%%%%%%%%%%%%%%%%%%%%%%%%%%%%%%%%%%%%%%%%%%%%%%%

An analogous procedure using genuine CTA data would also start by fitting.
On the one hand, we would fit a source with a model corresponding to each of the templates in the library considered on its own, taking either a set of rotations for each template or a rotation angle as a free parameter of the fit. 
Note that different rotations of a template can be treated, as we do, like any other template in the library (like a different PWN), which may simplify the fitting process. 
On the other hand, we would fit the same source with a model having different combinations of the templates of a putative library confused in pairs. The free parameters would be the separation between the templates and the relative orientation among the same.
Next, similarly to Equation \ref{eq:deltaTS}, we would compare the TS corresponding to the best-fitting model among the hypotheses regarding two confused templates to that of the best-fitting model involving the template of only one source. 
In our simulations, the fit to a Gaussian source involves three degrees of freedom more (i.e., position and width) than a fit to a generic template. However, when fitting the templates from the library to actual CTA data, both hypotheses will have the same degrees of freedom since the templates will not maintain a reference position (accounting also for the rotation angle of the template).

We have accounted for different orientations of the simulated templates, handling the rotation of the templates straightforwardly by considering each rotated template a separate template of the library. However, we noted that although the simulations are sensitive to a transformation of the templates such as blurring, degradation, or rotation, most templates of the library have a high morphological auto-correlation degree. Fitting the orientation of the templates from the observation simulations is challenging due to their high auto-similarity, mainly because we observe a degeneracy between the best-fitting angle and rotations close to $180\degr{}$, as expected for elliptical-like morphologies. 
In addition, the absence of small-scale structures (below the HGPS angular resolution, i.e., $\sim 0.08\degr{}$) that CTA may resolve in the H.E.S.S. maps used as templates could prevent us from detecting differences involving small sources in these simulations.

As we used H.E.S.S. observational data, the diffuse interstellar background is included (implicitly) in the simulations. However, it is only so in an idealized case since CTA may estimate a different level for the diffuse interstellar emission than H.E.S.S. at the position of the simulated sources. Our simulations may be sensitive to such differences. We leave the analysis of this component more in-depth to future studies with real CTA data since CTA will map with unprecedented precision the large-scale diffuse emission at VHE gamma rays.

Another limitation of this study is that we did not consider the physical distance of the sources fitted to the observation simulations. The same template can be artificially placed closer or further from a reference distance ($r$) by rescaling its flux (as $1/r^{2}$) and spatial axes (by $1/r$). The relative error for the latter approximation (from trigonometric arguments) is smaller than $0.01\%$ if $2r/D > 50$, where $D$ is the physical diameter of the template. Figure \ref{fig:VelaX_withdist} central and right panels, e.g., depict 25-hour simulations of HESS J0835-455 with the CTA southern array rescaled at twice and four times the distance of the reference template, respectively. We can handle the rescaled templates and the rotated ones alike. However, the number of templates in the library rapidly increases if considering different rotations at various distances for each source. 

The computation time required to cross-match all the templates with an observation simulation will be crucial for future studies with a more comprehensive template source library. Hence, it is helpful to have a good estimate based on this study.
We summarized the computation time to perform an observation simulation of two confused sources in Table \ref{tab:computtimeconf} for different input configurations using a regular commercial laptop. To compare with Table \ref{tab:computtimeconf}, note that we used a FoV of $1.5\degr{}$ of radius for the simulations presented in this paper. 
% i.e., the time to perform an observation simulation (in {\sc ctools}) from an input model with two sources characterized by two different morphological templates (plus background). 
This time does not include any likelihood fitting process.
The combination of large observation times (more than 50 hours) and low energy thresholds (E$_{th} < 100$ GeV) increases significantly (by a factor ten) the computational time of a single observation simulation. 
The number of simulations to perform at the end will also account for the two CTA arrays and multiple separations and orientations for each possible pair of sources from the sample, if not accounting also for various distances for each template. The cost in computing time of fitting different hypotheses to the observation simulations varies depending on the number of those considered and the number of free parameters involved. Still, the typical average computation time per fitted model (template) was $[1-2]$ minutes. Hence, one observation simulation can be fitted against, e.g., $10^{6}$ templates in one month if using a cluster with $\sim 30$ CPUs. The total computation time for the simulations we present in this paper is about a month per 132 simulations on a regular commercial laptop. Note that the number of templates needed to account for all possible pairings from a source library grows with the number of sources (N) as $\rm{N} \times (\rm{N}-1)/2$, not accounting for different orientations, separations, and/or distances.

% Some PWNe are known to present energy-dependent morphology, e.g., HESS J1303-631 or HESS J1825-137. However, we fixed the morphology of all sources at energies from 500 GeV to 300 TeV in this work. Accounting for energy-dependent morphologies in studies like the one we present supposes an additional complexity. We have not considered either the case of more than two sources being present in the region of interest or the contamination in the analysis of two confused sources from other nearby ones, which further limits this study. Again, these limitations emphasize considering the level of confusion obtained as a reasonable lower limit.

% We expect the limitations cited so far only to introduce additional complications in discerning the case of two (confused) sources from an isolated one. On the contrary, increasing the average separation among the sources ($d_{\rm PWN} > 0.5\degr{}$) would improve our prospects for resolving source confusion in CTA data (see Section \ref{subsec:seppwn}). For instance, the latter would occur if the Galactic Plane Survey does not achieve the integral sensitivity level assumed in the simulations (a few mCrab in flux). It is also likely to occur far outside the Galactic central regions.

Our simulations are based on {\sc ctools}, as explained in Section \ref{sec:simtools}. However, two packages have been developed independently to implement the CTA Science Tools, i.e., {\sc ctools} and {\sc gammapy} \citep{2017ICRC...35..766D,2019A&A...625A..10N}. The latter software was adopted as the Science Analysis Tools of the CTA by the CTA Observatory (CTAO), being this decision announced on 1 June 2021\protect\footnote{\url{https://www.cta-observatory.org/ctao-adopts-the-gammapy-software-package-for-science-analysis/}}. Both softwares were proven to be stable, providing nearly identical high-level analysis results (such as events sky maps or spectra) in different analyses \citep{2019A&A...632A..72M,2020MNRAS.492..708M}. Hence, we do not expect significantly different results if implementing our simulation scheme in {\sc gammapy}. 
To perform simulations of source templates, either in {\sc ctools} or {\sc gammapy}, is nearly a linear operation if not accounting for background. Hence, we can combine the templates before the simulation or stack the simulations of two templates afterward (with no background), as is shown in Figure \ref{fig:linearity}. To perform this test of linearity, we assigned an arbitrary spectrum to the confused template simulated in the left panel of Figure \ref{fig:linearity}. Next, we assigned to each template in the right panel of the cited figure the spectrum used for the confused one (at the left) with the normalization parameter divided by two. Both panels can be thus directly compared.

%%%%%%%%%%%%%%%%%%%%%%%%%%%%%%%%%%%%%%%%%%%%%%%%%%%%%%%%%%%%%%%%%%%
\section{Conclusions}
\label{sec:conclusions}
%%%%%%%%%%%%%%%%%%%%%%%%%%%%%%%%%%%%%%%%%%%%%%%%%%%%%%%%%%%%%%%%%%%
This paper presents the most detailed quantitative study up to date on source confusion of extended sources and identification capabilities with CTA.
The tools available to perform simulations of TeV sources with CTA allowed us to conduct simulations of two confused or closely located sources in a variety of configurations, involving different separations, relative orientations, flux levels, and extensions.
We showed how these simulations could be analyzed through a direct comparison to a library of extended and/or point-like TeV source templates and that it is possible (in some cases at very high confidence) to associate a CTA simulation with a combination of two templates from the library placed at a small distance ($\sim 0.5\degr{}$) of each other. 
We also demonstrated that if the latter association is reached and correct, the characteristics of both sources (despite lying one on top of the other) can be studied in detail. In these cases, we retrieve similar statistical errors to those obtained analyzing the sources in the isolated scenario.
% Note that the templates describe the intensity distribution of the sources in a particular region of interest around the same. Still, they are given in any arbitrary flux units, which facilitates the creation of the library.

For applying the method to real data we note that we would first need to obtain a library of morphological templates to cross-match with observations, representative of the expected population of sources. 
%
% Other limitations arise from the generally high degree of morphological auto-correlation of PWNe templates, which complicates the determination of the orientation of the simulation with respect to the template. 
%
Also, the templates used here are referred to a specific distance, and the source may present energy-dependent morphology in CTA data, which will further complicate the cross-matching of the simulations with the template library. We have not considered the source confusion cases regarding more than two sources, mainly affecting the central parts of the Galaxy, nor the contamination of other non-confused but nearby sources in the analysis of the simulations. Note that we have only considered non-variable sources. Additionally, the simulations are also limited by the uncertainty in the flux level of diffuse emission and the shapes and distribution of the extended sources. For instance, source confusion involving only small and/or point-like sources is marginally represented in our simulations.

Despite the cited limitations, we constrained the source confusion in CTA data, particularly regarding the future Galactic Plane Survey with CTA, limiting the amount of source confusion likely to be resolved (as a first approximation) to 18\% above 500 GeV with an upper limit of 23\% at 95\% CL (based on the currently available Galactic Plane Survey from the H.E.S.S. experiment). We also obtained an upper limit of 33\% (at 95\% CL) for the occurrence of strict source confusion, i.e., $\sigma_{1} + \sigma_{2} > d$. 
% Further studies are required to derive more precise constraints on source confusion in CTA data.
These numbers apply only if assuming that CTA will detect a similar population of PWNe to that detected by H.E.S.S.; which is optimistic for obvious reasons. 
The amount of presumably-resolvable source confusion may be limited to only $\sim 10$\% if more than half of the TeV PWNe population regarding the future CTA Galactic Plane Survey consists of faint and small sources (i.e., $\rm{F}_{\rm >1TeV} < 10^{-12}$ cm$^{2}$ s$^{-1}$ and $\sigma < 0.1\degr{}$), which is likely a more realistic scenario. 
% The latter case would limit the occurrence of strict source confusion to less than $\sim 20$\

The flawed cross-matches with the template library can be a problem of the approach. Approximately 8\% of the simulations performed, based on the Galactic Plane Survey from the H.E.S.S. experiment, presented a misidentification of sources. This problem can aggravate if accounting for a more realistic ratio of dim sources in the input library. For instance, this difficulty can appear in up to 26\% of (likely) resolved cases of source confusion for a 50\% (or higher) ratio of dim sources.

Even though, the general approach we present can provide handy information, facilitating the identification of numerous sources (as our results regarding isolated nebulae illustrate) and, to some extent, the resolution of source confusion cases. Even if achieving the latter is challenging according to our simulations, associating empirical or theoretical templates to CTA data could provide exhaustive information on multiple sources. Our approach matches the data with the knowledge applied to the template generation (handled by the user). Hence, given some data, it allows us to explore different hypotheses more in-depth than assuming a simple geometrical morphology for the sources as the usual Gaussian or point-like assumptions. 

A 3D binned likelihood fitting approach (newly introduced for VHE gamma-ray data analysis) has recently been demonstrated as an efficient technique to disentangle several regions with different spectral and morphological characteristics within a significantly extended emission (see \citealt{2021A&A...653A.152A} and references therein). This kind of analysis, together with the (also 3D) approach we tested in this work, can potentially alleviate the source confusion furthermore.

To conclude, at the methodological level, we note that the general approach developed here can be applied too at other frequencies, allowing similar template comparisons with X-rays and radio data. Likewise, the methodology we present can be straightforwardly generalized to source confusion regarding other extended Galactic sources (e.g., between SNRs and PWNe) or extragalactic ones.

%%%%%%%%%%%%%%%%%%%%%%%%%%%%%%%%%%%%%%%%%%%%%%%%%%%%%%%%%%%
\section*{Acknowledgements}
%%%%%%%%%%%%%%%%%%%%%%%%%%%%%%%%%%%%%%%%%%%%%%%%%%%%%%%%%%%

This research made use of {\sc ctools}, a community-developed analysis package for Imaging Air Cherenkov Telescope data. {\sc ctools} is based on GammaLib, a community-developed toolbox for the scientific analysis of astronomical gamma-ray data. We made use of the CTA IRFs provided by the CTA Consortium and Observatory, see https://www.cta-observatory.org/science/cta-performance/ (version {\sc prod5 v0.1}; \citealt{cherenkov_telescope_array_observatory_2021_5499840}). We made use of R: A language and environment for statistical computing \citep{Rmanual}.
This paper has been peer reviewed internally by the CTA Collaboration. We acknowledge Gerrit Spengler, Jean Ballet, Luigi Tibaldo, Heide Costantini, and Barbara Olmi for providing us with useful comments that lead to an improvement of the paper. We also thank an anonymous referee. 
This work has been supported by the Spanish grants PID2021-124581OB-I00. DFT was also supported by the Spanish program Unidad de Excelencia ``María de Maeztu" CEX2020-001058-M. EM acknowledges support by grant P18-FR-1580 from the   Consejer\'{\i}a de Econom\'{\i}a y Conocimiento de la Junta de Andaluc\'{\i}a under the Programa Operativo FEDER 2014-2020. EdOW acknowledges the support of the Alexander von Humboldt fundation and DESY (Zeuthen), a member of the Helmholtz Association HGF.

\section*{Data Availability}

The empirical data used to gather the templates for the sources considered are public and available in the additional materials regarding the H.E.S.S. Galactic Plane Survey \citep{HESSGPS_paper}, and were obtained from \url{https://www.mpi-hd.mpg.de/hfm/HESS/hgps/}. The CTA IRFs are public and available in \url{https://www.cta-observatory.org/ science/cta-performance/}.



\bibliographystyle{mnras}
\bibliography{example} % if your bibtex file is called example.bib

%\begin{thebibliography}{99}
%\end{thebibliography}

%%%%%%%%%%%%%%%%%%%%%%%%%%%%%%%%%%%%%%%%%%%%%%%%%%

%%%%%%%%%%%%%%%%% APPENDICES %%%%%%%%%%%%%%%%%%%%%

\appendix

\section{Akaike's Information Criterion compared to the Test Statistic}
\label{AppendixA}

The AIC statistic of a given model with $k$ degrees of freedom is defined by
\begin{equation}
        {\rm AIC} = 2k - 2\ln(L)
\label{eq:AIC}
\end{equation}
Similarly to Equation \ref{eq:deltaTS}, we can then define:
\begin{equation}
\begin{split}
        {\rm \Delta AIC} = \min({\rm AIC}_{\rm Gauss}, {\rm AIC}_{\rm Temp.\ 1, \alpha},
        {\rm AIC}_{\rm Temp.\ 2, \alpha}, \\
        {\rm AIC}_{\rm Temp. i, \alpha}) - \min({\rm AIC}_{\rm Conf.})
\end{split}
    \label{eq:deltaAIC}
\end{equation}
A $\Delta {\rm AIC}$ large and positive can thus be related (as in the case of $\Delta {\rm TS}$) to better prospects for unraveling source confusion. When comparing two different models, we obtain, according to equations \ref{eq:TSdif} and \ref{eq:AIC}:
\begin{equation}
\begin{split}
        {\rm AIC}_{\rm{H}_{i}} - {\rm AIC}_{\rm{H}_{j}} = 2(k_{\rm{H}_{i}}-k_{\rm{H}_{j}}) - ({\rm TS}_{\rm{H}_{i}} - {\rm TS}_{\rm{H}_{j}}) 
\end{split}
    \label{eq:AICdif}
\end{equation}

\begin{figure}
    \centering
    \includegraphics[width=\linewidth]{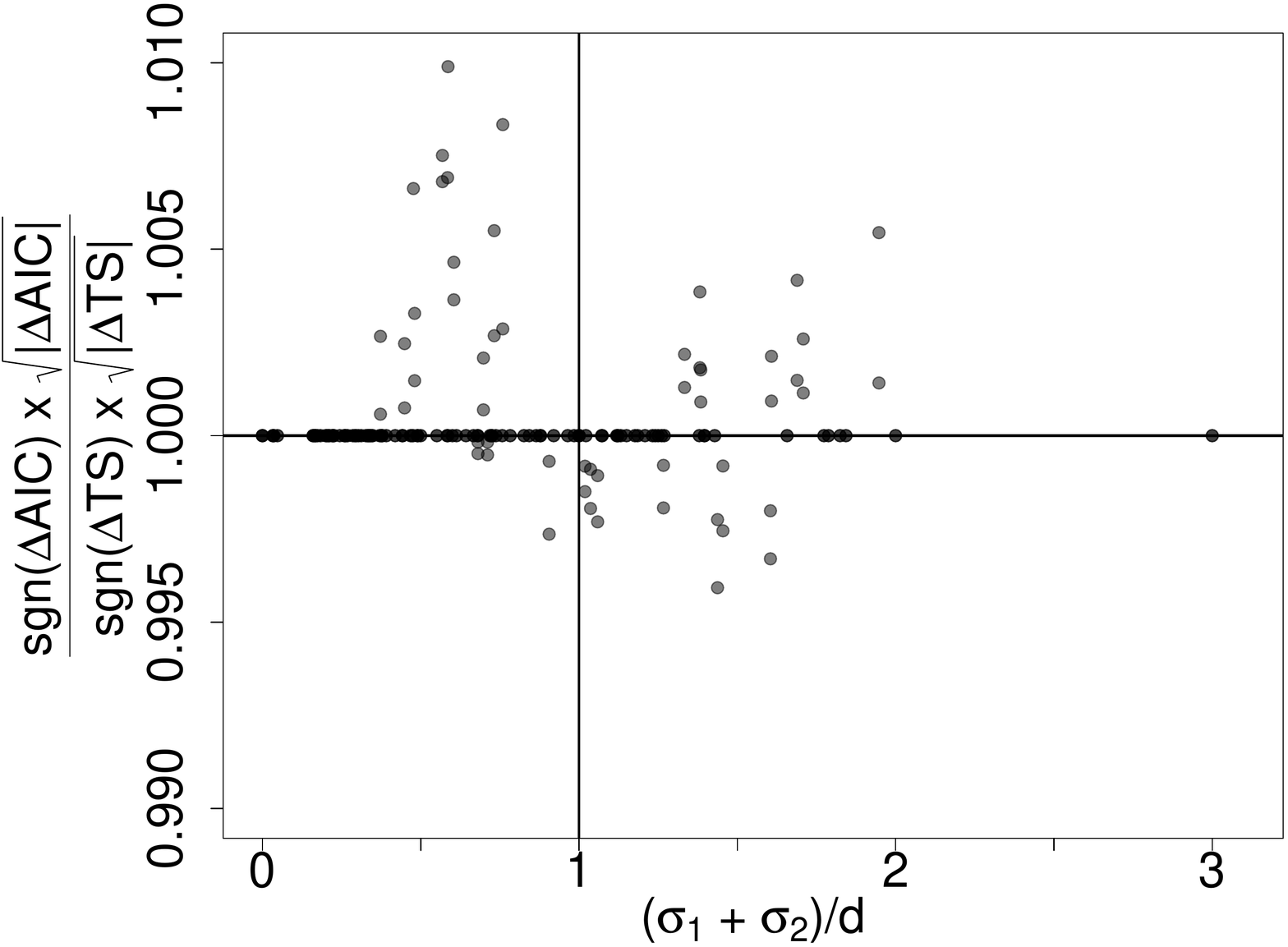}
    \caption{The ratio of $\Delta {\rm AIC}$ and $\Delta {\rm TS}$ defined by Equations \ref{eq:deltaAIC} and \ref{eq:deltaTS} is depicted for different observation simulations.}
    \label{fig:AICvsTS}
    %AIC_SC_plots.R
\end{figure}

The latter translates into ${\rm \Delta AIC} \approx {\rm \Delta TS}$ for the simulations we present. This is due to two reasons.
First, because the ${\rm H}_{\rm Conf.}$ model of maximum TS (i.e., log-likelihood) results in all cases to be the ${\rm H}_{\rm Conf.}$ model of minimum AIC, and the same occurs for the rest of the source templates (i.e., for ${\rm H}_{\rm Temp.\ 1, \alpha}, {\rm H}_{\rm Temp.\ 2, \alpha}$, and ${\rm H}_{\rm Temp. i, \alpha}$ hypotheses). 
Secondly, it is due to the difference of free parameters between models ($k_{\rm{H}_{i}}-k_{\rm{H}_{j}}$), which is generally negligible compared to the term concerning the TS and, in most cases (in absolute value), either one or zero. $k_{\rm{H}_{i}}-k_{\rm{H}_{j}} = 1$ occurs for the case of an exponentially cutoff power-law spectral model compared to a simple power-law one (the spatial templates do not have free parameters associated). In any case, $k_{\rm{H}_{i}}-k_{\rm{H}_{j}} \leq 4$ and $k_{\rm{H}_{i}}-k_{\rm{H}_{j}} > 1$ only occur if the Gaussian source hypothesis is that of the smallest AIC compared to all templates (models) of isolated sources in the library, which is highly unusual. From this point, the reader may consider that $\Delta {\rm AIC}$ and $\Delta {\rm TS}$ are virtually switchable throughout this paper. Figure \ref{fig:AICvsTS} illustrates how $\Delta {\rm AIC}$ and $\Delta {\rm TS}$ compare for the different simulations performed.

\section{Tables and plots}

\begin{figure*}
    \centering        \includegraphics[width=0.45\linewidth]{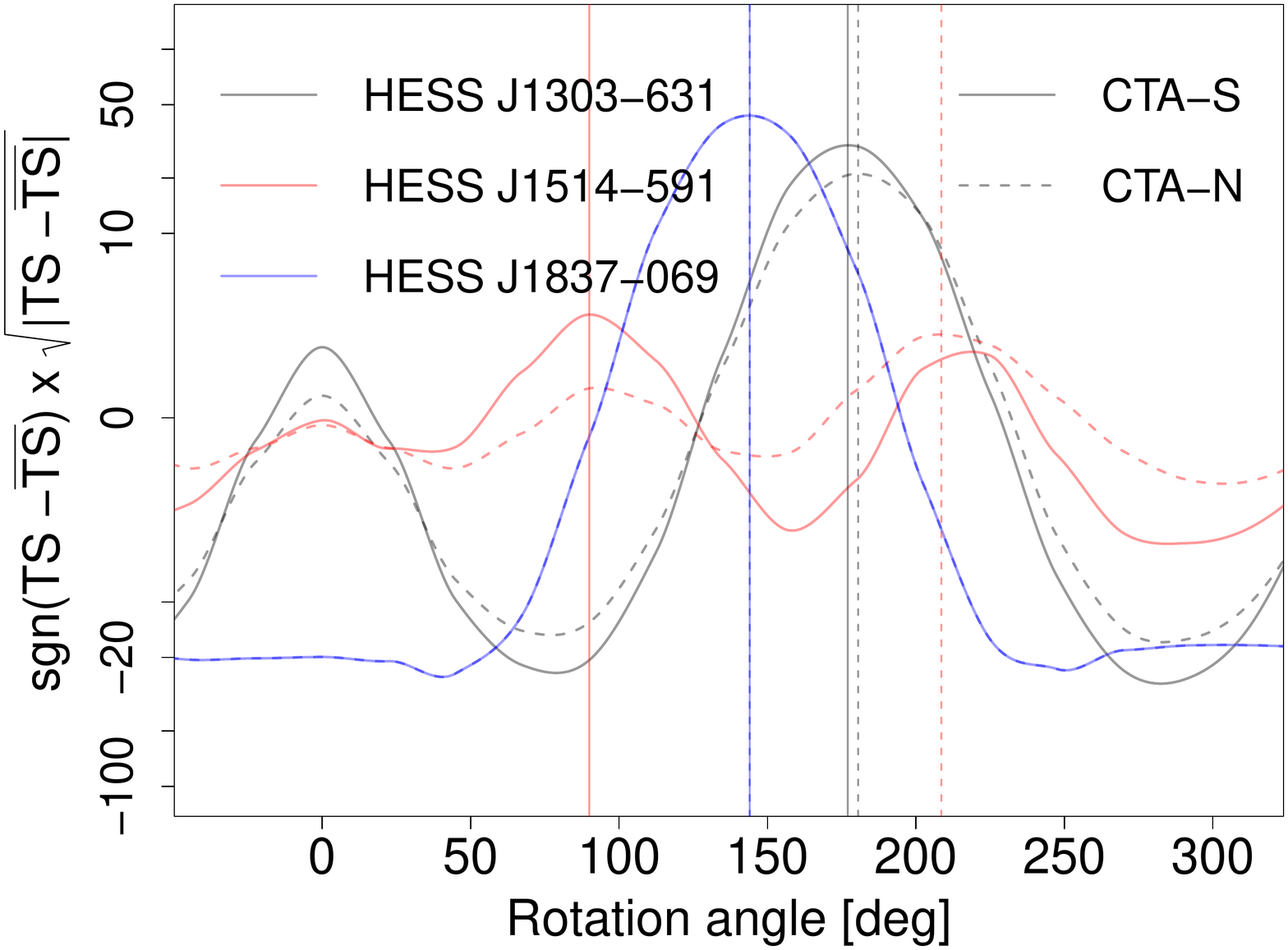}
    \includegraphics[width=0.45\linewidth]{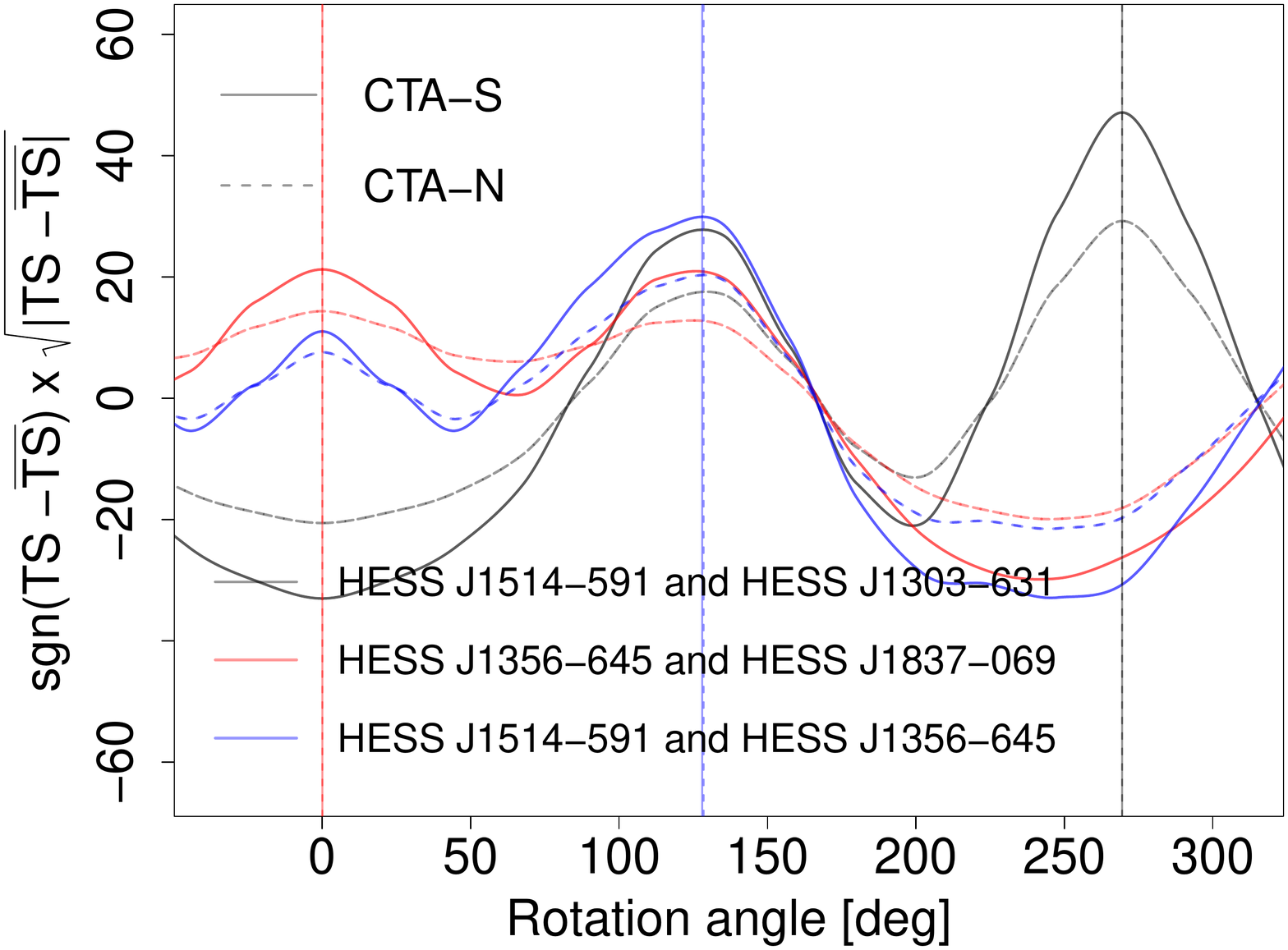}
    \caption{In the left panel, the detection significance retrieved for observation simulations of three sources when isolated (compared to their average over different rotation angles) versus the orientation of the template fitted to the observation simulation. The vertical solid and dashed lines correspond to the absolute maxima for CTA southern and northern arrays, respectively. The input orientations for HESS J1303-631, HESS J1514-591, and HESS J1837-069 in this case are $180\degr{}$, $216\degr{}$, and $144\degr{}$, respectively. In the right panel, the same for different artificially confused sources. The input orientation in the HESS J1514-591 confused with HESS J1303-631 case (in black) is $252\degr{}$, while in the other two cases (in blue and red) it is $36\degr{}$.}
    \label{fig:angldeg}
    %orients_resolv.R
\end{figure*}

\begin{figure*}
    \centering
    \includegraphics[width=0.45\linewidth]{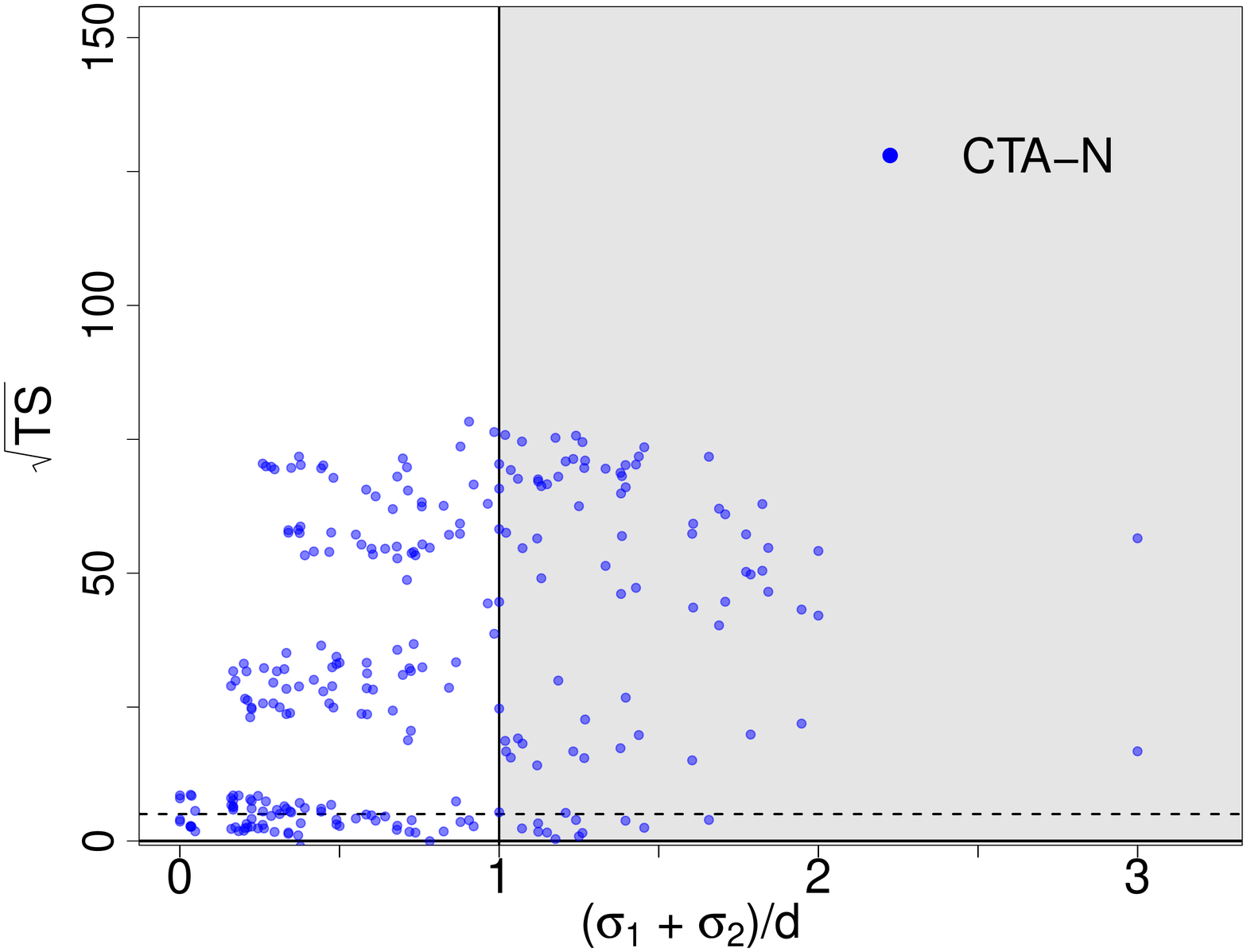}
    \includegraphics[width=0.45\linewidth]{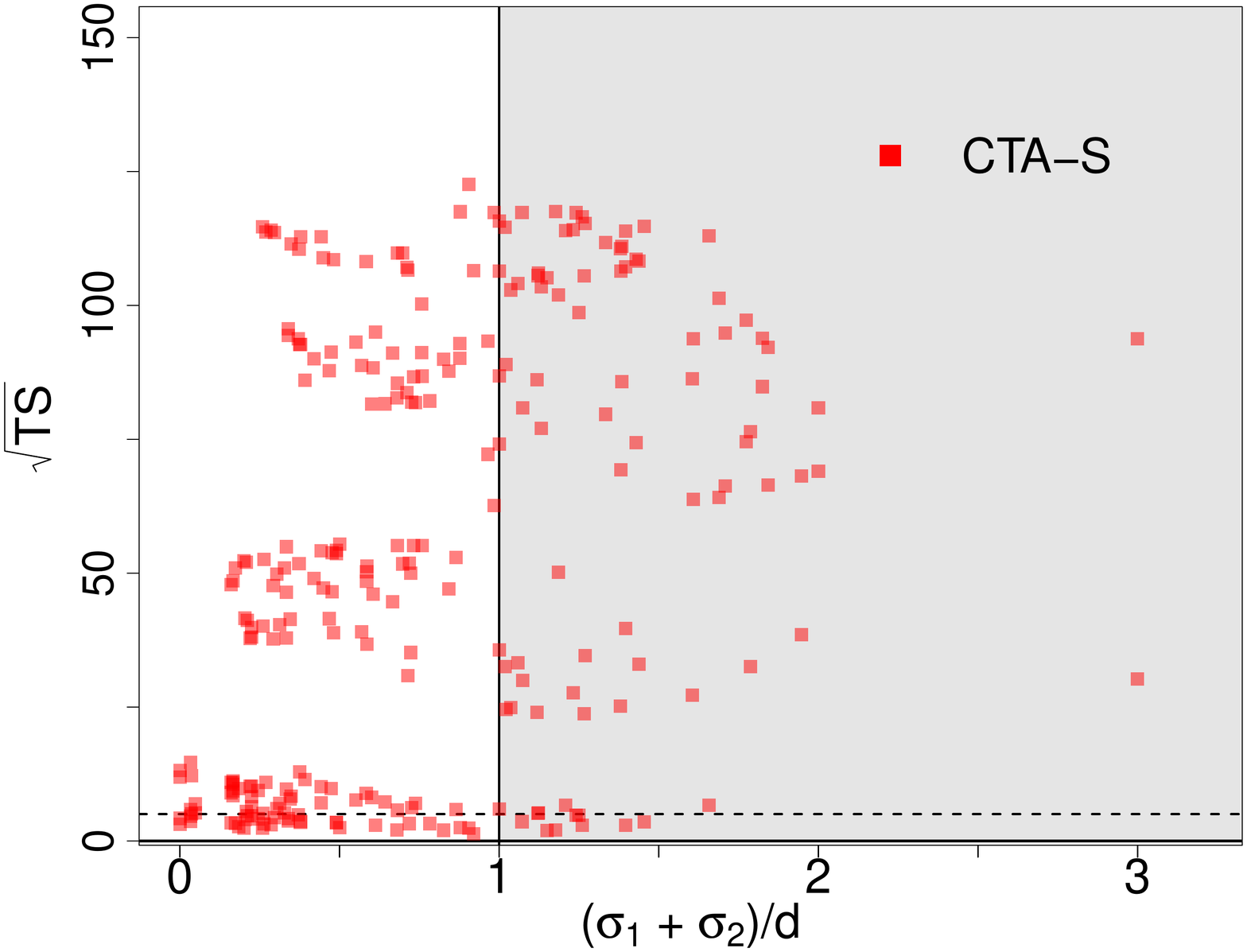}\\
    \includegraphics[width=0.45\linewidth]{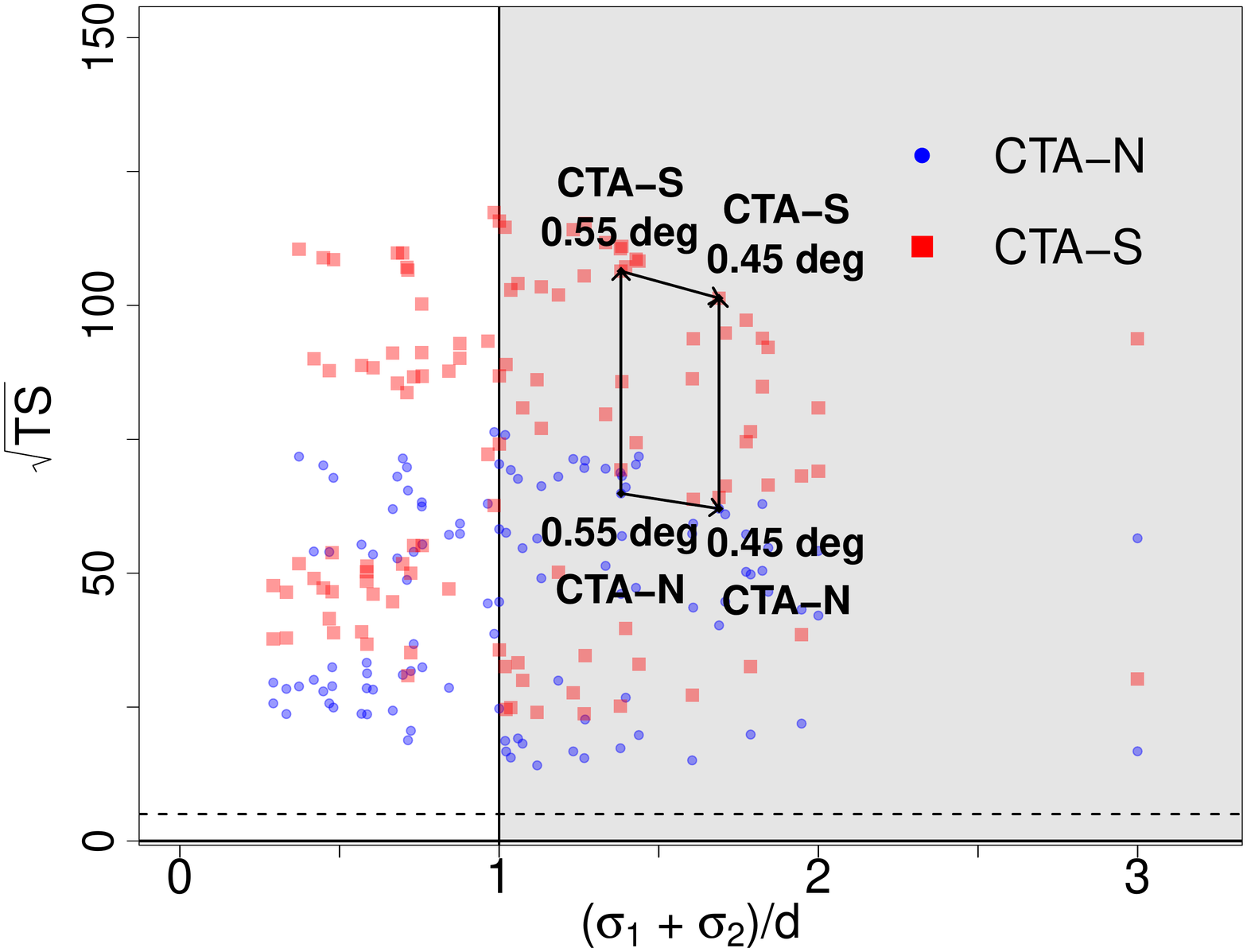}
    %TS_twosrchyp.R
        \caption{Similarly to  Figure \ref{fig:eq1_stat}, the square root of the TS retrieved for the sources from the best-fitting models to the simulations under the hypothesis of two confused sources, i.e., TS$_{\rm 2src,1}$ and TS$_{\rm 2src,2}$ in the notation of Section \ref{sec:simulanals}, compared to the degree of confusion between the sources. We excluded HESS J1554-550 and HESS J1849-000 and the two point-like sources for the lower panel. The dashed horizontal line correspond approximately to a $5\sigma$ detection.}
    \label{fig:two_src_TS}
\end{figure*}

\begin{figure}
    \centering
    \includegraphics[width=\linewidth]{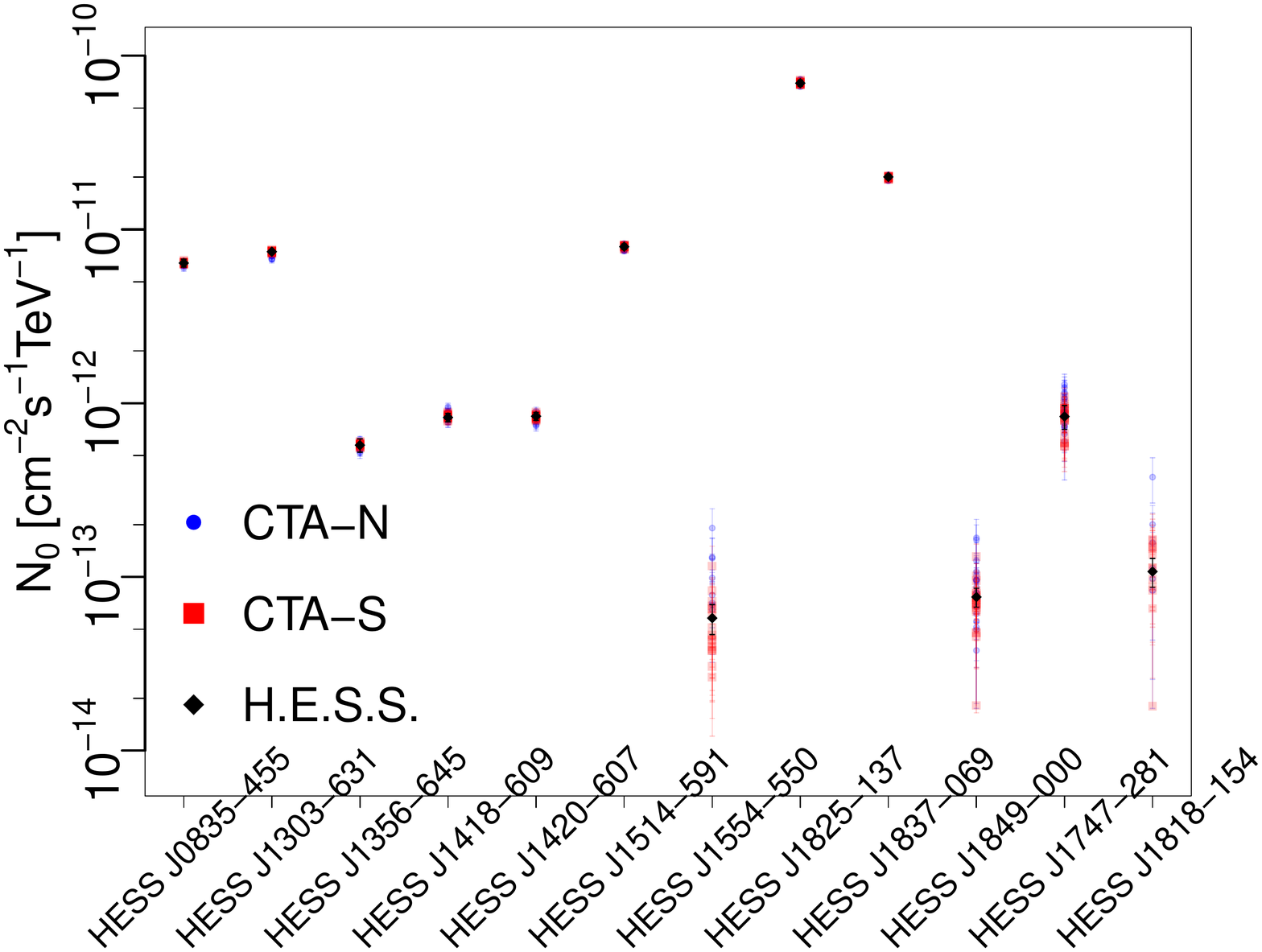}
    \caption{The reconstructed amplitude for the spectrum of each source from the best-fitting models to the simulations under the hypothesis of two confused sources. We exclude the results corresponding to a source detection of $\rm{TS} < 9$. The black diamonds represent the H.E.S.S. measurements in Table \ref{tab:HESS_sources}.}
    \label{fig:amplitudes}
    %amplitudestests.R
\end{figure}

\begin{figure}
    \centering
    \includegraphics[width=\linewidth]{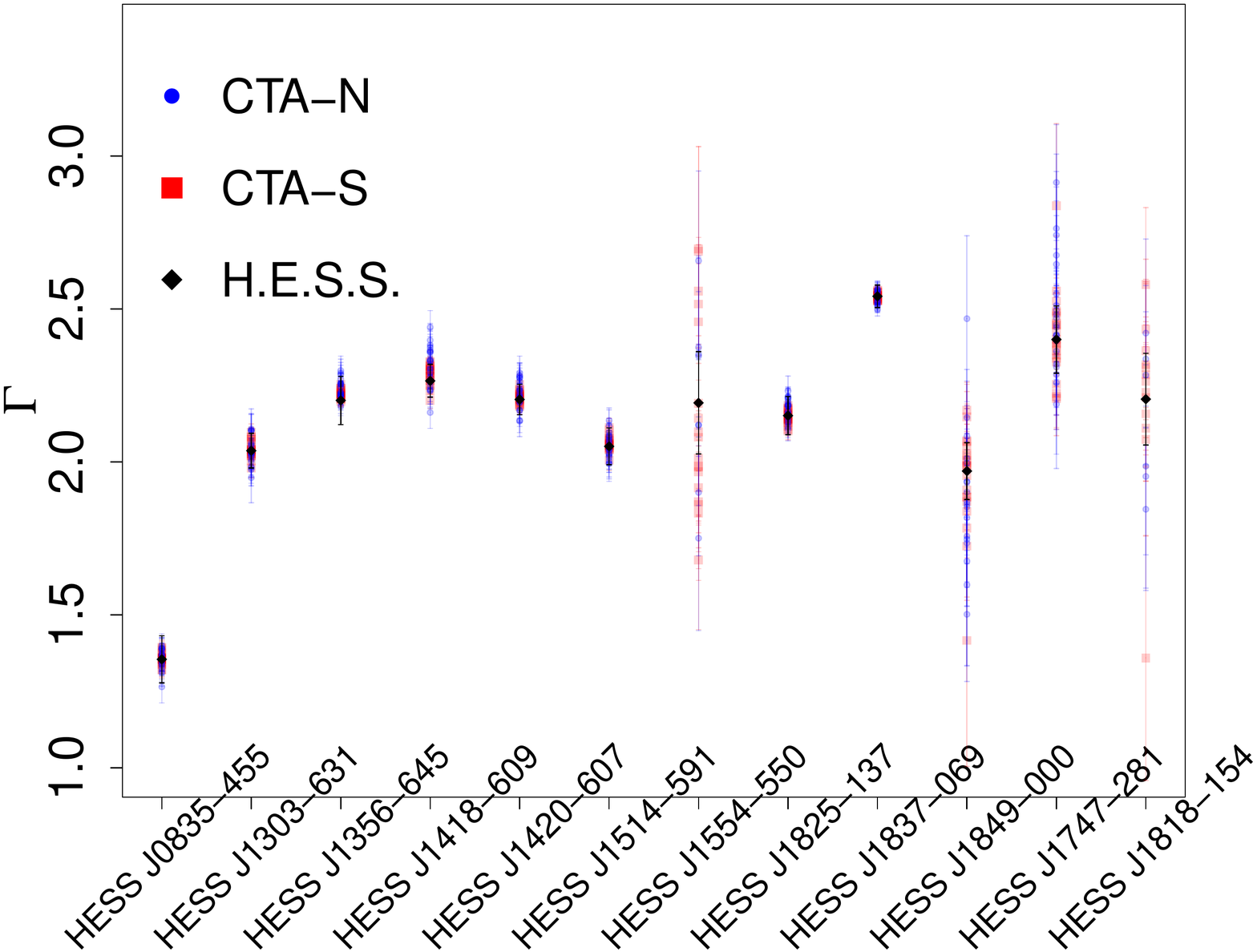}
    \caption{The same as Figure \ref{fig:amplitudes}, but for the spectral index.}
    \label{fig:indices}
    %indextests.R
\end{figure}

\begin{figure}
    \centering
    \includegraphics[width=\linewidth]{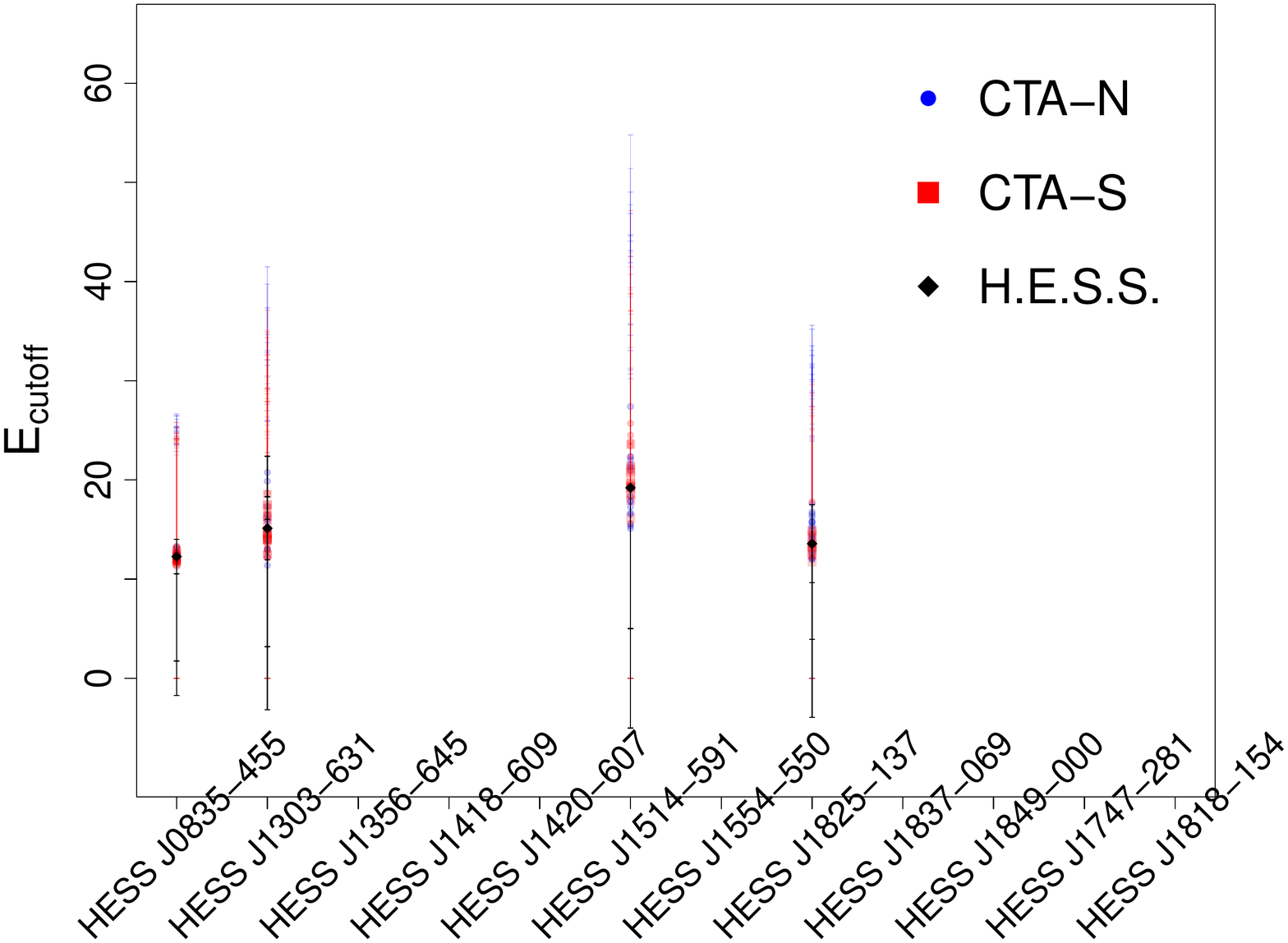}
    \caption{The same as Figure \ref{fig:amplitudes}, but for the cutoff energy.}
    \label{fig:Ecuts}
    %ecutstests.R
\end{figure}

\begin{figure}
    \centering
    \includegraphics[width=\linewidth]{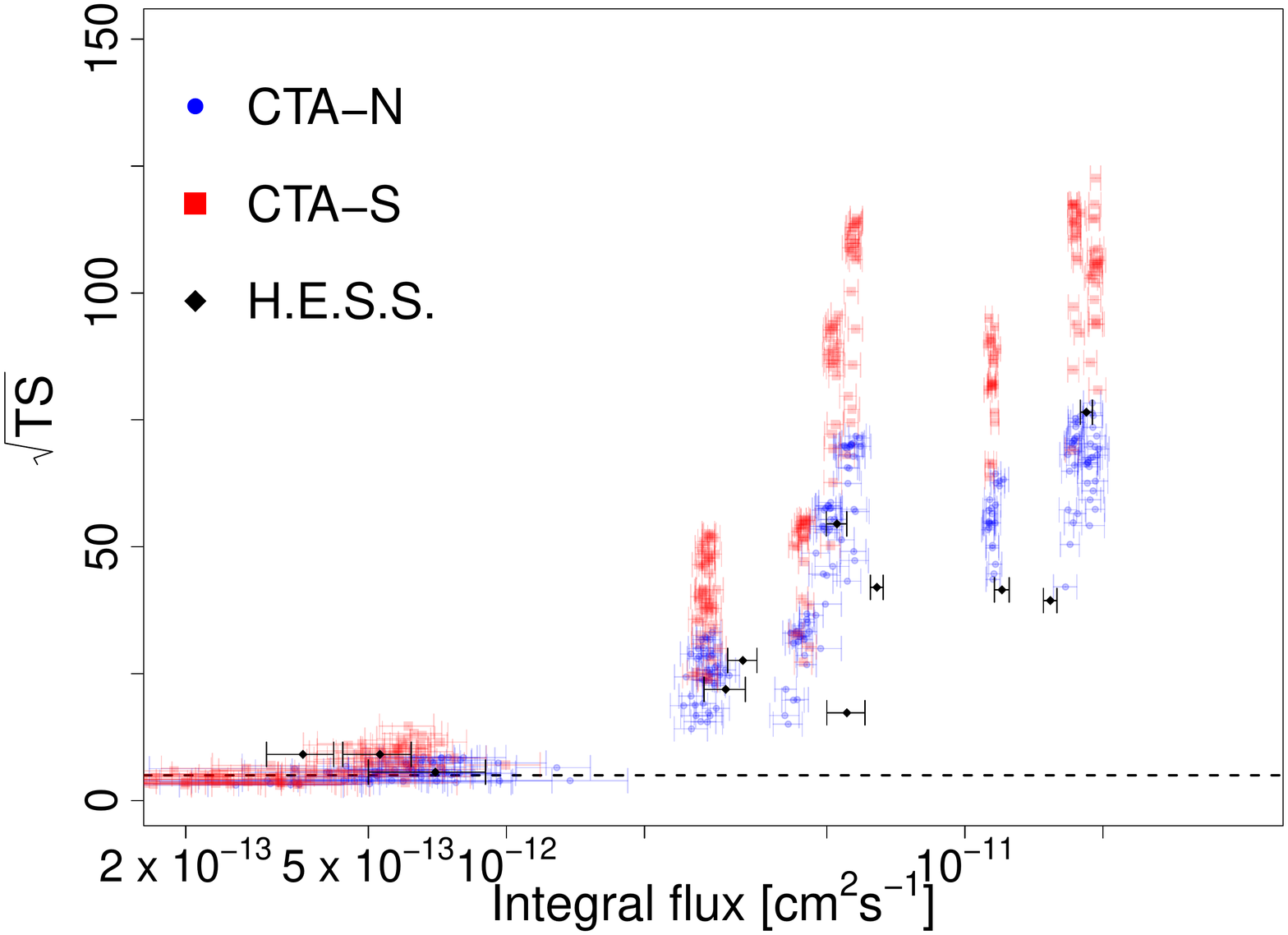}
    \caption{The square root of TS versus the reconstructed integral flux above 1 TeV for each source from the best-fitting models to the simulations under the hypothesis of two confused sources, i.e., TS$_{\rm 2src,1}$ and TS$_{\rm 2src,2}$ in the notation of Section \ref{sec:simulanals}. The black diamonds represent the H.E.S.S. measurements in Table \ref{tab:HESS_sources}.}
    \label{fig:TS_vs_flux}
    %plot_TS_flux.R
\end{figure}

\begin{figure*}
    \centering
    \includegraphics[width=0.498\linewidth]{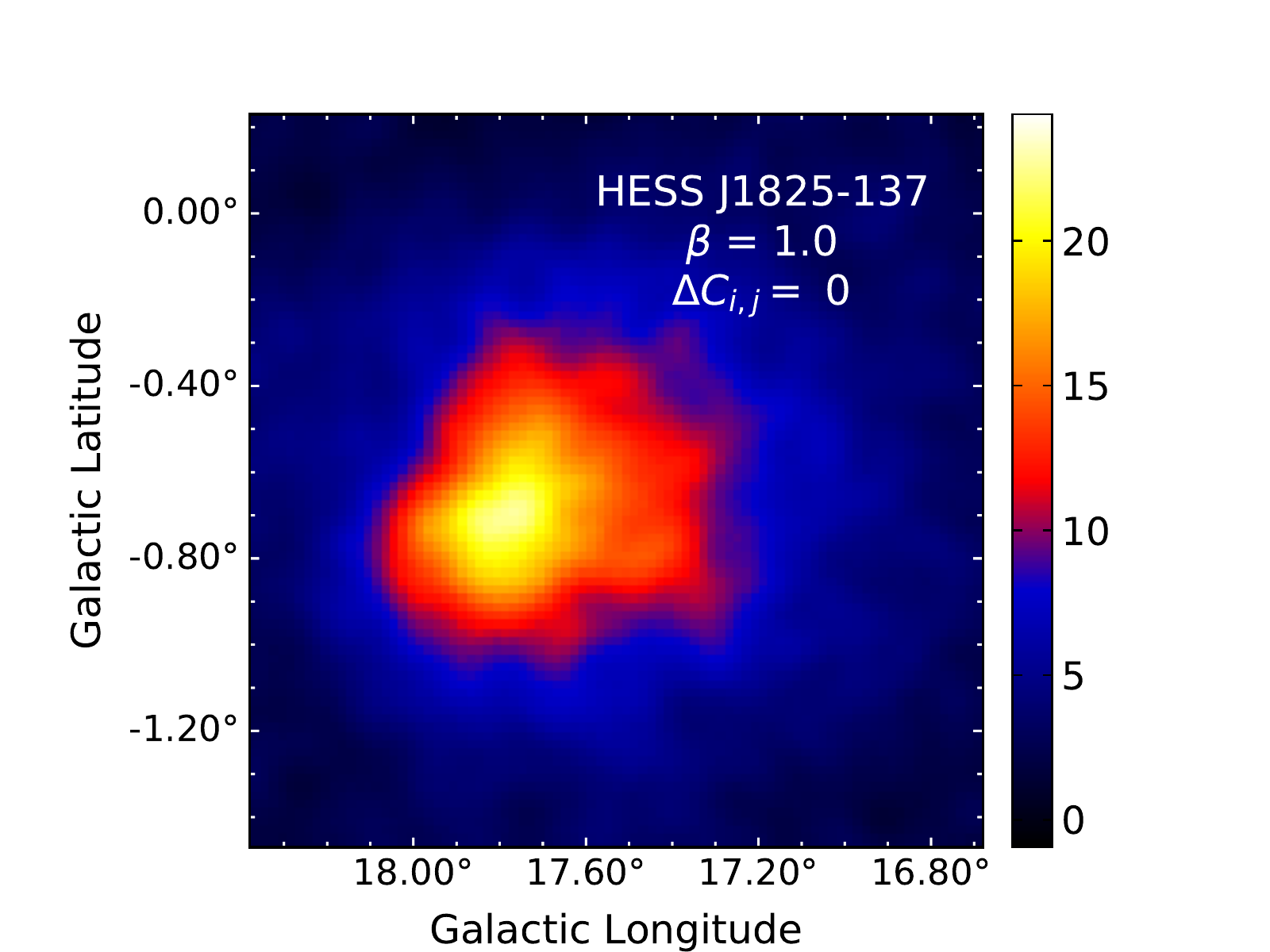}
    \includegraphics[width=0.498\linewidth]{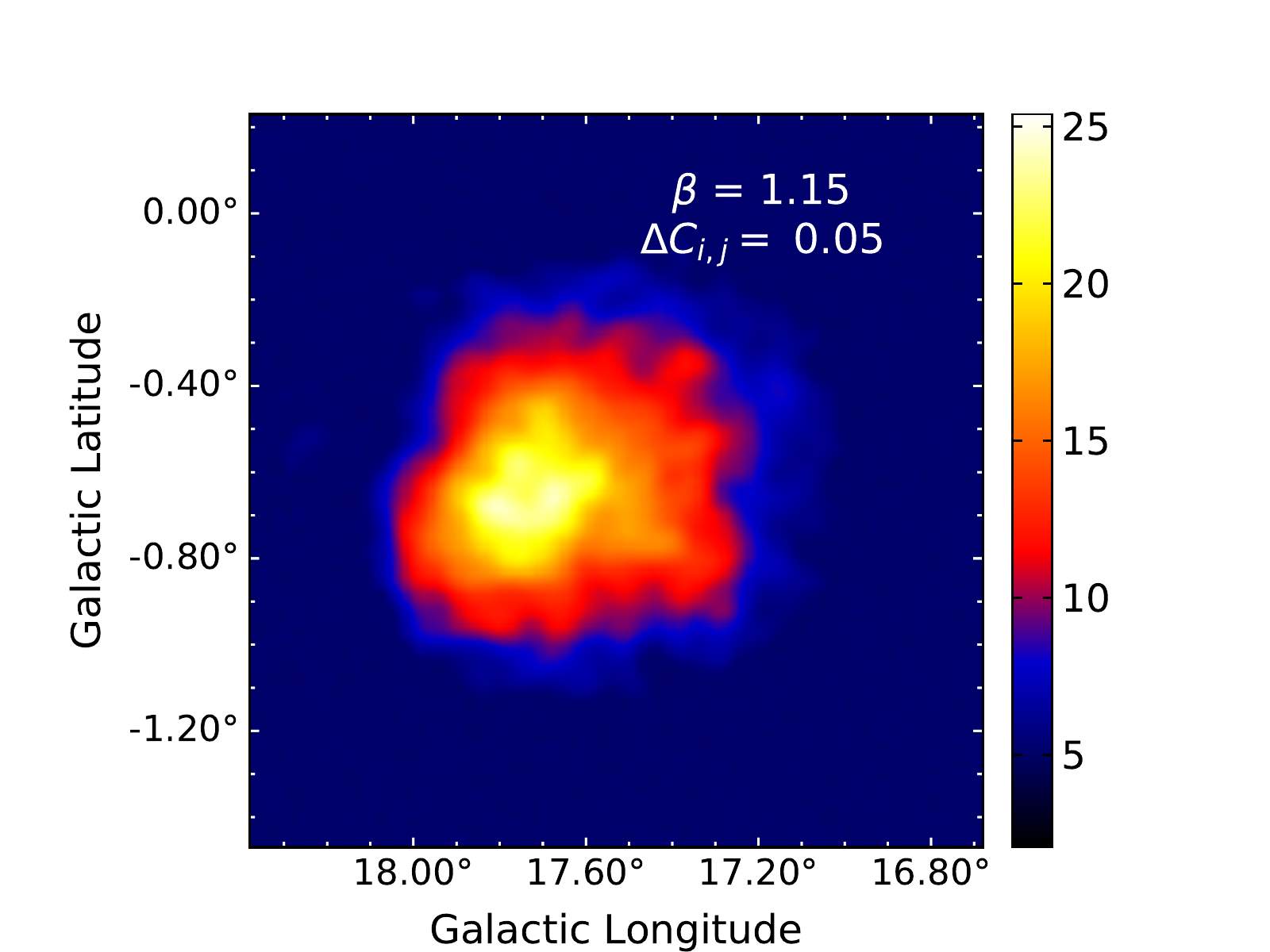}
    \includegraphics[width=0.498\linewidth]{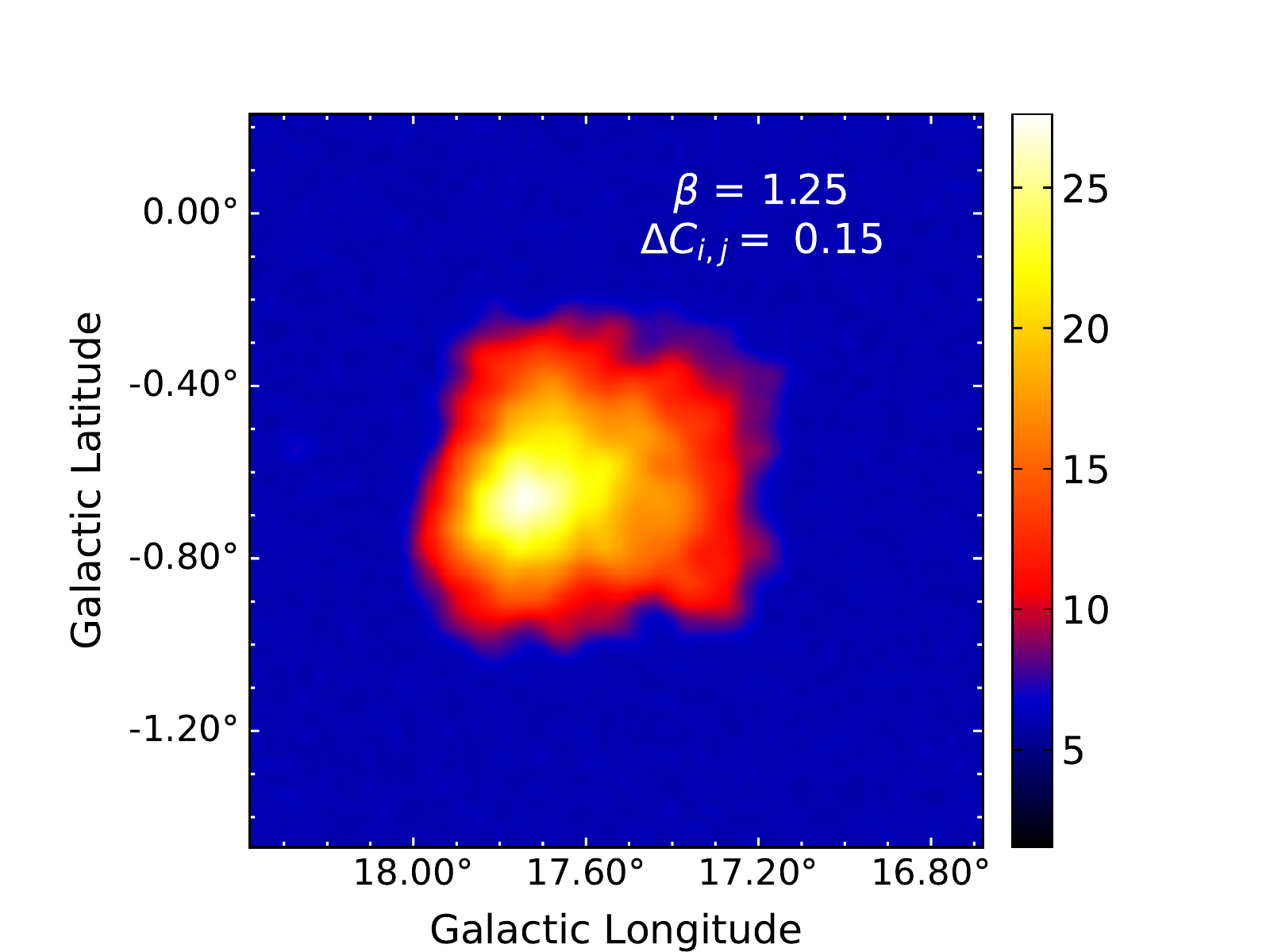} 
    \includegraphics[width=0.498\linewidth]{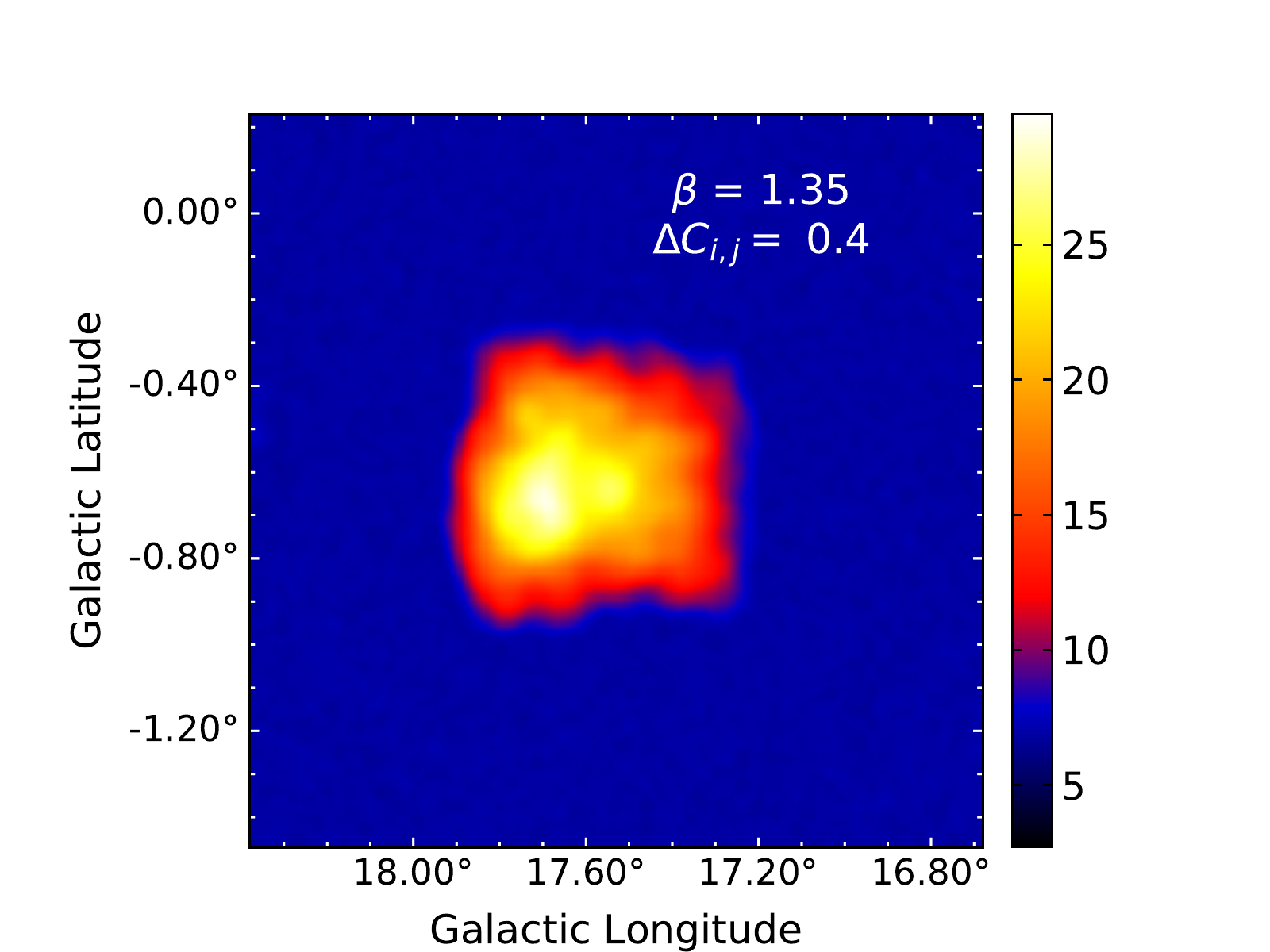}
    \includegraphics[width=0.498\linewidth]{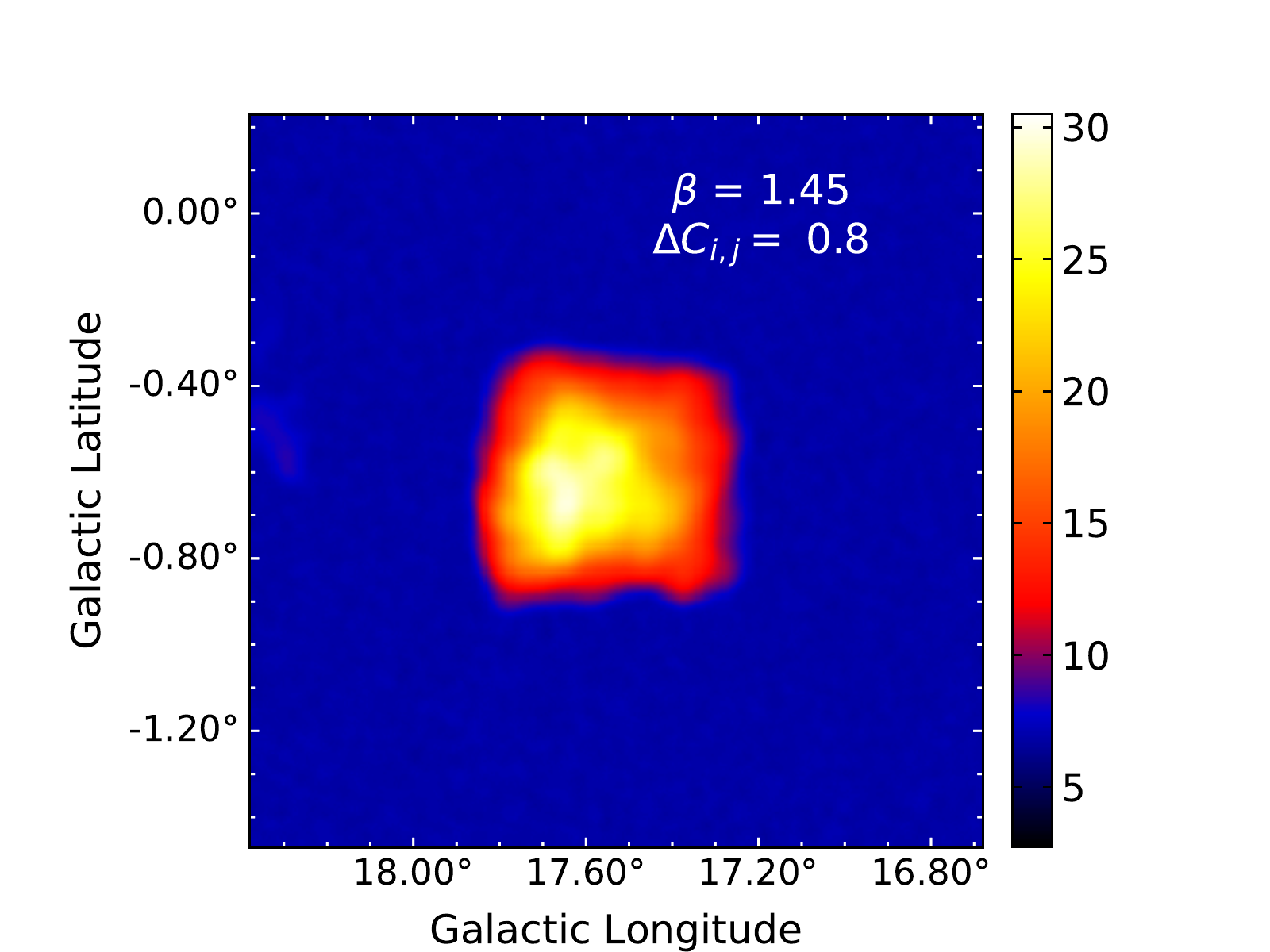}
    \caption{The top left panel depicts an observation simulation of the HESS J1825-137 template blurred with $\beta = 1$, i.e., the softest blurring applied, using 25\,h of CTA-S (in events counts without background subtraction). Similarly, the adjacent panels show observation simulations for various deformations of the HESS J1825-137 template corresponding to different values of the $\beta$ parameter (see Section \ref{sec:autocor}), with Pearson's correlation coefficient relative to the unaltered template ($\Delta C_{i,j}$) noted. The plots have been smoothed with a small Gaussian kernel ($\sigma \sim 0.04\degr{}$) applied.}
    \label{fig:templateblurring}
\end{figure*}

\begin{figure*}
    \includegraphics[width=0.498\linewidth]{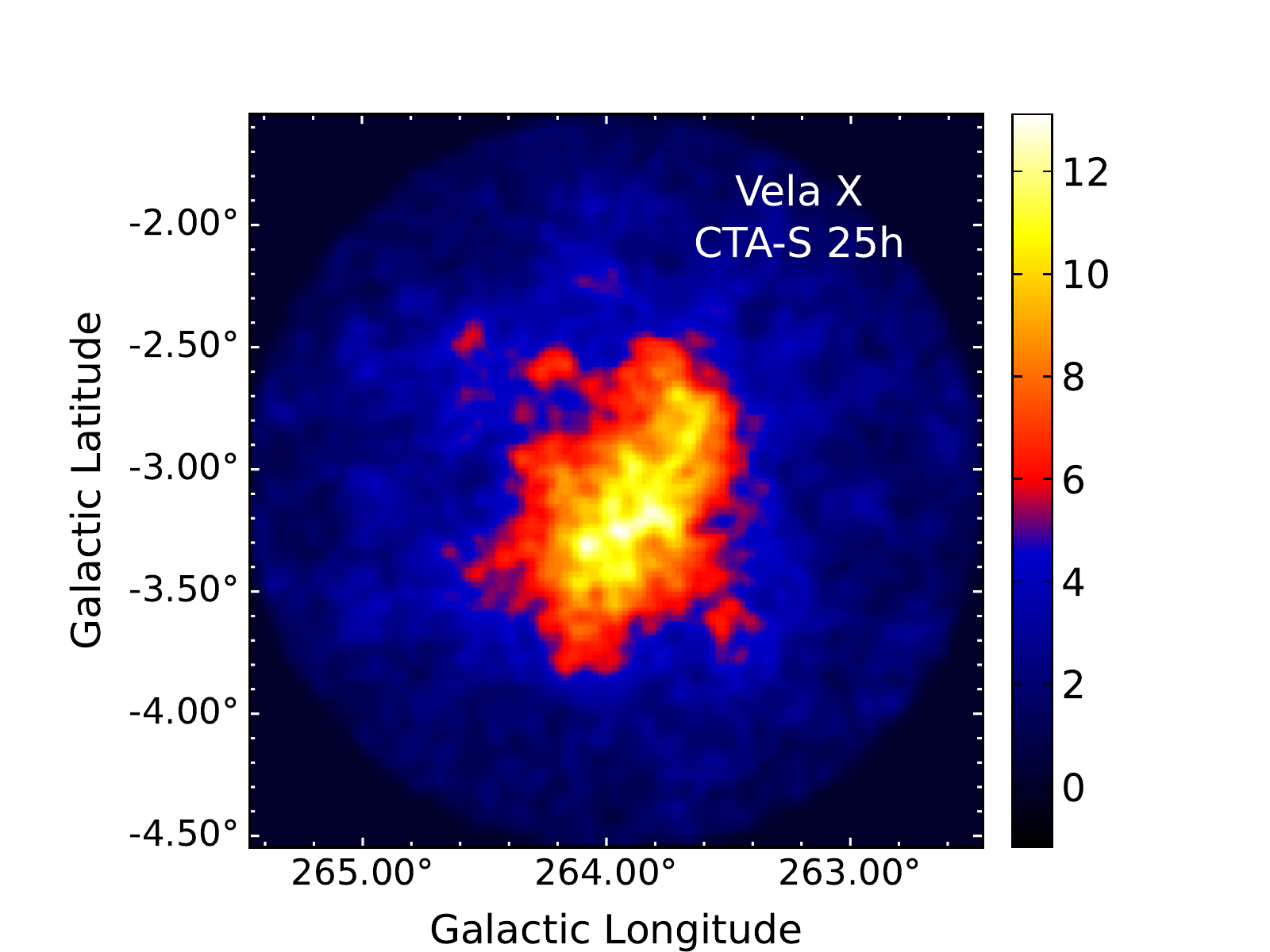}
    \includegraphics[width=0.498\linewidth]{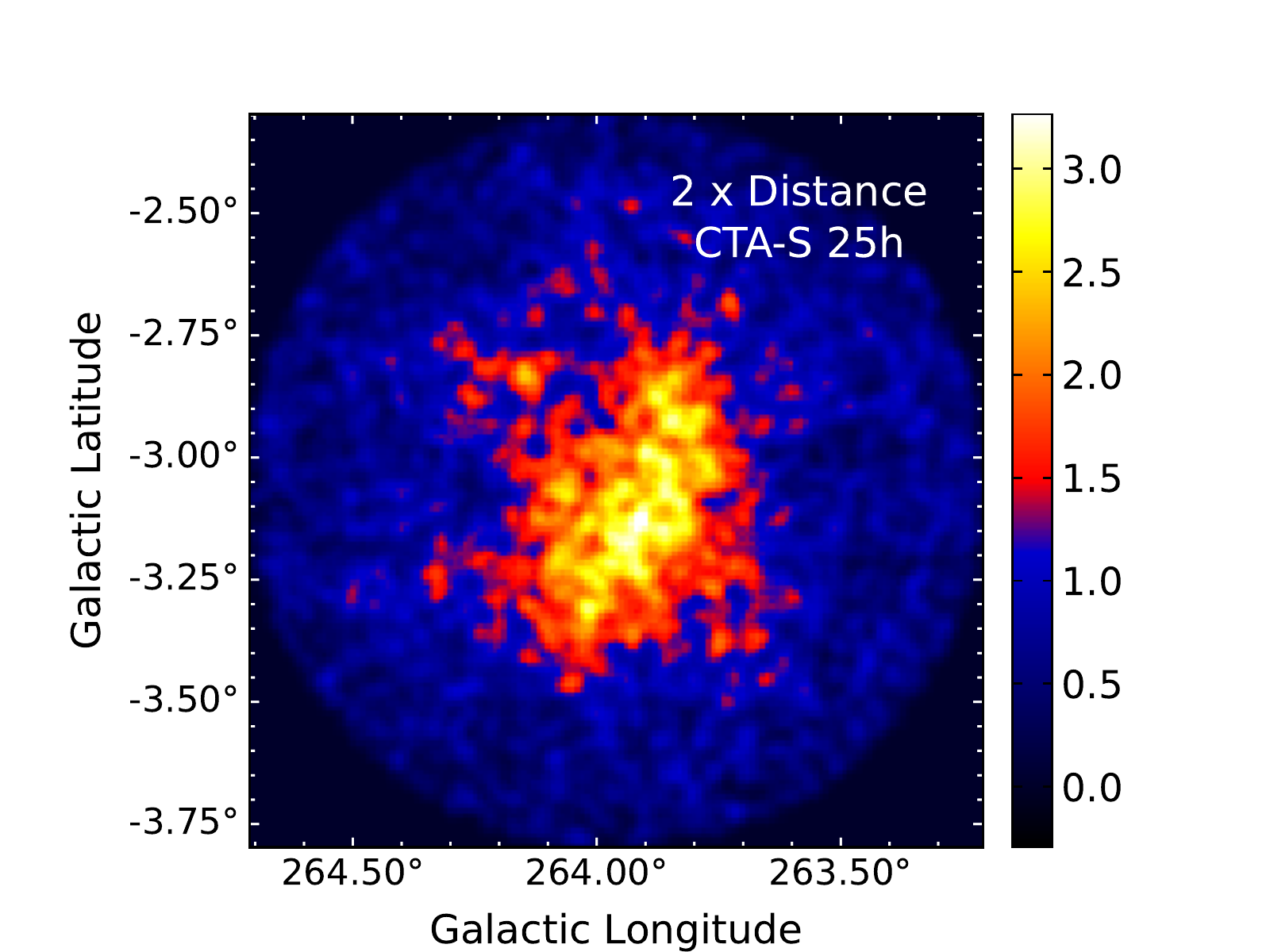}
    \includegraphics[width=0.5\linewidth]{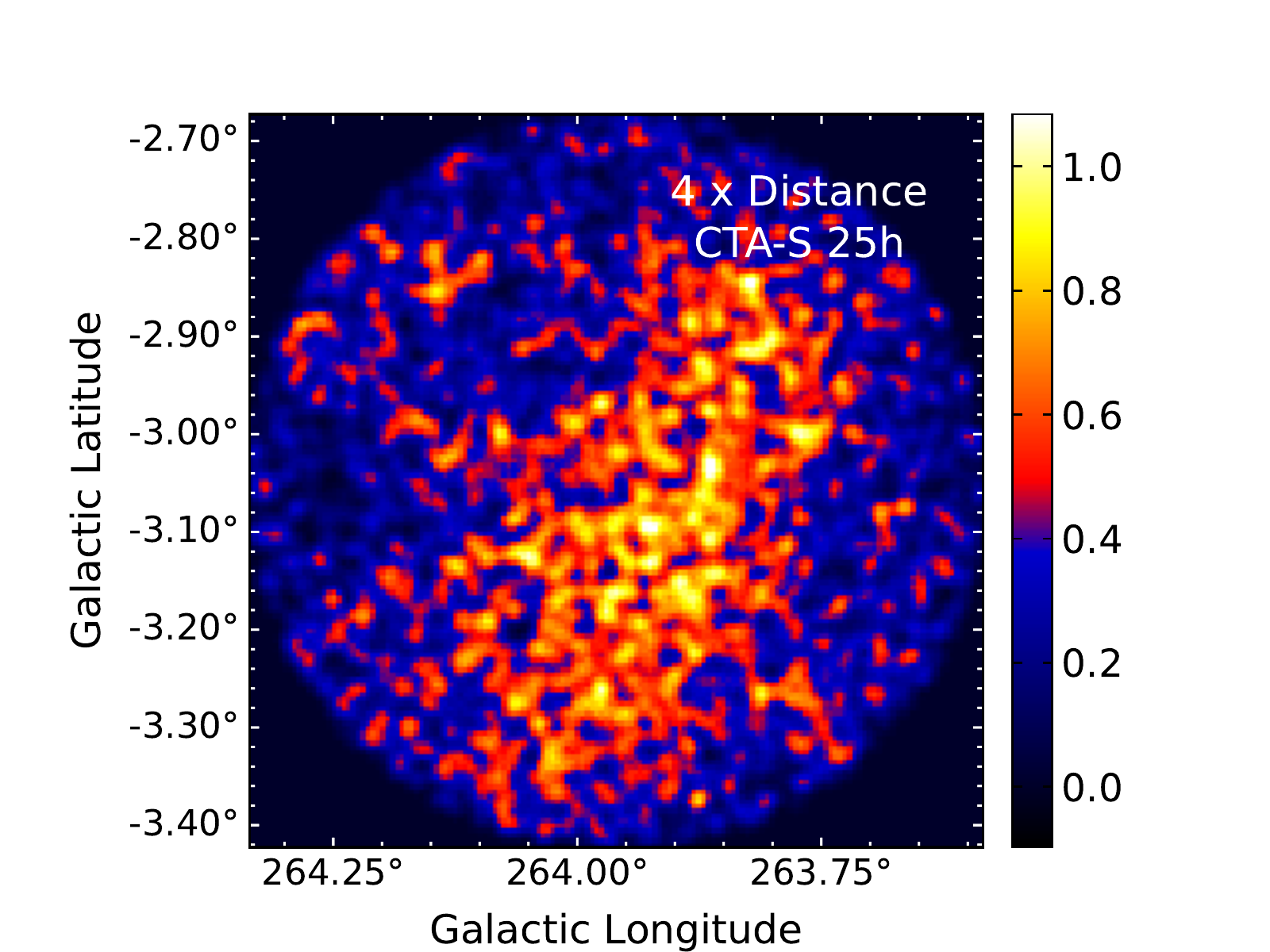}
    \caption{Left panel: sky map resulting from the simulation of the 25-hour Vela X observation (see the right panel in Figure \ref{fig:VelaXmaps}) with the CTA southern array above 0.5 TeV. In the central and right panels, we show the same but with the Vela X template rescaled to place it at two and four times its distance. The plots have been smoothed with a small Gaussian kernel ($\sigma \sim 0.04\degr{}$) applied.}
    \label{fig:VelaX_withdist}
\end{figure*}

\begin{figure*}
    \centering
    \includegraphics[width=0.498\linewidth]{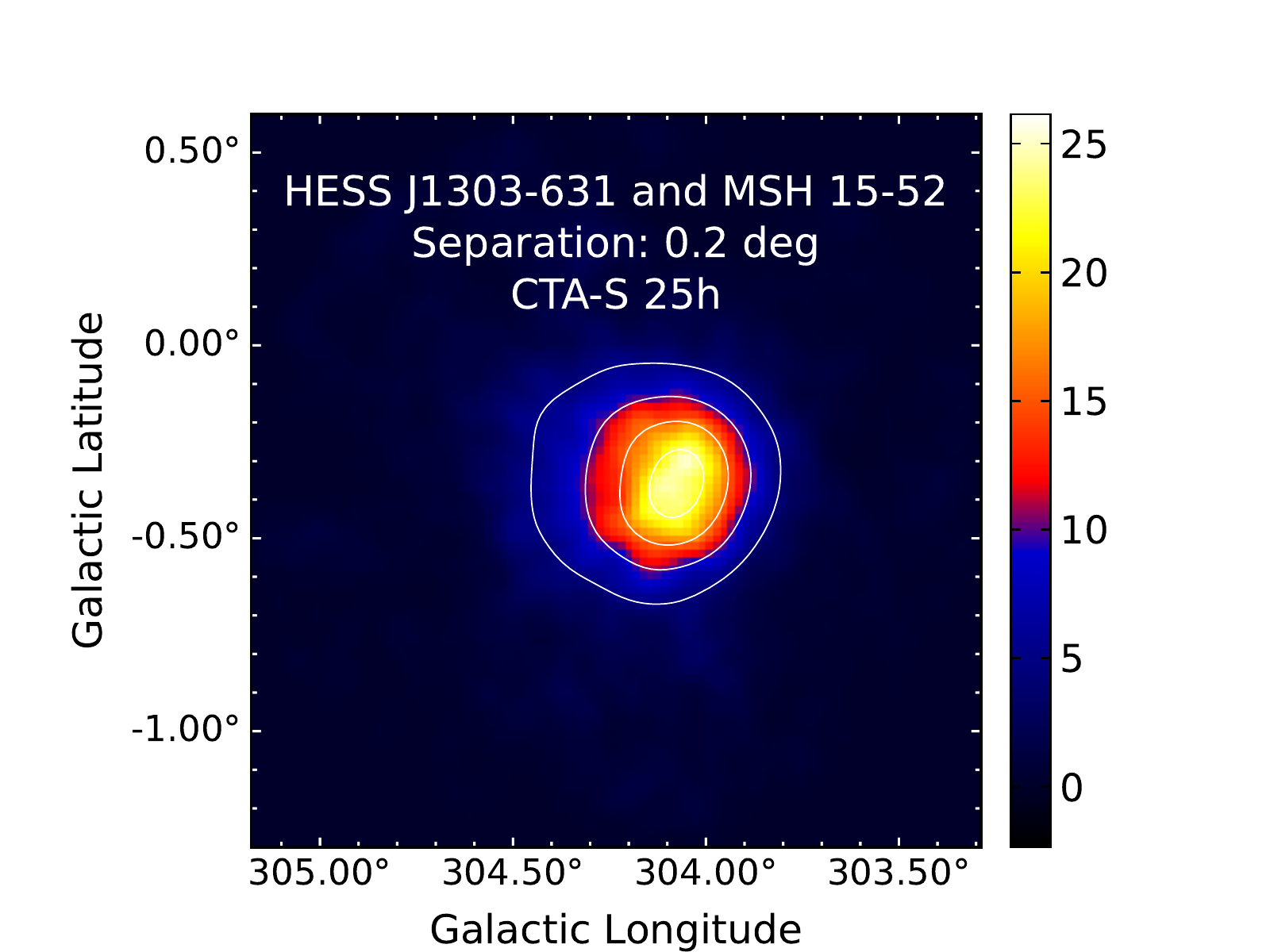} 
    \includegraphics[width=0.498\linewidth]{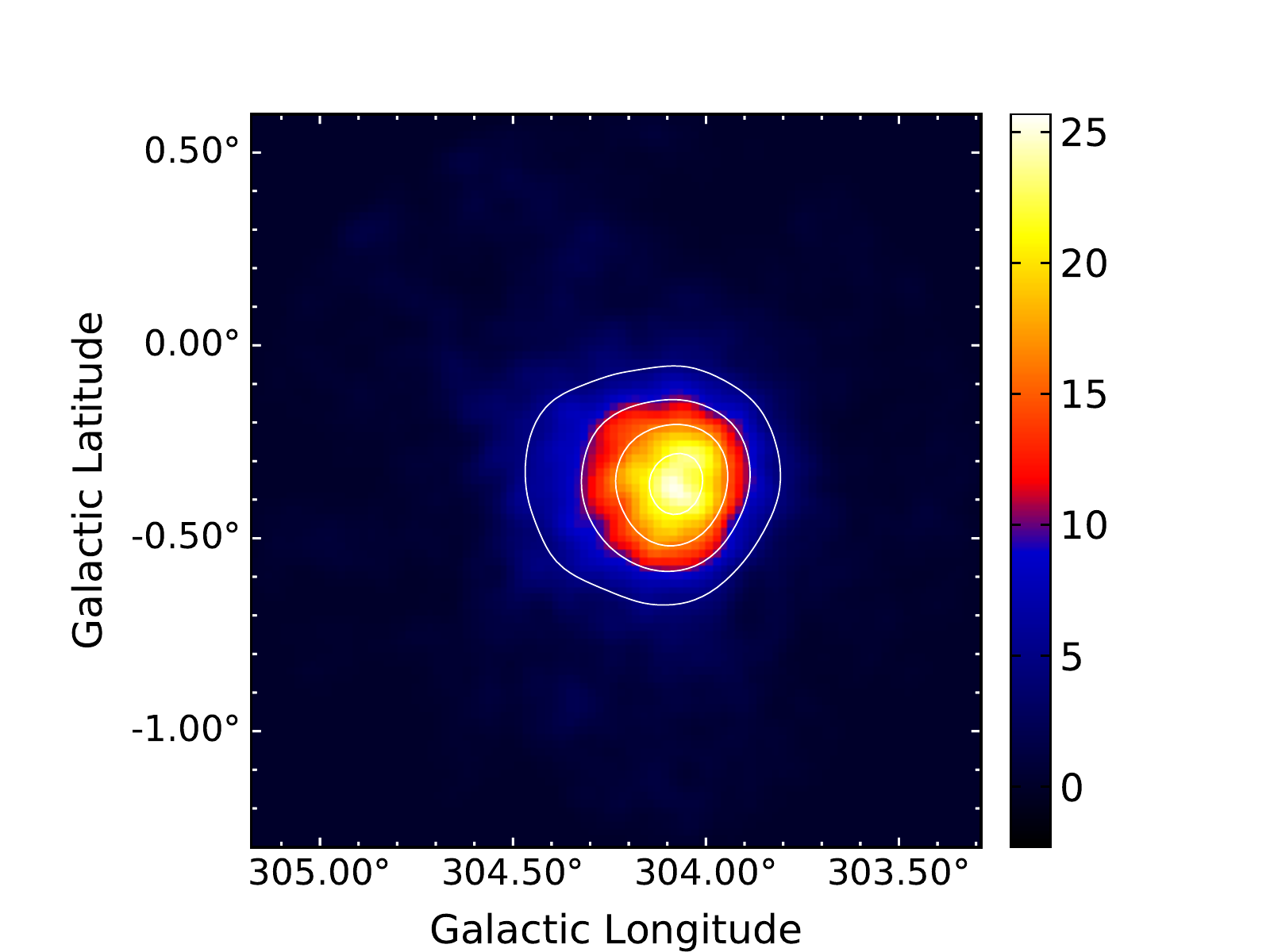}
    \caption{Left panel: a simulation of the HESS J1303-631 and MSH 15-52 templates confused, with $0.2\degr{}$ of separation under 25 h of observations with the CTA southern array (with no background simulated). The simulation is centered at the position of HESS J1303-631. On the right, HESS J1303-631 and MSH 15-52 templates were simulated individually (with no background), and next, the simulations were summed. The white contour lines correspond to 5, 10, 15, 20, and 25 events (counts). The plots have been smoothed with a small ($\sigma \sim 0.04\degr{}$) Gaussian kernel applied.}
    \label{fig:linearity}
\end{figure*}

 This appendix presents different figures and tables to offer a more profound insight into the simulation results and performance. In particular, we illustrate the results for the orientation of the templates (Fig. \ref{fig:angldeg}) and the detection significance (Fig. \ref{fig:two_src_TS}), recovered spectral parameters (Figs. \ref{fig:amplitudes}, \ref{fig:indices}, \ref{fig:Ecuts}), and flux (Fig. \ref{fig:TS_vs_flux}) of the sources. Also, we exemplify the performance of the simulations if blurring the source templates (Fig. \ref{fig:templateblurring}) or modifying the distances to the sources (Fig. \ref{fig:VelaX_withdist}). Figure \ref{fig:linearity} demonstrates the linearity of the simulations (without background). Finally, Table \ref{tab:computtimeconf} summarizes the computation time required for the simulations.  

\begin{figure}
    \centering
    \includegraphics[width=\linewidth]{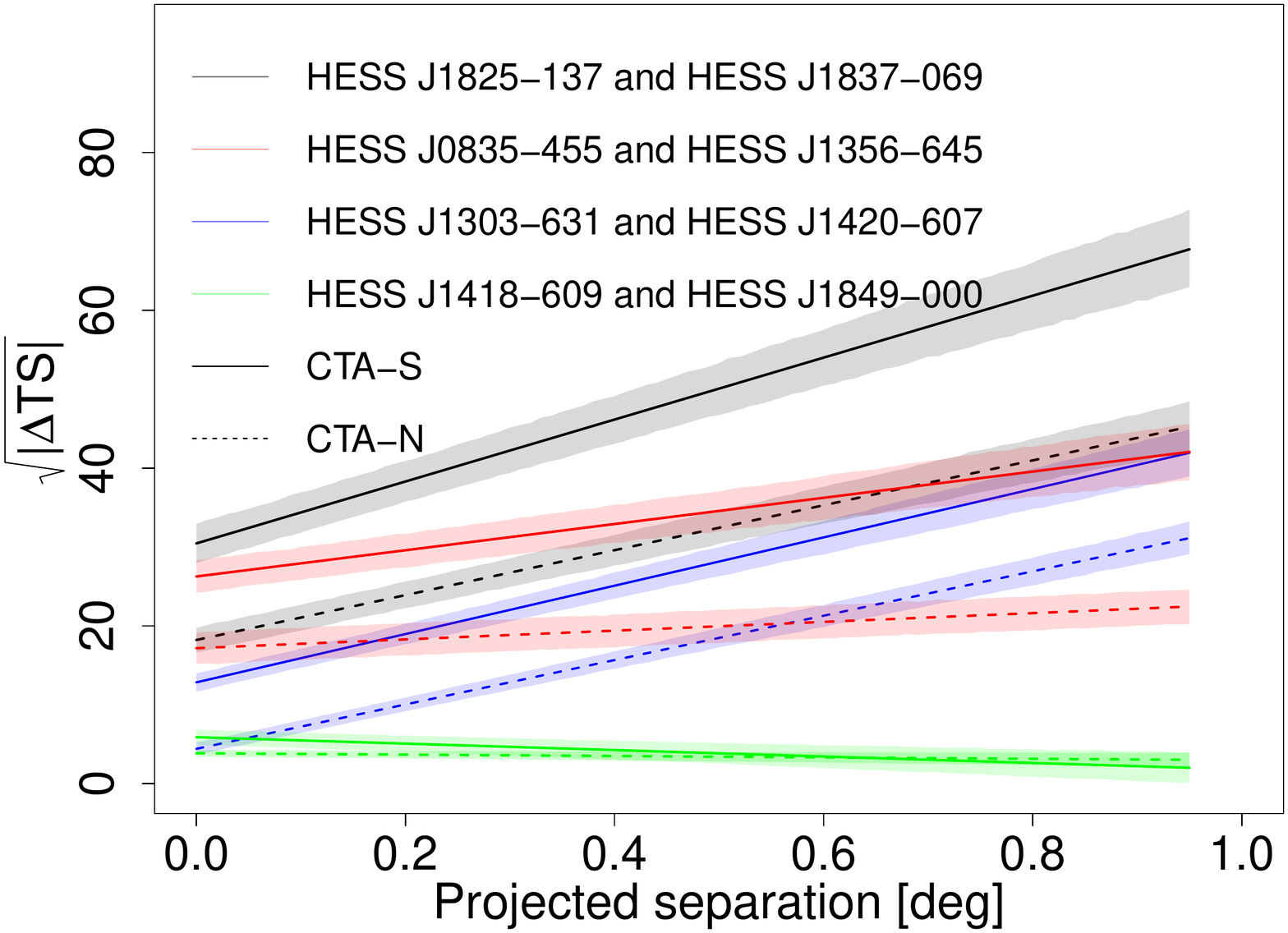}
    \caption{The square root of $\Delta \rm{TS}$ (in absolute value) from Equation \ref{eq:deltaTS} computed for simulations of several artificially confused PWNe placed at different projected separations from each other.}
    \label{fig:sqrtTS_vs_separation}
    %DeltaTS_withseparation.R
\end{figure}

\begin{table}
\caption{The computation time ($t_{\rm{simul}}$) required for a \textsc{ctools} simulation of two sources (from spatial templates) plus background for different input parameters (using a regular commercial laptop)}. The latter are the radius of the field-of-view, observation time ($t_{\rm{obs}}$), energy threshold (E$_{\rm{th}}$), and total flux of the sources above 1 TeV of energy ($\rm{F}_{>1\rm{TeV}}$). The bottom part of the table combine increasing observation times and lower energy thresholds, boosting the cost in computation time.
\centering
\begin{tabular}{lllll}
\hline
Input parameters & & & & Computation time [s] \\
\hline
Radius (FoV) & $t_{\rm{obs}}$ & E$_{\rm{th}}$ & $\rm{F}_{>1\rm{TeV}}$ & $t_{\rm{simul}}$ \\

[deg] & [h] & [TeV] & [$10^{-12}\ \times$\ cm$^{-2}$ s$^{-1}$] & [s] \\
\hline
\hline
1 & 1 & 1 & 1 & $2.24 \pm 0.01$ \\
\hline
2 & 1 & 1 & 1 & $2.24 \pm 0.01$ \\
10 & 1 & 1 & 1 & $2.23 \pm 0.01$ \\
20 & 1 & 1 & 1 & $2.27 \pm 0.04$ \\
\hline
1 & 2 & 1 & 1 & $2.25 \pm 0.02$ \\
1 & 10 & 1 & 1 & $2.38 \pm 0.02$ \\
1 & 100 & 1 & 1 & $3.95 \pm 0.05$ \\
\hline
1 & 1 & 0.5 & 1 & $2.27 \pm 0.01$ \\
1 & 1 & 0.1 & 1 & $2.69 \pm	0.02$ \\
1 & 1 & 0.03 & 1 & $3.60	\pm 0.07$ \\
\hline
1 & 1 & 1 & 2 & $2.26 \pm 0.04$ \\
1 & 1 & 1 & 10 & $2.2 \pm 0.1$ \\
1 & 1 & 1 & 100 & $2.26 \pm 0.05$  \\
\hline
10 & 25 & 0.5 & 2 & $3.45 \pm 0.03$ \\
10 & 25 & 0.1 & 2 & $14.2 \pm 0.5$ \\
10 & 25 & 0.03 & 2 & $34.6 \pm 0.9$ \\
\hline
\hline
\end{tabular}
\label{tab:computtimeconf}
\end{table}

%%%%%%%%%%%%%%%%%%%%%%%%%%%%%%%%%%%%%%%%%%%%%%%%%%

% Don't change these lines
\bsp	% typesetting comment
\label{lastpage}
\end{document}